\documentclass[letterpaper,12pt]{article}
\usepackage[margin=1in]{geometry}
\usepackage[english]{babel} 
\usepackage{graphicx, lipsum, textcomp, wrapfig, float} 
\usepackage{hyperref} 
\usepackage[utf8]{inputenc} 
\usepackage{chemfig} 
\usepackage[version=4]{mhchem} 
\usepackage{advdate, datenumber} 
\usepackage[nomessages]{fp} 
\usepackage{amsmath} 
\usepackage{amsfonts}
\usepackage{listings} 
\usepackage[style=ieee, isbn=false]{biblatex}
\usepackage{multirow}
\usepackage{comment}
\usepackage{longtable}

\usepackage{caption}
\usepackage{tocbasic}
\DeclareTOCStyleEntry[
  beforeskip=.2em plus 1pt,
  pagenumberformat=\textbf
]{tocline}{section}

\usepackage{abstract}
\usepackage{titling}

\definecolor{mypurple}{RGB}{140,54,140}
\definecolor{homered}{RGB}{127, 0, 10}
\definecolor{officeorange}{RGB}{204, 75, 0}
\definecolor{mauroblue}{RGB}{53, 48, 217}
\definecolor{citegreen}{RGB}{15, 133, 13}
\hypersetup{
    colorlinks=true,
    linkcolor=blue,
    filecolor=mypurple,      
    urlcolor=teal,
    citecolor=citegreen
}

\title{
\textbf{
Extensible Structure-Informed Prediction of Formation Energy with Improved Accuracy and Usability employing Neural Networks
}
}
\author{
\textbf{Adam M. Krajewski\textsuperscript{a}}, Jonathan W. Siegel \textsuperscript{b}\\ Jinchao Xu \textsuperscript{b}, Zi-Kui Liu\textsuperscript{a}\\
 \footnotesize{
 a. Department of Materials Science and Engineering, The Pennsylvania State University, USA}\\
 \footnotesize{b. Department of Mathematics, The Pennsylvania State University, USA}
 \vspace{6pt}\\
 \footnotesize{Corresponding author: Adam M. Krajewski, 216-456-1534, ak@psu.edu}
}

\renewcommand{\maketitlehookd}{%
\begin{abstract}
    
    In the present paper, we introduce a new neural network-based tool for the prediction of formation energies of atomic structures based on elemental and structural features of Voronoi-tessellated materials. We provide a concise overview of the connection between the machine learning and the true material-property relationship, how to improve the generalization accuracy by reducing overfitting, how new data can be incorporated into the model to tune it to a specific material system, and preliminary results on using models to preform local structure relaxations.
    
    The present work resulted in three final models optimized for (1) highest test accuracy on the Open Quantum Materials Database (OQMD), (2) performance in the discovery of new materials, and (3) performance at a low computational cost. On a test set of 21,800 compounds randomly selected from OQMD, they achieve a mean absolute error (MAE) of 28, 40, and 42 meV/atom, respectively. The second model provides better predictions in a test case of interest not present in the OQMD, while the third reduces the computational cost by a factor of 8.
    
    We collect our results in a new open-source tool called SIPFENN (Structure-Informed Prediction of Formation Energy using Neural Networks). SIPFENN not only improves the accuracy beyond existing models but also ships in a ready-to-use form with pre-trained neural networks and a GUI  interface. By virtue of this, it can be included in DFT calculations routines at nearly no cost.
    
    \rule{0.88\textwidth}{0.5pt}
\end{abstract}
}

\begin{document}

\maketitle

\tableofcontents

\pagebreak
\section{Introduction} \label{sec:Introduction}

\label{ssec:Motivation}
In recent years the field of material data informatics has been growing in importance thanks to the proliferation of open-access databases \cite{Saal2013MaterialsOQMD,Kirklin2015TheEnergies, vandeWalle2018TheDatabase,Jain2013Commentary:Innovation,Curtarolo2013AFLOW:Discovery,Toher2018TheDiscovery,Pizzi2016AiiDA:Science} and new methods being implemented to predict a wide variety of material properties \cite{Isayev2017UniversalCrystals, Legrain2017HowSolids, Pilania1987MachineSuperlattices, Jung2019BayesianSteels, Ouyang2020ComputationalConductors,Bucior2019Energy-based,Chandrasekaran2019SolvingLearning, Kim2018Machine-learning-acceleratedCompounds,Wen2019MachineProperty, Scime2019UsingProcess}. Within these methods, machine learning (ML) and, more broadly, artificial intelligence (AI) is becoming dominant, as noted in two recent reviews \cite{Schmidt2019RecentScience, Vasudevan2019MaterialsPhysics}, which listed a total of around 100 recent studies that attempted to solve material science problems using ML and AI techniques. These studies report benefits such as a 30-fold increase in material discovery rate when guided by an ML-model \cite{Kim2018Machine-learning-acceleratedCompounds}, or the ability to create new state-of-the-art materials in highly complex design spaces like 6-component alloys \cite{Wen2019MachineProperty}. They also dive into new paradigms of materials science by handling previously unthinkable amounts of data, allowing the creation and analysis of an energy convex-hull calculated for all elements \cite{Aykol2019NetworkDiscovery, I.Hegde2020TheMaterials}, or a concurrent analysis of all available literature texts to find paths for material synthesis \cite{Kononova2019Text-minedRecipes}. In addition, some studies promise to solve significant industrial challenges such as detection of additive manufacturing flaws with relatively simple and accessible data, but above-human pattern recognition quality and speed \cite{Scime2019UsingProcess}.

A common approach is to  focus on the discovery of candidate materials promising a new state-of-the-art performance, which must then be validated by experiment. The mismatch between the predictions and experiment measures the quality of the model, and reducing this gap is a major challenge due to the newly designed materials often being far from known materials, combined with attention placed on regions with extraordinary predictions.However, even if design models were perfectly accurate, many predicted materials cannot be physically made in the lab. An increasing number of studies attempt to solve this challenge by focusing not only on predicting how the material will perform but also on whether it can be manufactured \cite{Alberi2019TheRoadmap}. Generally, these include predicting materials' stability \cite{Balachandran2018PredictionsTheory, Li2019ThermodynamicLearning, I.Hegde2020TheMaterials, Im2022ThermodynamicModeling, Shang2021FormingJoints} and synthesizability \cite{Hattrick-Simpers2018AMaterials,Kononova2019Text-minedRecipes, Aykol2019NetworkDiscovery} with the stability being the more constraining parameter, as it determines whether the material could be stable or metastable in the use conditions, and therefore whether it can be synthesizable. Thus, predicting stability through prediction of fundamental thermodynamic properties such as formation energy is of special importance.

In the present work, new ML models and a tool to quickly use them are developed to improve the process of materials discovery by efficient prediction the formation energy and streamlined incorporation into materials discovery frameworks that aim to screen billions rather than hundreds of candidates available with cost-intensive calculations like first-principles calculations based on the density functional theory (DFT). 

\label{ssec:currentapproach}
In simple terms, every ML model is composed of three essential elements: a database, a descriptor, and an ML technique (also known as ML algorithm). The first element, databases, contain prior knowledge and are becoming increasingly shared between many studies, thanks to being open-access and often containing orders of magnitude more experimental or computational data than could be feasibly collected for a single study \cite{Saal2013MaterialsOQMD,Kirklin2015TheEnergies, vandeWalle2018TheDatabase,Jain2013Commentary:Innovation,Curtarolo2013AFLOW:Discovery,Toher2018TheDiscovery,Pizzi2016AiiDA:Science}. Databases used within the present paper are detailed in Section \ref{sssec:Data}.

The second element of an ML model is the descriptor (i.e., feature vector describing the material) which determines a representation of knowledge (data from the database) in a way relevant to the problem. It is typically built from many features, also known as attributes or vector components, which usually are determined through domain knowledge to be relevant or selected through correlation analysis. All combined, these features are a representation of some state whose meaning will be problem-specific.

When treating materials on the single atomic configuration level, descriptors can be generally divided into composition-based (also known as stoichiometric, structure-invariant, or elemental) \cite{Jha2018ElemNet:Composition, Ward2016AMaterials, , Legrain2017HowSolids} and structure-informed \cite{Ward2017IncludingTessellations, Seko2017RepresentationProperties,Schutt2014HowProperties}. The first type usually provides a more compact representation at a much lower computational cost, as calculating a composition-based descriptor often needs to involve only simple linear algebra operations such as matrix multiplication \cite{Ward2016AMaterials}, or prior-knowledge-incorporating attention-based analysis of a graph representation of the composition \cite{Goodall2020PredictingStoichiometry}. In cases where deep neural networks (DNNs) are employed, descriptor calculation can be skipped altogether by passing a composition vector directly \cite{Jha2018ElemNet:Composition}.

It is important to recognize that the descriptor choice impacts both the performance and applicability of the model. In the case of prediction of material properties, such as formation energy, selecting a composition-based descriptor, no matter how complex, limits the model to either a specific arrangement of atoms, such as BCC or amorphous, or some defined pattern of structures, such as the convex hull of lowest-energy structures. Such limitation of the problem domain, given a comparable amount of data, allows to quickly achieve much lower prediction error at a cost of fundamentally changing the problem, making a comparison between methods impossible. Furthermore, a composition-only representation is inherently unsuitable for the direct prediction of most material properties that depend on the atomic structure. The structure-informed descriptors can include much more information related to interatomic interactions, making them more robust and more physics-relevant. They also, implicitly or explicitly, include symmetries present in the material, which can be used to predict certain properties, such as zero piezoelectric response, with high confidence. Furthermore, such descriptors often include extensive composition-based arguments within them \cite{Ward2017IncludingTessellations}, making it possible to both recognize patterns in the property coming from different chemical species occupying the same structure and structural effects in the case of a single composition.

At the same time, it is important to consider that physically existing materials are rarely described by a single atomic configuration, usually requiring considerations for defects and coexisting configurations. Thus, like a traditional DFT-based modeling, in order to reproduce real material behavior, a structure-informed model will often require utilization of a method such as CALPHAD \cite{Kaufman1970ComputerMetals, Liu2018OceanLearning}. One of such methods, recently developed by authors and named "zentropy theory" shows the potential to connect individual configurations to predict macroscopic properties, such as colossal positive and negative thermal expansions \cite{Liu2021ZentropyExpansions}.

In some cases however, investigating all configurations can be a very challenging task (e.g., for high entropy alloys), necessitating the use of an elemental-only model trained to give predictions assuming future observations to be consistent with the past ones \cite{Debnath2021GenerativeAlloys}. 

\label{ssec:specificapproach}

The structure-informed representation which was the ground for the present work has been developed by Ward et al. based on information from the Voronoi tesselation of a crystal structure \cite{Ward2017IncludingTessellations}. Ward's descriptor contains 271 features that combine information from elemental properties of atoms, such as shell occurrences, with information about the their local environments, such as coordination number or bond lengths to neighbours. This approach was demonstrated to work excellently when comprehensively compared to two previous approaches based on the Coulomb matrix (CM) \cite{Schutt2014HowProperties} and on the partial radial distribution function (PRDF) \cite{Seko2017RepresentationProperties}, when trained on the same data from the Open Quantum Materials Database (OQMD) and with the same machine learning algorithm. A more detailed overview is given in \ref{ssec:descriptorused}.

Ward et al. used an automated Random Forest ML algorithm \cite{Ward2017IncludingTessellations} set to a fully automatic parameter selection. While fairly common, that approach without complexity limit for the model, and when trained on over 400,000 materials, resulted in a forest composed of 100 trees with approximately 700,000 nodes each. Such model requires over 27 GB of RAM memory to run, making it unusable on a typical personal or lab computer. Such size also results in a relatively low efficiency, requiring over 100 ms to run on a high-performance lab computer \cite{Ward2017IncludingTessellations}.

In the present work, aforementioned issues are resolved through a targeted design of the ML algorithm to fully utilize the data and its representation. This is done by consideration of the problem formulation and the deep neural network technique (see \ref{ref:machinelearningoverview}), combined with iterative model design (see \ref{sssec:NetDesign}), and by designing and testing over 50 neural networks belonging to around 30 designed architectures. Notably, in the time between Ward's work and the present paper, neural networks have been used in this application, e.g., \cite{Jha2019IRNet}, which uses residual neural networks. However, as we show in Section \ref{ssec:oqmdperformance}, the present paper provides more accurate predictions than both Ward's model and the state-of-the-art neural network model \cite{Jha2019IRNet}.

Additionally, the present work brings two further improvements. The first one is good transfer learning ability, described in \ref{ssec:transferlearningresults} allowing other researchers, at a relatively small cost, to adjust the model to small problem-specific databases, typically consisting of tens of DFT calculations or less. This method substantially improves predictions for similar materials while retaining the general knowledge learnt from the large data set and demonstrates that the model learns features related to underlying physics. The second  improvement is the end-user usability. While most of the materials-related ML model are reported in a reproducible way with an evaluation of the performance \cite{Ward2017IncludingTessellations, Schutt2014HowProperties, Schutt2018SchNetMaterials, Seko2017RepresentationProperties}, only a fraction goes beyond to make models accessible to the community. the present work has been focused on creating a Findable, Accessible, Interoperable, and Reusable tool, inspired by FAIR principles \cite{FAIRFAIR}, created open-source with common and convertible data formats as is described in more detail in \ref{sssec:SoftwareUsed}. This lead to many standalone components combined into an end-user tool, described in \ref{ssec:SIPFENN}, that is ready to use without any costly computation to create the model and can be run on any modern computer, as low-power as smartphones.

\section{Methodology} \label{sec:methodology}

\subsection{Descriptor Used} \label{ssec:descriptorused}
A descriptor of a material is a point in a well-defined multidimensional property space that can be used to represent knowledge associated with entries in a database in vector form. Within the present work, the property space has 271 dimensions (corresponding to 271 features) related to elemental properties and atomic structure of an arbitrary crystalline material, as designed by Ward et al. \cite{Ward2016AMaterials, Ward2017IncludingTessellations} utilizing the voro++ code \cite{rycroft2009voro++}. These features can be categorized as:

\begin{itemize}
    \item \textbf{Elemental Attributes} (145 total): Attributes which only depend upon the elements present and their stoichiometry.
    \begin{itemize}
        \item \textbf{Stoichiometric Attributes} (6): Describe the components fractions.
        \item \textbf{Elemental Properties Attributes} (132): Contain statistics taken over the various elemental properties, weighted by the stoichiometry of the structure.
        \item \textbf{Attributes based on Valence Orbital Occupation} (4): Depend upon the distribution of valence electrons across different orbitals, i.e. on the total number of valence electrons in each orbital across the structure.
        \item \textbf{Ionic Character Attributes} (3): Attributes which encode whether the material is ionically bonded. 
    \end{itemize}
    \item \textbf{Structural Attributes} (126 total): Attributes which depend on the precise structural configuration, i.e. exactly how the atoms are arranged in space.
    \begin{itemize}
        \item \textbf{Geometry Attributes} (16): Attributes which depend upon the spatial configuration of atoms only.
        \item \textbf{Physical Property Differences Attributes} (110): Contain statistics taken over the differences between elemental properties of neighboring sites in the structure, weighted by the size of the Voronoi cell face between the neighbors.
    \end{itemize}
\end{itemize}

A complete table list of features is given in Table \ref{feature-table}. Further details can be found in \cite{Ward2016AMaterials, Ward2017IncludingTessellations}.

\begin{table}[H]
    \footnotesize
    \centering
    \caption{List of Features with Descriptions. \textbf{Site Statistics} refers the mean, range, mean absolute error, maximum, minimum, and mode unless otherwise stated in the description. \textbf{Difference Statistics} refers to the mean, mean absolute error, minimum, maximum and range of the differences between neighboring sites in a structure, weighted by the size of the face between them in the Voronoi tessellation. }
    \begin{tabular}{|p{2cm}|p{2cm}|c|c|}
        \hline
        \textbf{Site \hspace{0.3cm} Statistics} & \textbf{Difference \hspace{0.3cm} Statistics} & \textbf{Name} & \textbf{Description} \\
        \hline
        1-4 & - & Effective Coordination Number & mean, mean abs error, min, max\\
        \hline
        5-7 & - & Mean Bond Length & mean abs error, min, max\\
        \hline
        8-11 & - & Bond Length Variation & mean, mean abs error, min, max \\
        \hline
        12 & - & Cell Volume Variation & Variation in the voronoi cell volume\\
        & & & no statistics \\
        \hline
        13-15 & - & Mean WC Magnitude & shells 1-3, global non-backtracking \\
        \hline
        16 & - & Packing Efficiency & no statistics \\
        \hline
        133-138 & 17-21 & Atomic Number & \\
        \hline
        139-144 & 22-26 & Mendeleev Number & \\ 
        \hline
        145-150 & 27-31 & Atomic Weight & \\ 
        \hline
        151-156 & 32-36 & Melting Temperature & \\ 
        \hline
        157-162 & 37-41 & Column & Group in Periodic Table \\ 
        \hline
        163-168 & 42-46 & Row & Period in Periodic Table\\ 
        \hline
        169-174 & 47-51 & Covalent Radius & \\ 
        \hline
        175-180 & 52-56 & Electronegativity & \\ 
        \hline
        181-210 & 57-81 & Valence Electron Count & Listed for s,p,d,f orbitals and total \\ 
        \hline
        211-240 & 82-106 & Unfilled Count & Number of unfilled orbitals \\
         & & & Listed for s,p,d,f orbitals and total \\
         \hline
        241-246 & 107-111 & Ground State Volume & \\ 
        \hline
        247-252 & 112-116 & Ground State Band Gap & \\ 
        \hline
        253-258 & 117-121 & Ground State Magnetic Moment & \\ 
        \hline
        259-264 & 122-126 & Space Group Number & Index of Space group\\
        \hline
        127 & - & Number of Components & no statistics \\
        \hline
        128-132 & - & $\ell^p$-norms of Component Fractions & $p \in \{2,3,5,7,10\}$ \\
        \hline
        265-268 & - & Fraction of Valence Electrons & \\
        & & in s,p,d,f orbitals & no statistics\\
        \hline
        269 & - & Can Form Ionic Compound & boolean, no statistics\\
        \hline
        270-271 & - & Ionic Character & max, mean over pairs of species\\
        \hline
    \end{tabular}
    \label{feature-table}
\end{table}

\subsection{Machine Learning Techniques Overview} \label{ref:machinelearningoverview}
This section gives a brief overview of the employed machine learning techniques and terminology, described in more detail in the Appendix \ref{appedix1}. The interest is placed on the statistical problem of regression, whose goal is to learn a functional relationship $f:X\rightarrow Y$ which minimizes the risk (also known as loss or expected error) \cite{vapnik1999overview} given by
\begin{equation}\label{true_risk}
    R(f) = \mathbb{E}_{x,y\sim \mathcal{P}} l(y,f(x)).
\end{equation}
Here $X$ denotes a space of input features, $Y$ denotes an output space, the expectation above is taken over an unknown distribution $\mathcal{P}$ on $X\times Y$ (representing the true relationship between inputs and outputs), and $l$ is a given loss function. 

In the specific application considered here, the function $f$ which is to be learned, maps input material structures (arrangements of atoms) $x$ to the predicted formation energy $y$. The distribution $\mathcal{P}$ is unknown, but samples $(x_i,y_i)$ are given, consisting of structures $x_i$ and corresponding predictions $y_i$ which are used to learning $f$. In the present case, this data comes from the OQMD and other smaller materials databases.


In order to learn the relationship $f$ from the data, the empirical risk
\begin{equation}\label{empirical_risk}
    L(f) = \frac{1}{n}\displaystyle\sum_{i=1}^n l(y_i, f(x_i)),
\end{equation}
is minimized over a class of functions defined by a neural network architecture. A neural network architecture consists of a sequence of alternating linear functions and point-wise non-linear functions defined by an activation function (see \cite{goodfellow2016deep} for more information about neural networks). As the loss function $l$ in \eqref{empirical_risk} the $\ell^1$-loss function $l(y,x) = |x-y|$ is used. The neural networks are trained on this loss \eqref{empirical_risk} using the common ADAM optimizer \cite{kingma2014adam}. 

An important issue when training complex statistical models is the overfitting, which occurs when a model accurately fits the training data but fails to generalize well to new examples. In order to detect overfitting, the standard practice of dividing the data into training, validation, and test datasets \cite{hastie2009elements} is used. In order to mitigate overfitting, dropout \cite{srivastava2014dropout} and weight decay, two standard methods for regularizing neural networks, are used. In Section \ref{sssec:DesignedModels}, Figure \ref{fig:trainingvalidation-body} illustrates overfitting mitigation effects on the training process of neural networks designed in the present paper.

\subsection{Software Used} \label{sssec:SoftwareUsed}
The choice of software for the machine learning portion of this project was Apache MXNet \cite{ChenMXNet:Systems} due to it's open source nature, model portability, and state-of-the-art scalability, allowing the same code to run on a laptop with a low-power CPU/GPU and a supercomputer (e.g., ORNL Summit) with hundreds of powerful GPU's. It's portability allows trained networks to be converted and used with other popular frameworks such as Google Tensorflow, PyTorch, or even Apple Core ML, making results of the present paper highly accessible.

MXNet framework was used through Wolfram and Python languages. Wolfram Language was used primarily for the network architecture design, training, and testing, as it provides an excellent interface with detailed training results shown in real-time during the training process. It also provides good out-of-the-box performance due to its well-optimized memory handling when training on a single GPU setup. 

Python, on the other hand, was used when writing the end-user tool for running previously trained networks. This choice was made so that the software is completely open-source and can be easily reused for specific purposes or incorporated within other packages. Furthermore, Python allowed quick implementation of a Graphical User Interface (GUI) through the wxpython package.

\subsection{Data Acquisition and Curation} \label{sssec:Data}

Four sets of data were used within the present work. The largest by volume and significance was the Open Quantum Materials Database (OQMD) \cite{Kirklin2015TheEnergies, Saal2013MaterialsOQMD}, which contains the results of DFT calculations performed by the Vienna Ab Initio Simulation Package (VASP) \cite{Kresse1993AbMetals} for a broad spectrum of materials. The snapshot used here was extracted from the database by Ward et al. in 2017 and contained 435,792 unique compounds \cite{Ward2017IncludingTessellations}. The choice of 2017 snapshot rather than the current one was made to ensure direct performance comparison between new and previously reported methods. The second database was a part of the Inorganic Crystal Structure Database (ICSD), a subset of the OQMD with only experimentally obtained structures containing around 30,000 entries. ICSD was primarily used for the quick design of simple neural network architectures at the beginning, and OQMD used for more complex models designed later. 

Two smaller data sets were used, in addition to these large databases. The first small dataset contained DFT-calculated formation energies of Fe-Cr-Ni ternary $\sigma$-phase endmembers in the 5-sublattice model \cite{Feurer2019Cr-Fe-NiCalculations}. As this model contains 5 chemically distinct positions (Wyckoff positions), populated by one of 3 elements, in total it included 243 ($3^5$) structures with 30-atom basis each. This data served as an example of a relatively complex structure that was not included in the OQMD. Furthermore, it was a test case of a material that is highly industry-relevant, as it causes steel embrittelment \cite{Hsieh2012OverviewSteels} and is costly to investigate using traditional methods due to compositional and configurational complexity. The second small dataset included 13 Special Quasirandom Structures (SQS), which are the best periodic supercell approximations to the true disordered state of metal alloys \cite{Zunger1990SpecialStructures, Jiang2004First-principlesStructures, Shin2006ThermodynamicStructures}. SQS structures in this set were binary alloys containing Fe, Ni, Co, and V, laying on deformed FCC (A1), BCC (A2), or HCP (A3) lattices. The main purpose of these smaller datasets was to test the performance in extrapolation from OQMD, in a particular case of interest for the author's.

During the network design process described in \ref{sssec:NetDesign}, it was found that a small fraction of the OQMD dataset (under 0.03\%) contains anomalous values of formation energy above 10 eV/atom. In the extreme case of $CuO_2$ (OQMD ID: 647358) this value was 1123 eV/atom or 108350 kJ/mole. Since the source database contains hundreds of thousands of data points reported by many scientists, it can be expected that a small fraction of the data may contain some sort of errors and typos. In the present work, they were removed from all datasets used for training and evaluation.

\subsection{Neural Network Design Process} \label{sssec:NetDesign}

This section conceptually outlines the network design process leading to the final models. All essential details regarding the design and performance of intermediate models, useful for better understanding changes and for applying the similar approach in different problems, can be found in the Appendix \ref{appendix2}.

The design started with the simplest single-layer neural network (perceptron) with the Sigmoid activation function, trained on the ICSD and its smaller subset, to provide a baseline for the design. Then, the process was conducted in the following steps:

    \begin{wrapfigure}{r}{0.57\textwidth}
    \vspace{-12pt}
    \centering
    \includegraphics[width=0.56\textwidth]{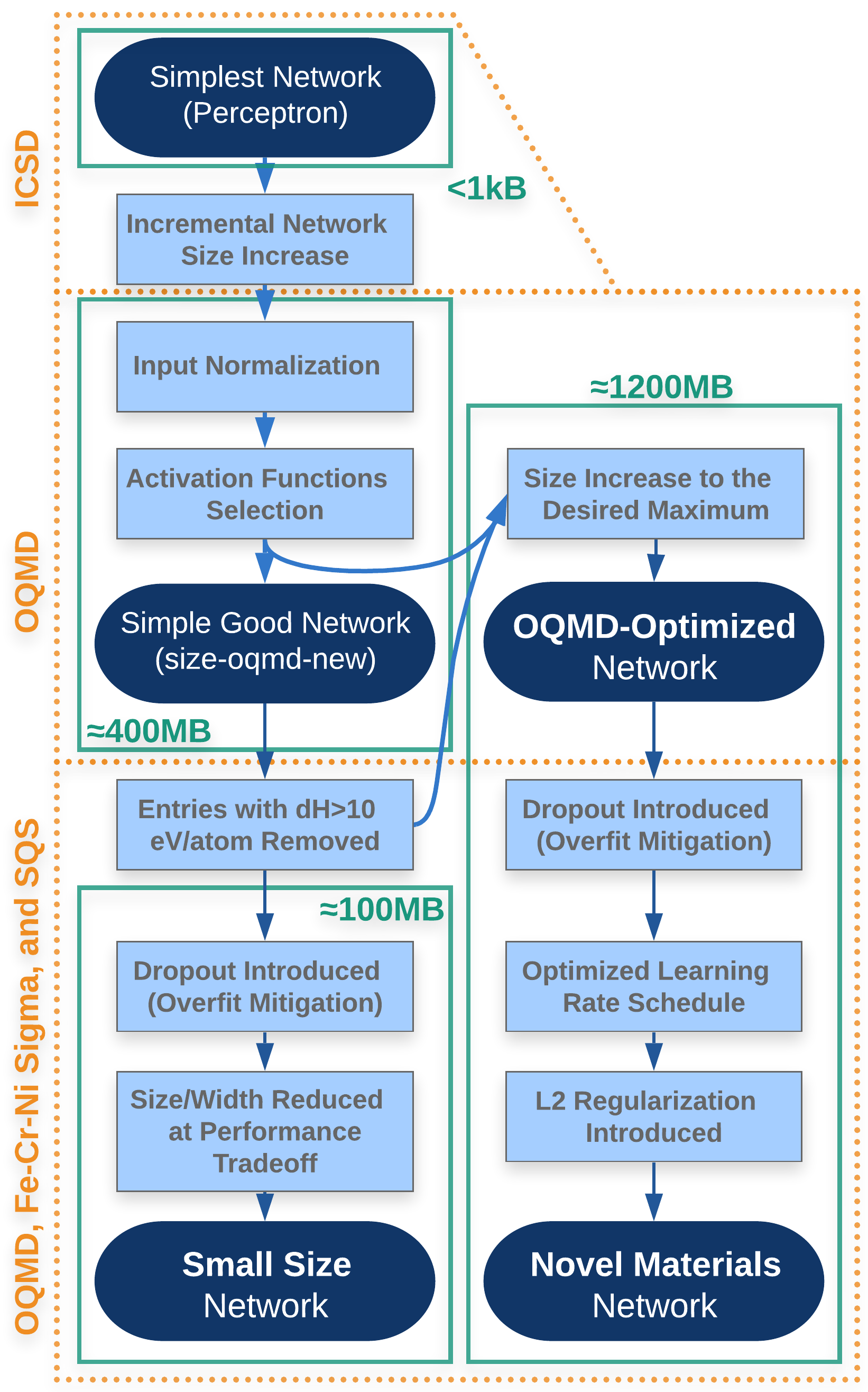}
    \caption{The model design process schematic.}
    \label{fig:designprocess}
    \vspace{-18pt}
    \end{wrapfigure}

\textbf{1. }The network size has been increased step-wise while training on the ICSD dataset (30k+ entries). Results were extrapolated to estimate network size suitable for larger OQMD (400k+) to be 4 hidden layers in a (10000, 10000, 1000, 100) configuration.

\textbf{2. }To improve convergence during the training, descriptor features values were normalized to their maximum values present in the OQMD dataset.

\textbf{3. }Performance and time to convergence were improved by moving from Sigmoid activation function to a mix of Soft Sign, Exponential Linear Unit, and Sigmoid. This relatively simple model has improved performance over the existing Random Forest model \cite{Ward2017IncludingTessellations}, achieving MAE of 42 meV/atom on the same dataset.

\textbf{4. }At this step, it was noticed that a small fraction (around 0.03\%) of data points exhibits extreme errors, as high as over 1,000,000 meV/atom causing some instability during the training process, despite the large batch size of 2048. They also caused a high deviation in test MAE values across repeated model training rounds. As describes in \ref{sssec:Data}, these were identified to be a few rare errors in the dataset and removed during later model design. 

\textbf{5. }The network size was increased to around 1GB limit (maximum size target) by the addition of two more 10,000-width layers. This \textbf{OQMD-optimized} network has achieved the best performance on the OQMD out of all designed in the present paper, with an MAE of 28 meV/atom. Performance analysis can be found in \ref{ssec:oqmdperformance} and in Figure \ref{fig:oqmdperformance}.

\textbf{6. }After the good performance on the OQMD was achieved, the design goals shifted to (1) reducing the training-set-to-validation-set error mismatch during the network training, while (2) keeping the test MAE on the OQMD on a suitable level (below 50 meV/atom), and (3) improving performance on datasets not presented to network before (see \ref{sssec:Data}). The first step was the introduction of Dropout layers\cite{srivastava2014dropout}, described in more detail in Appendix \ref{appedix1}, which allow for better distribution of knowledge across the network.

\textbf{7. }The introduction of strong Dropout\cite{srivastava2014dropout} made the network prone to falling in local minima, which was solved by the introduction of a changing learning rate schedule.

\textbf{8. }With optimized network architecture, lastly, the descriptor interpretation by the network has been modified through the introduction of L2 regularization \cite{L2Regularization}, a technique which assigns an error penalty for "attention" (input layer weights) to each of the descriptor features, effectively refining features in the descriptor to only the most significant ones. Figure \ref{fig:squaredweights} ranks them. The resulting \textbf{Novel Materials} model achieved a much lower training-set-to-validation-set error mismatch (1.15 vs 1.57 after 240 rounds), presented in Figure \ref{fig:trainingvalidation} as a function of training progress. On the OQMD test set, it achieved a higher, yet suitable 49 meV/atom.

\textbf{9. }To cater to applications requiring very high throughput or low memory consumption, an additional \textbf{Small Size} network was designed by adding Dropout to one of the earlier networks, designed before the size increase step, and then reducing its size to the desired level. It was found that after reduction of total size from around 400MB to around 100MB, the network retained MAE of 42 meV/atom on an OQMD test set and further reduction was possible if needed for the application.

\section{Results} \label{sec:Results}

\subsection{Final Predictive Models} \label{sssec:DesignedModels}

Throughout the architecture design process described in \ref{sssec:NetDesign}, detailed in Appendix \ref{appendix2}, and depicted in Figure \ref{fig:designprocess}, new networks were designed and tested in various ways, leading to about 50 predictive models (trained neural networks) with varying training parameters and training data. The majority of the intermediate networks were stored for the record, and are available upon request. Details regarding hyper-parameters and training routines used to obtain three resulting models can be found in the Appendix \ref{appedix1}.

\begin{figure}[H]
    \centering
    \frame{\includegraphics[width=0.30\textwidth]{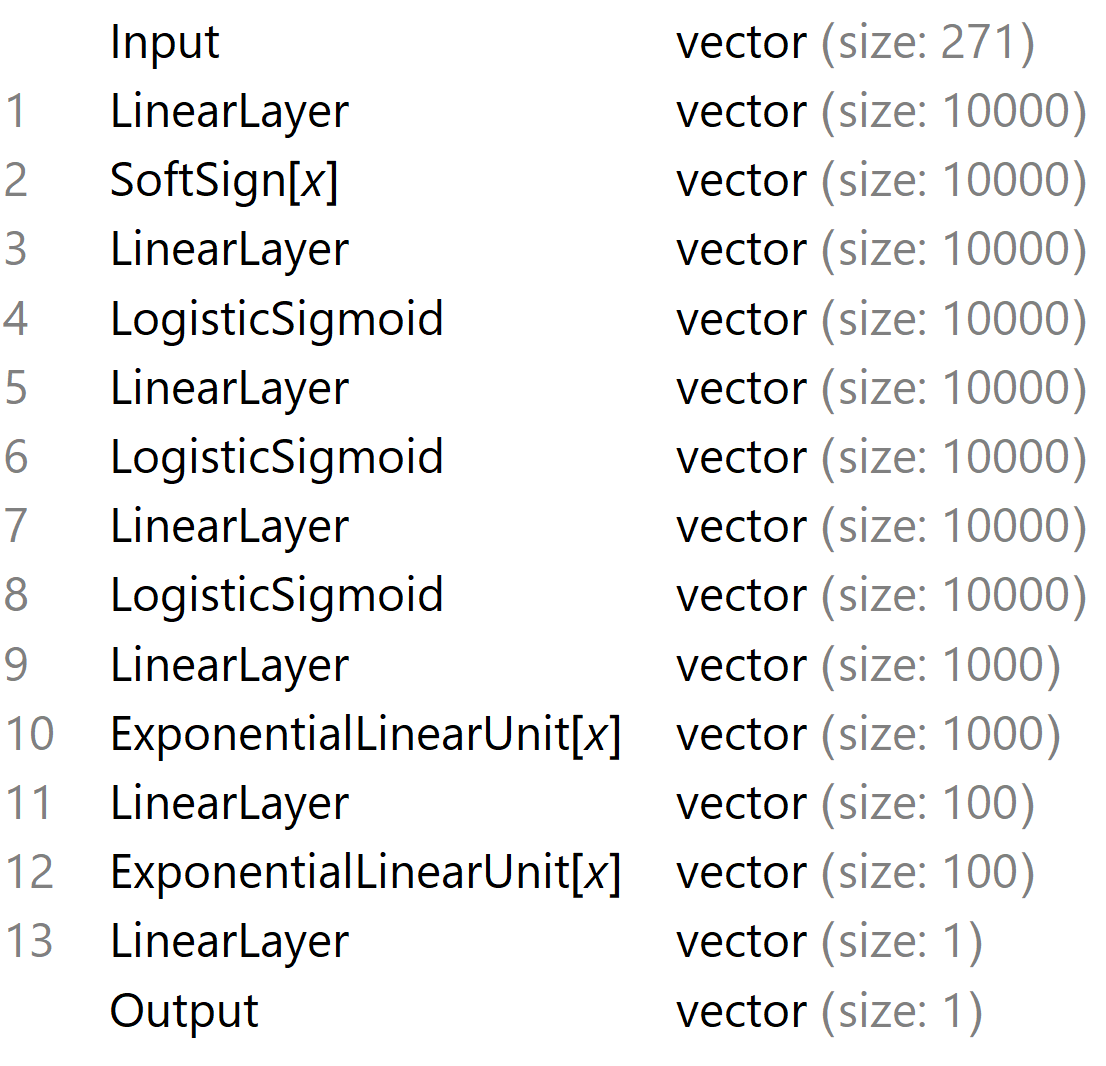}}
    \hspace{6pt}
    \frame{\includegraphics[width=0.30\textwidth]{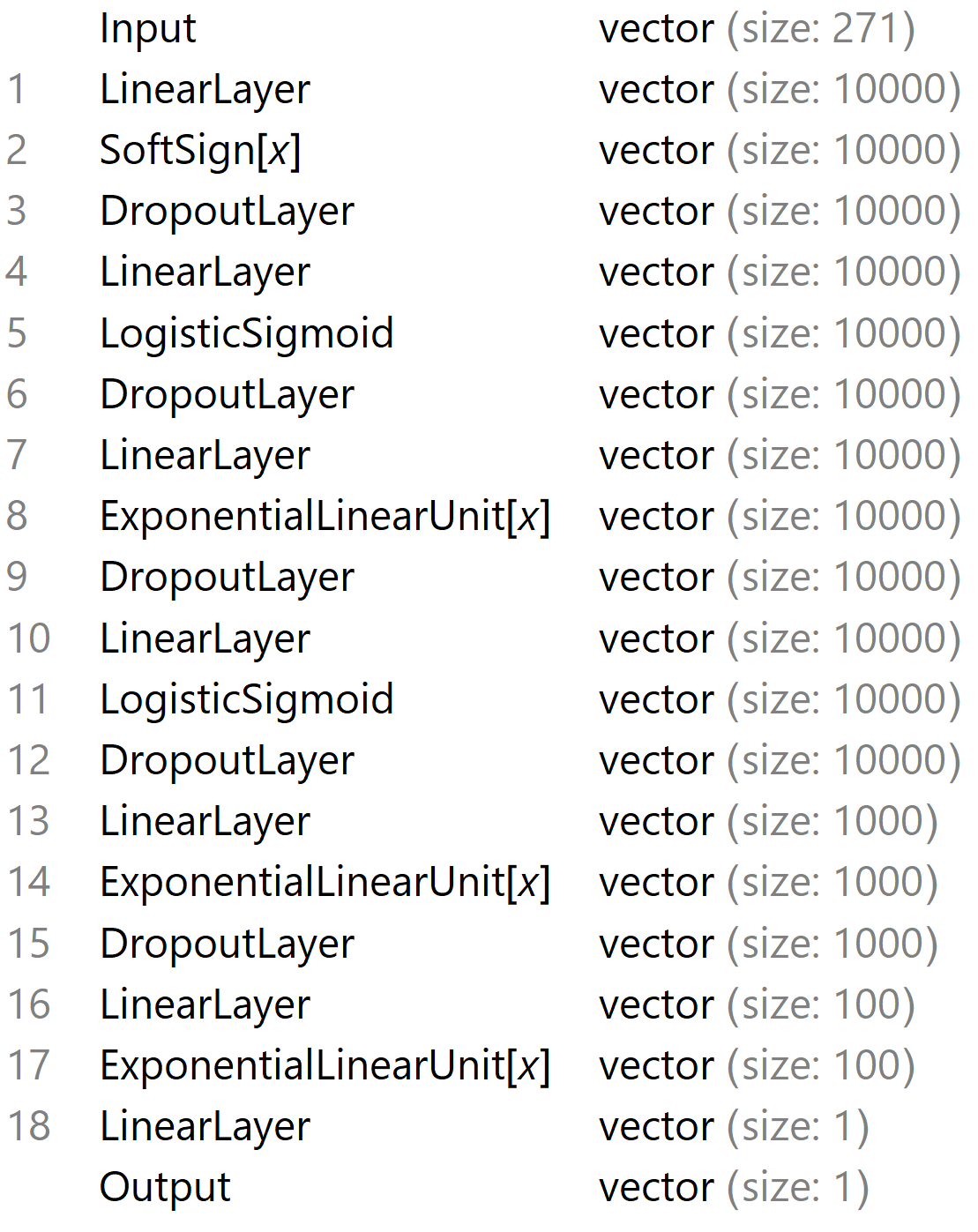}}
    \hspace{6pt}
    \frame{\includegraphics[width=0.30\textwidth]{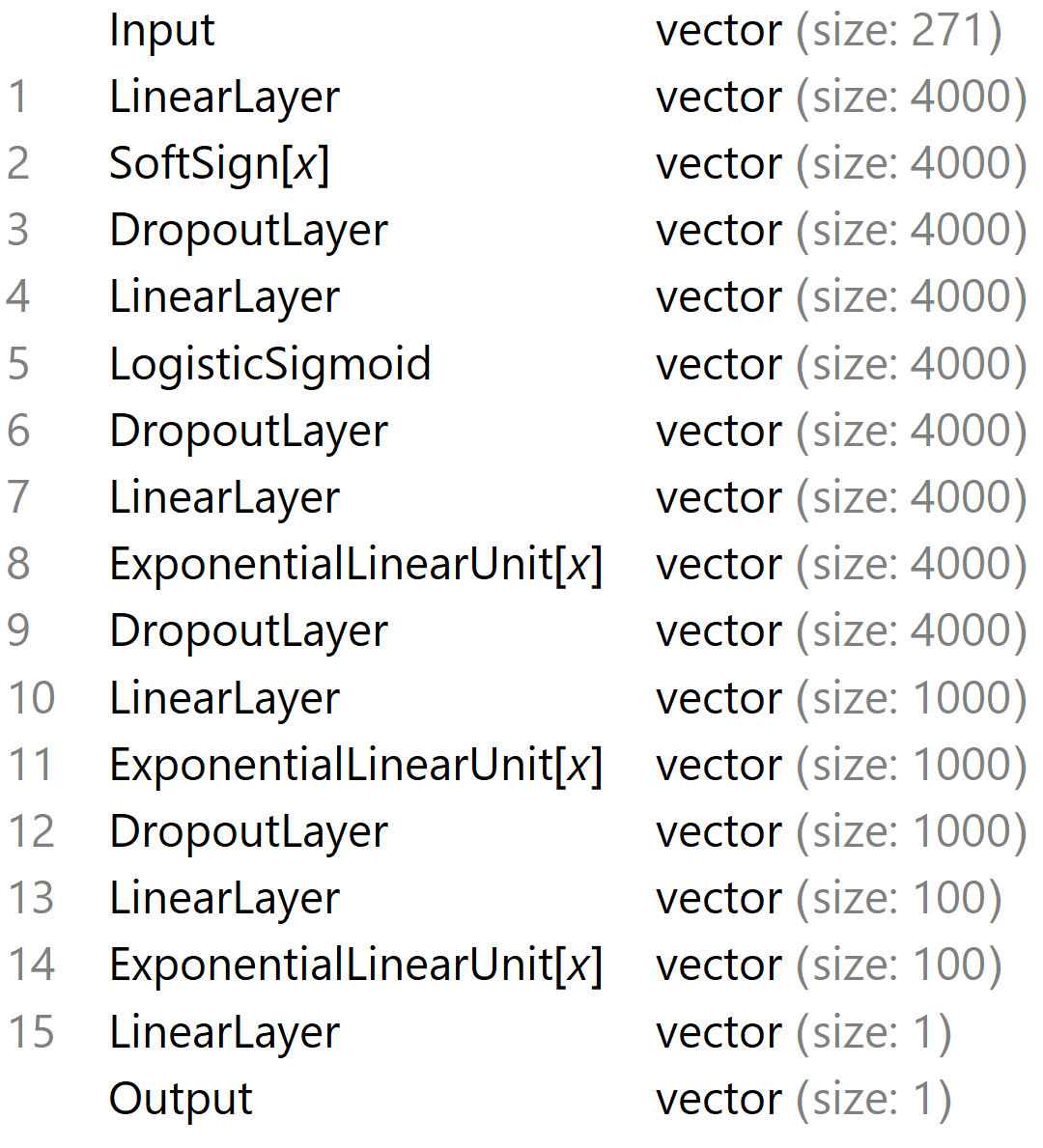}}
    \caption{Three selected architectures designed within the present work. Optimized for: (Left) OQMD performance, (Middle) predicting new materials, (Right) small size at good performance. Internally in the code, they are designated as NN9, NN20, and NN24.}
    \label{fig:architectures}
    \vspace{-6pt}
\end{figure}

Out of all trained neural networks, three were selected and can be considered final outcomes of the design process, optimized for different objectives. Their architectures are presented in Figure \ref{fig:architectures}. The first one, denoted NN9, was created specifically for the OQMD performance. This was the same objective as in the study by Ward et al. \cite{Ward2017IncludingTessellations} and its performance serves as a direct comparison to the Random Forest method employed in that paper \cite{Ward2017IncludingTessellations} and other works \cite{Schutt2014HowProperties, Seko2017RepresentationProperties}.

\begin{wrapfigure}{r}{0.53\textwidth}
    \vspace{-12pt}
    \centering
    \includegraphics[width=0.52\textwidth]{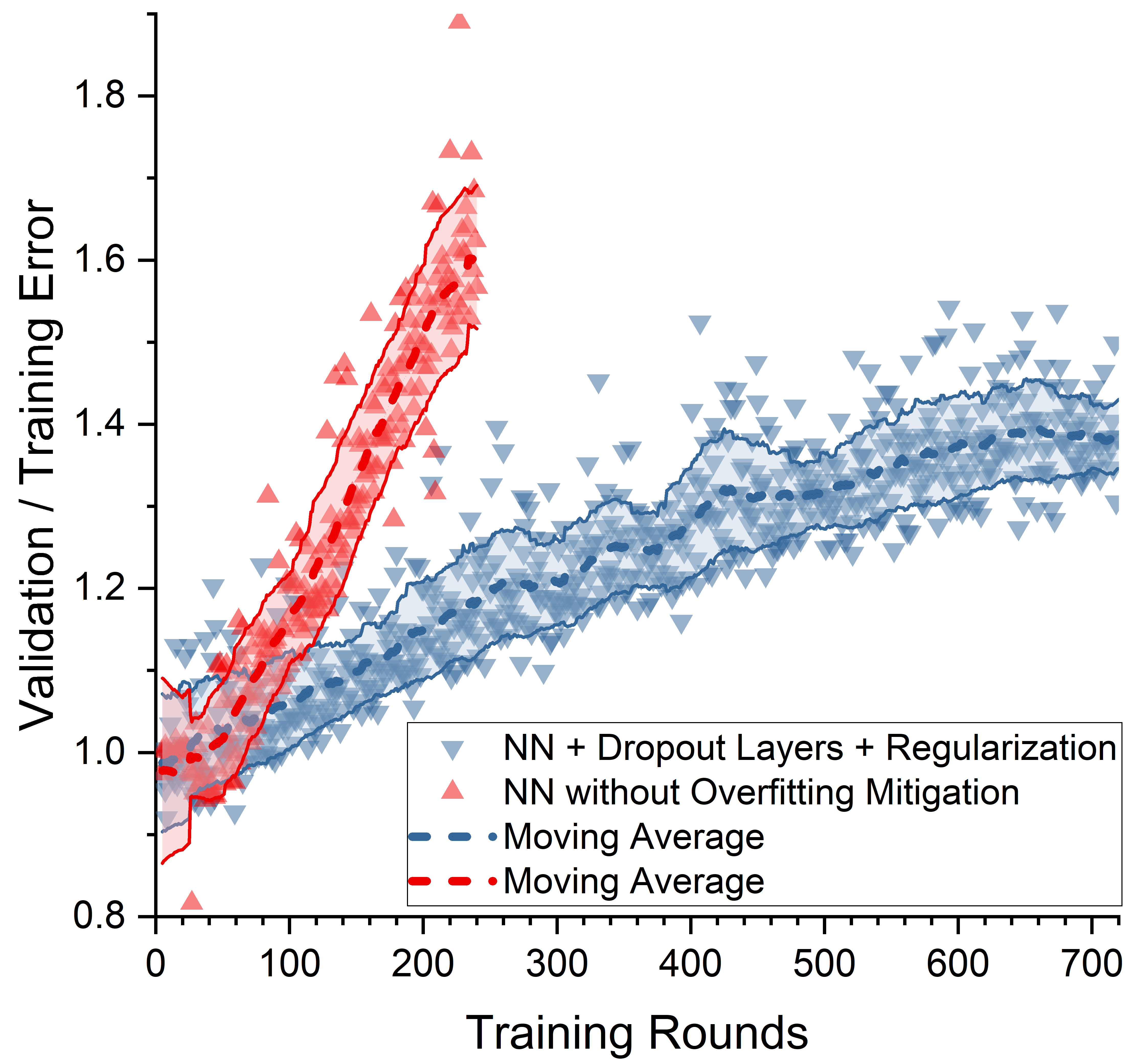}
    \caption{Training Loss to Validation Loss in a model that does without (NN9) and with overfitting mitigation (NN20), plotted versus training progress.}
    \label{fig:trainingvalidation-body}
\end{wrapfigure}

\begin{figure}[h]
    \centering
    \begin{minipage}[c]{0.65\textwidth}
    \includegraphics[width=\textwidth]{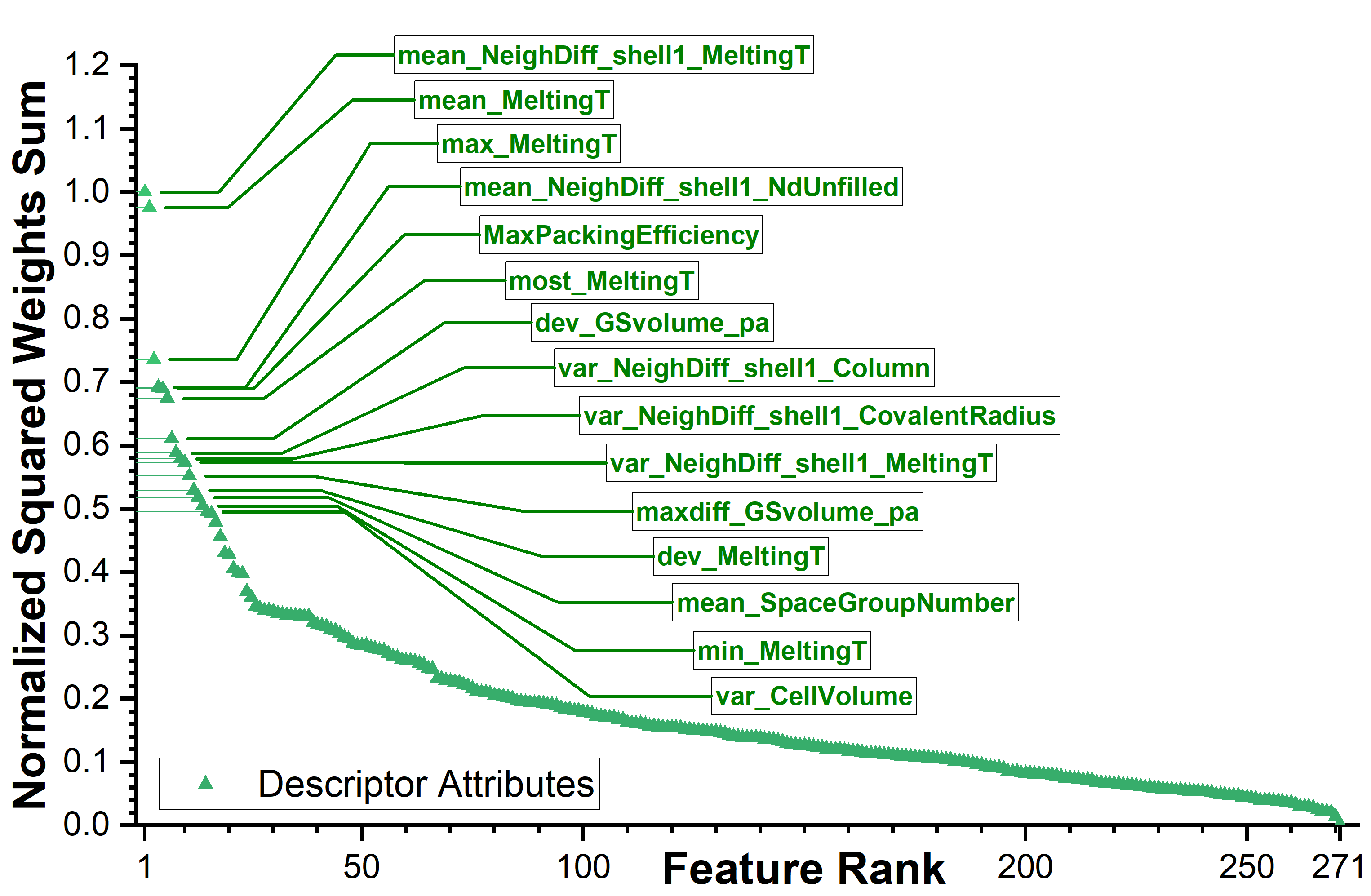}
    \end{minipage}\hfill
    \begin{minipage}[c]{0.33\textwidth}
    \caption{Distribution of sums of squared input weights. High values correspond to attributes that were not lowered due to their contribution to pattern recognition of the model. 15 attributes with the highest values are labeled. The labels are taken from the descriptor definition in  \cite{Ward2016AMaterials}.}
    \label{fig:squaredweights}
    \end{minipage}
\end{figure}

The second network was optimized for improved pattern recognition on OQMD and improved performance on non-OQMD datasets used in the present work (i.e. SQS/$\sigma$-phase datasets). This was achieved primarily through extensive overfitting mitigation, applied during design and training (see Figure \ref{fig:trainingvalidation-body}), which leads to a network with improved generalization/materials-discovery capability. Furthermore, one fo the overfitting mitigation methods, namely the regularization described in \ref{ref:machinelearningoverview}, have allowed identification of descriptor attributes that contributed the most to the predictive capability and the ones that were almost completely discarded once the penalty for considering them was assigned. Figure \ref{fig:squaredweights} presents the distribution of sums of squared weights between each neuron in the input layer (each of the 273 descriptor features) and all 10,000 neurons in the first hidden layer. 

Feature rankings, such as presented in Figure \ref{fig:squaredweights}, allow a more efficient selection of input features in future studies looking into the same problem; thus both reducing the number of features that need to be computed for each atomic configuration and the total number of weights in the network. Furthermore, it can be used to gain an insight into the model interpretability. Looking at the specific ranking for NN20, the high-impact features present a mix of elemental features, likely allowing the model to establish some formation energy baseline for a given composition, and structure-informed features allowing to distinguish between polymorphic configurations. High impact elemental features include different statistics on elemental melting temperatures and ground-state structure volume per atom. The structural features extend them by considering how they differ between neighboring atoms and also include purely structural features such as packing efficiency and variance in Wigner–Seitz cells volumes. A complete ranking of features is included in Appendix \ref{appendix3}.

The third network, denoted NN24, was created for memory/power-constrained applications requiring a balance between OQMD performance and memory intensity and processing power required. Model parameters contained in this architecture occupy only 145MB, over 8 times less than two other models and around 200 times less than the model reported by Ward et al. \cite{Ward2017IncludingTessellations}.

\subsection{OQMD Data Performance} \label{ssec:oqmdperformance}
\begin{figure}[H]
    \centering
    \includegraphics[width=0.31\textwidth]{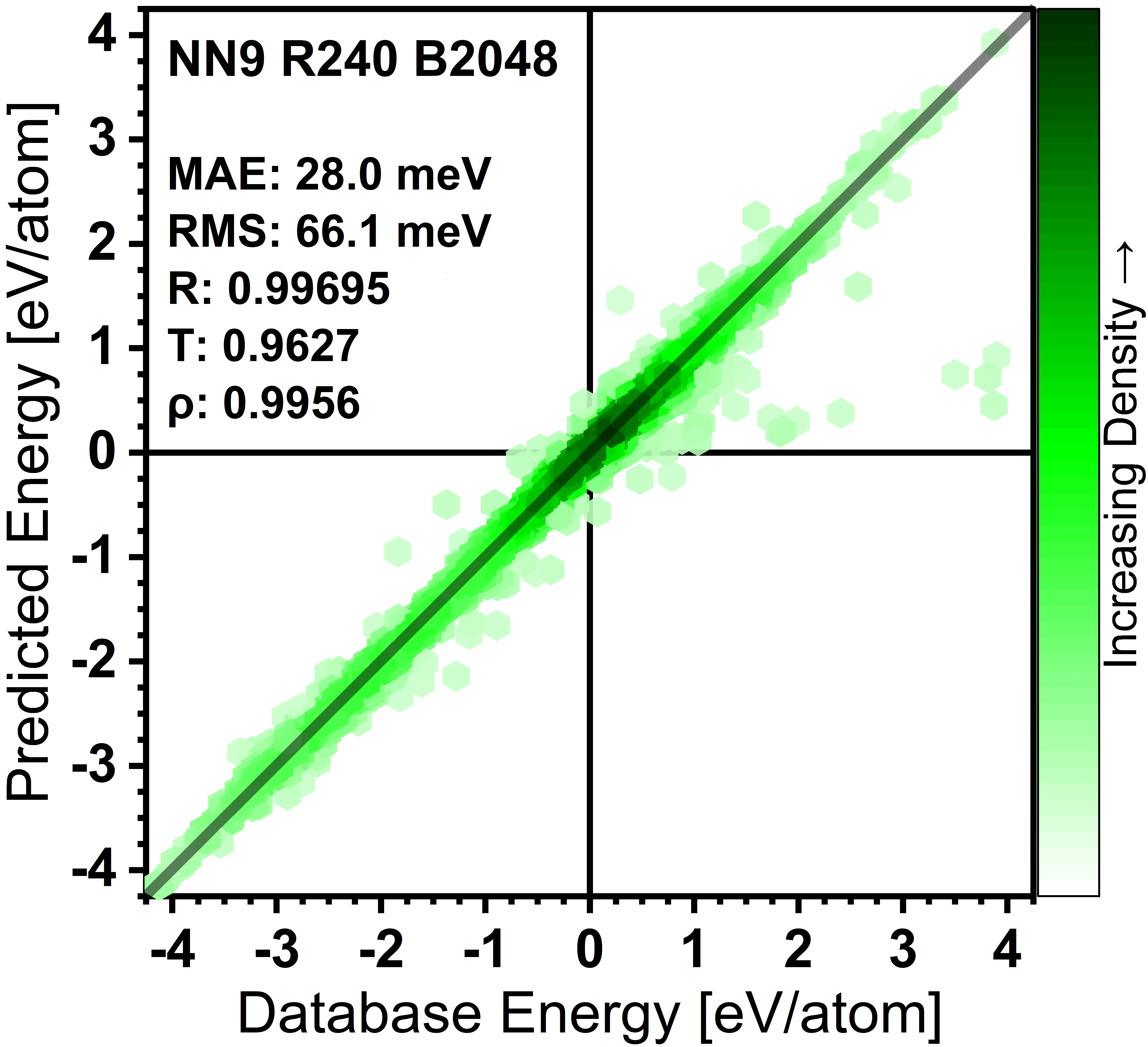}
    \hspace{0.01\textwidth}
    \includegraphics[width=0.31\textwidth]{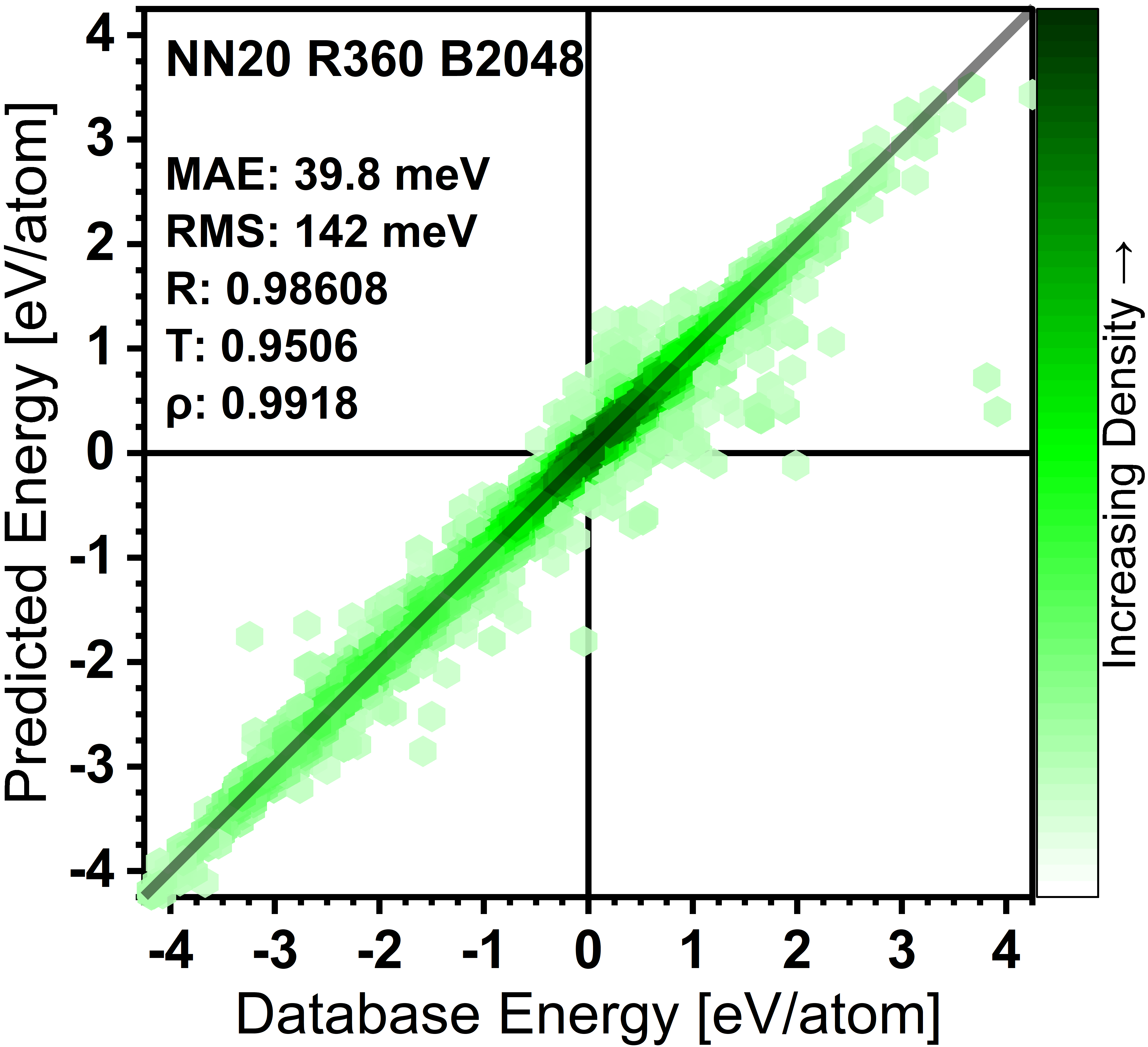}
    \hspace{0.01\textwidth}
    \includegraphics[width=0.31\textwidth]{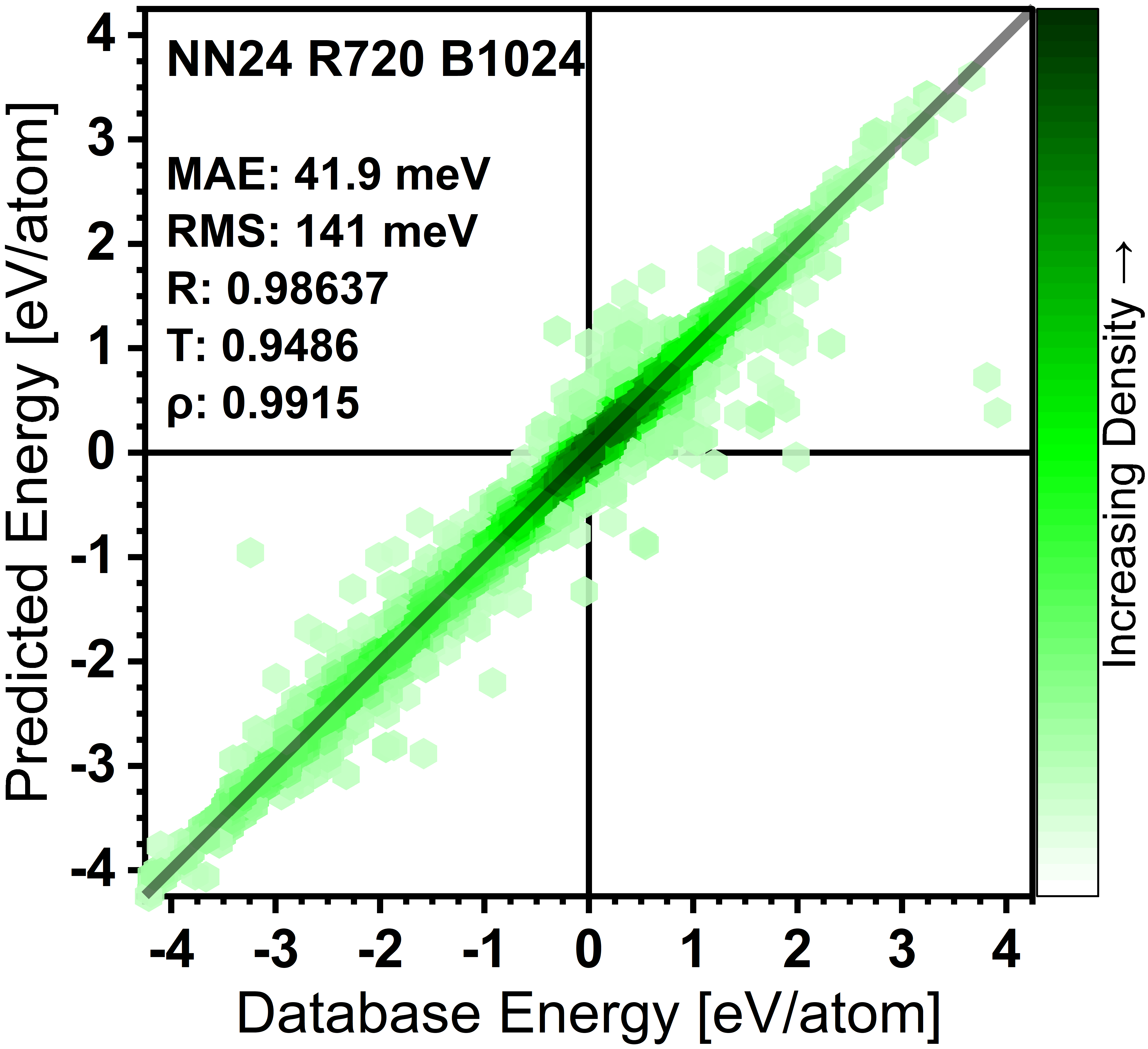}
    \includegraphics[width=0.31\textwidth]{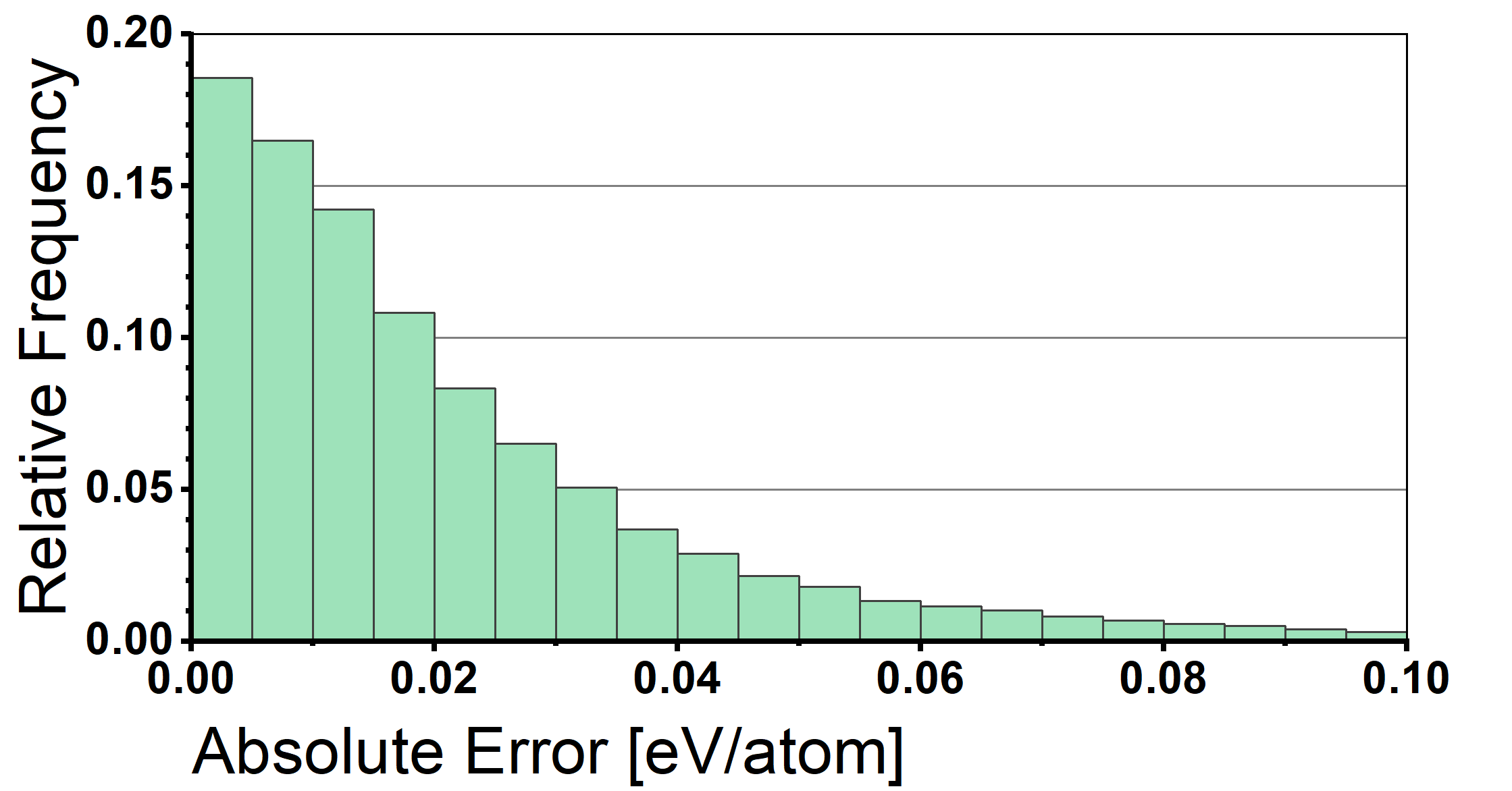}
    \hspace{0.01\textwidth}
    \includegraphics[width=0.31\textwidth]{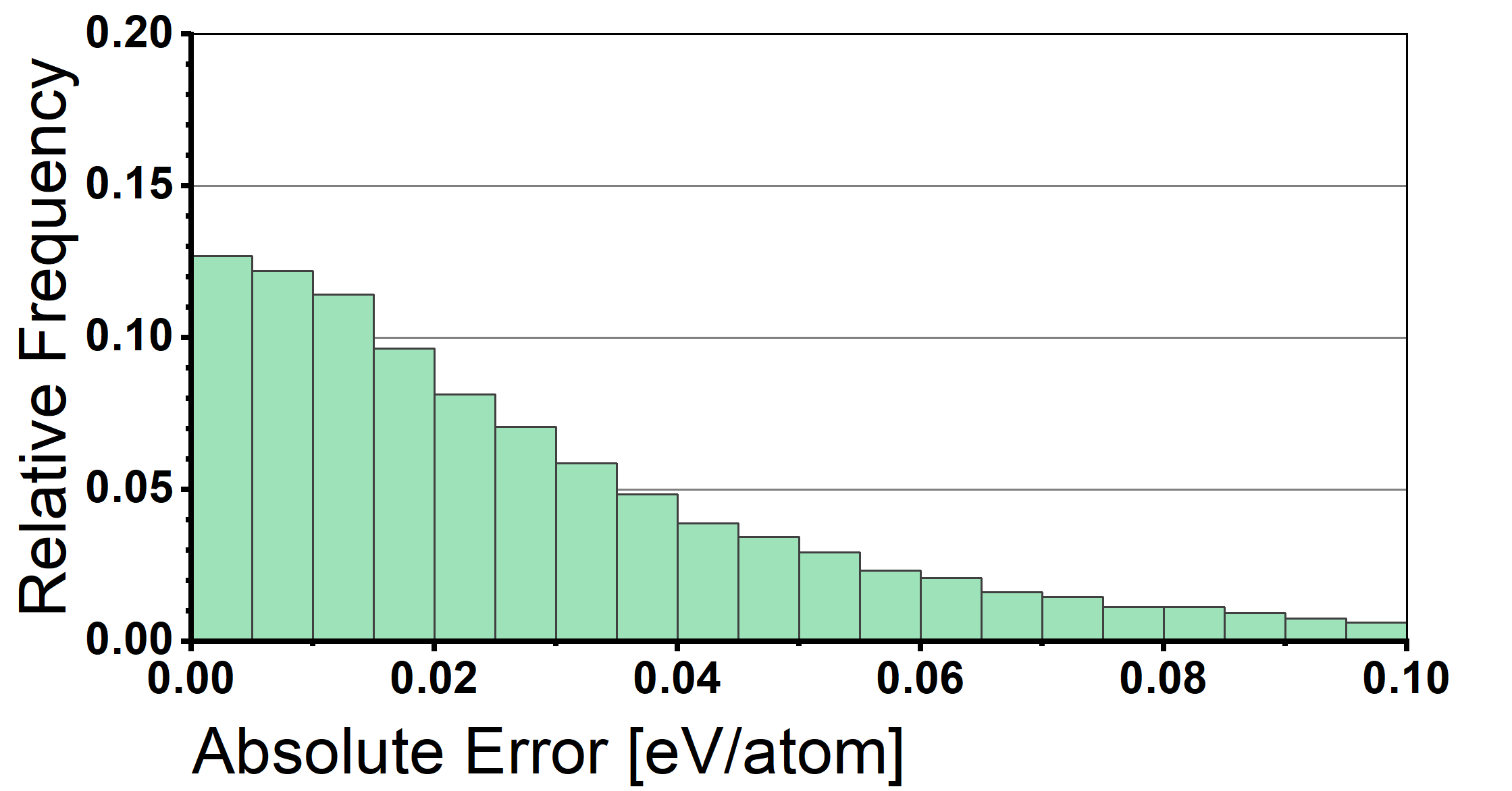}
    \hspace{0.01\textwidth}
    \includegraphics[width=0.31\textwidth]{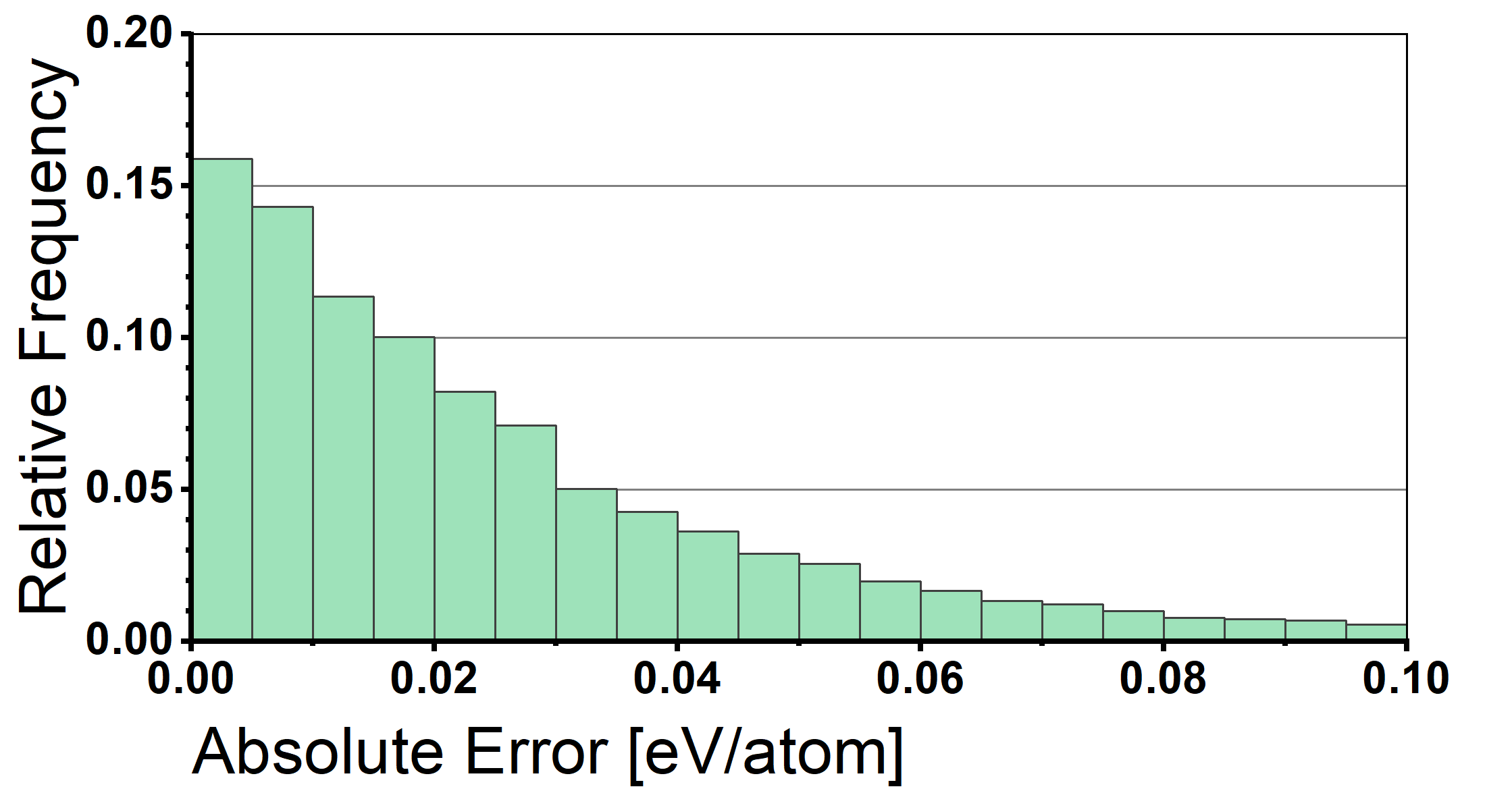}
    \caption{Performance of 3 selected neural networks on a random subset of 21,800 entries from OQMD. (Left) OQMD performance, (Middle) predicting new materials, (Right) small size at good performance. Internally in the code, they are designated as NN9, NN20, and NN24.}
    \label{fig:oqmdperformance}
\end{figure}

As described in \ref{sssec:NetDesign}, all three final networks were evaluated on a randomly selected subset of the OQMD to give a comparison between the state-of-the-art model presented by Ward et al. \cite{Ward2017IncludingTessellations} and the present ML method. This random subset consisted of 21,800 OQMD entries, constituting approximately $5\%$, which were not presented to the network, nor used for evaluation at any stage of the training process. This sample size was considered to be representative of the whole dataset once the small fraction ($0.026\%$) of likely incorrect entries were removed from the dataset as described in \ref{sssec:Data}. The random selection itself was initially performed separately for each training process and recorded after completion. Later, when networks were modified to mitigate overfitting, a single random subset was used for all of them to allow more careful design and more accurate comparative analysis of results. Figure \ref{fig:oqmdperformance} gives (1) prediction vs OQMD values of formation energy plot, (2) statistics related to the error in predictions relative to the OQMD values, and (3) a histogram of the absolute error in predictions relative to the OQMD values.

\subsection{Existing Methods Comparison} \label{ssec:existing}
In this section, the performance of the models is compared with a few similar existing approaches based on the OQMD dataset, when formation energy of a structure is predicted \cite{Ward2016AMaterials, Ward2017IncludingTessellations, Jha2019IRNet}, or its subset of the convex-hull structures, when formation energy of the most stable structure is predicted \cite{Jha2018ElemNet:Composition, Goodall2020PredictingStoichiometry}.  This division is made based on the reasoning presented in \ref{ssec:currentapproach}. While the latter type cannot be used to predict the formation energy of any arbitrary structure, the structure-informed models like SIPFENN (the present work) can be tested on the convex hull structures.

\begin{table}[H]
\begin{center}
\begin{tabular}{|c|c|c|}
\hline
 Method & Formation Energy MAE & Convex Hull MAE \\
 \hline
 SIPFENN (This Work) & \textbf{28.0 meV/atom} (OQMD Opt.) & 32meV/atom (Novel. Mat.) \\
 Ward2017 \cite{Ward2016AMaterials, Ward2017IncludingTessellations} & 80 meV/at & N/M \\
 ElemNet \cite{Jha2018ElemNet:Composition} & N/A & 50 meV/at\\  
 IRNet \cite{Jha2019IRNet} & 38 meV & N/M \\
 Roost \cite{Goodall2020PredictingStoichiometry} & N/A & 29 meV/at | \textbf{24 meV/at}\\
 \hline
\end{tabular}
\caption{Comparison of our method with existing state-of-the-art methods. N/A and N/M respectively stand for not applicable and not measured.}
\label{comparison-results}
\end{center}
\vspace{-24pt}
\end{table}

The results are shown in Table \ref{comparison-results}. The SIPFENN convex hull MAE has been reported based on using the Novel Materials Model limiting the original test set to structures laying within 50meV/atom from the convex hull. From these results, we can see that the SIPFENN neural networks approach outperforms existing state-of-the-art methods for predicting the formation energy of any material. At the same time, while not being the best, it is capable of reaching performance levels of specialized models in predicting the formation energies of structures laying on the convex hull.

\subsection{Non-OQMD Data Performance} \label{ssec:sigmasqsperformance}

Models created in the present work, specifically the ones optimized for predicting the formation energy of new materials, were designed and implemented to serve as tools for materials discovery. Evaluating their performance on data from the same source as the training set done in \ref{ssec:oqmdperformance} is inherently biased towards favoring models that provide the best fit to the prior (training) knowledge. This is amplified by the fact that many entries in the database are reported in groups that come from common studies and span similar materials, causing high domain clustering, which in some cases effectively makes such evaluation more akin to interpolation than extrapolation of knowledge.

To partially mitigate the described issue, the performance of the models was also evaluated on two smaller non-OQMD databases, described in \ref{sssec:Data}, representing an example of chemistries and structures that were of interest to the authors project on Ni-based superalloys. At the same time, they were not directly presented to the network in any capacity during the training process.

\begin{figure}[H]
\centering
    \includegraphics[width=0.3\textwidth]{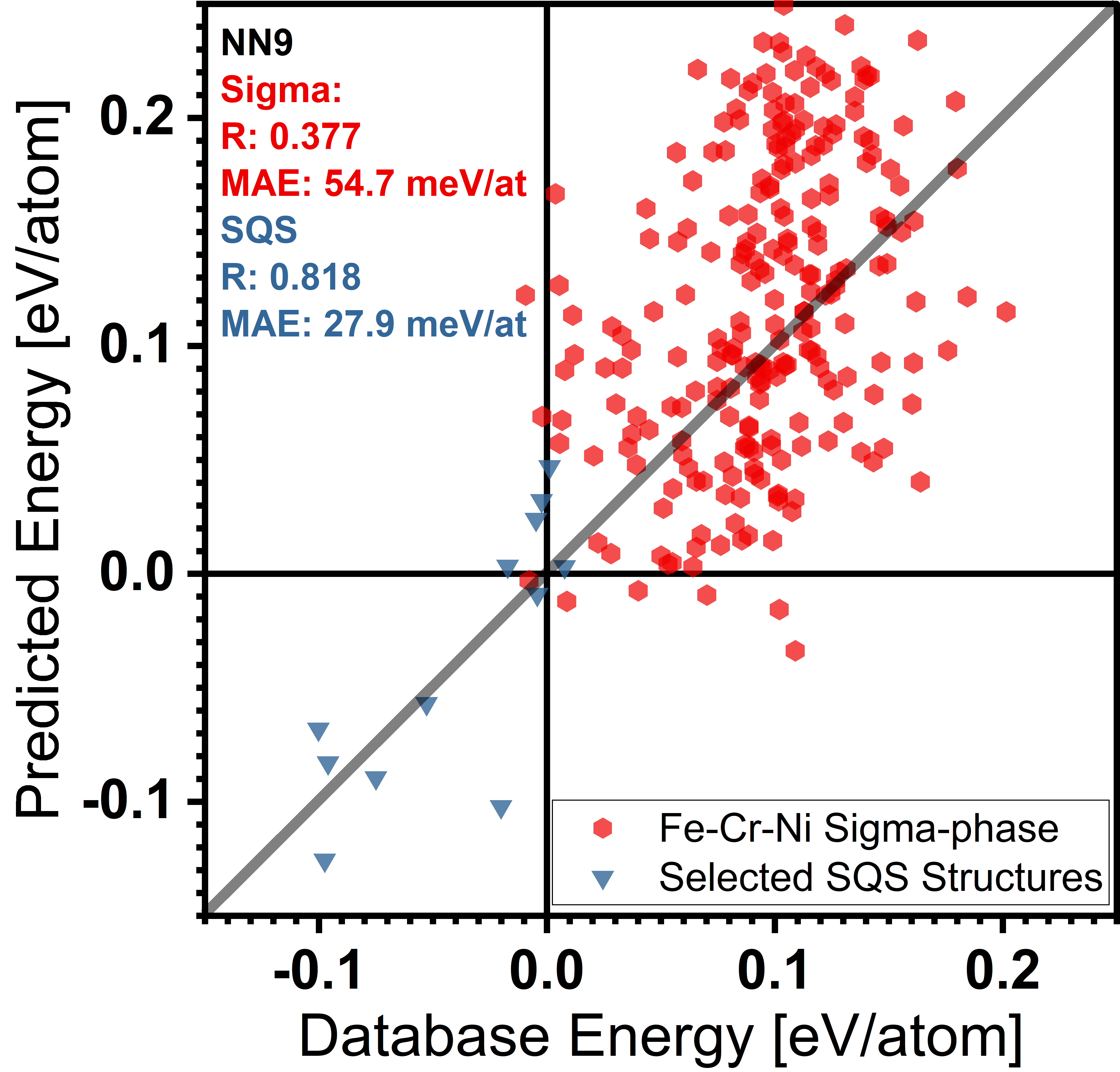}
    \hspace{0.03\textwidth}
    \includegraphics[width=0.3\textwidth]{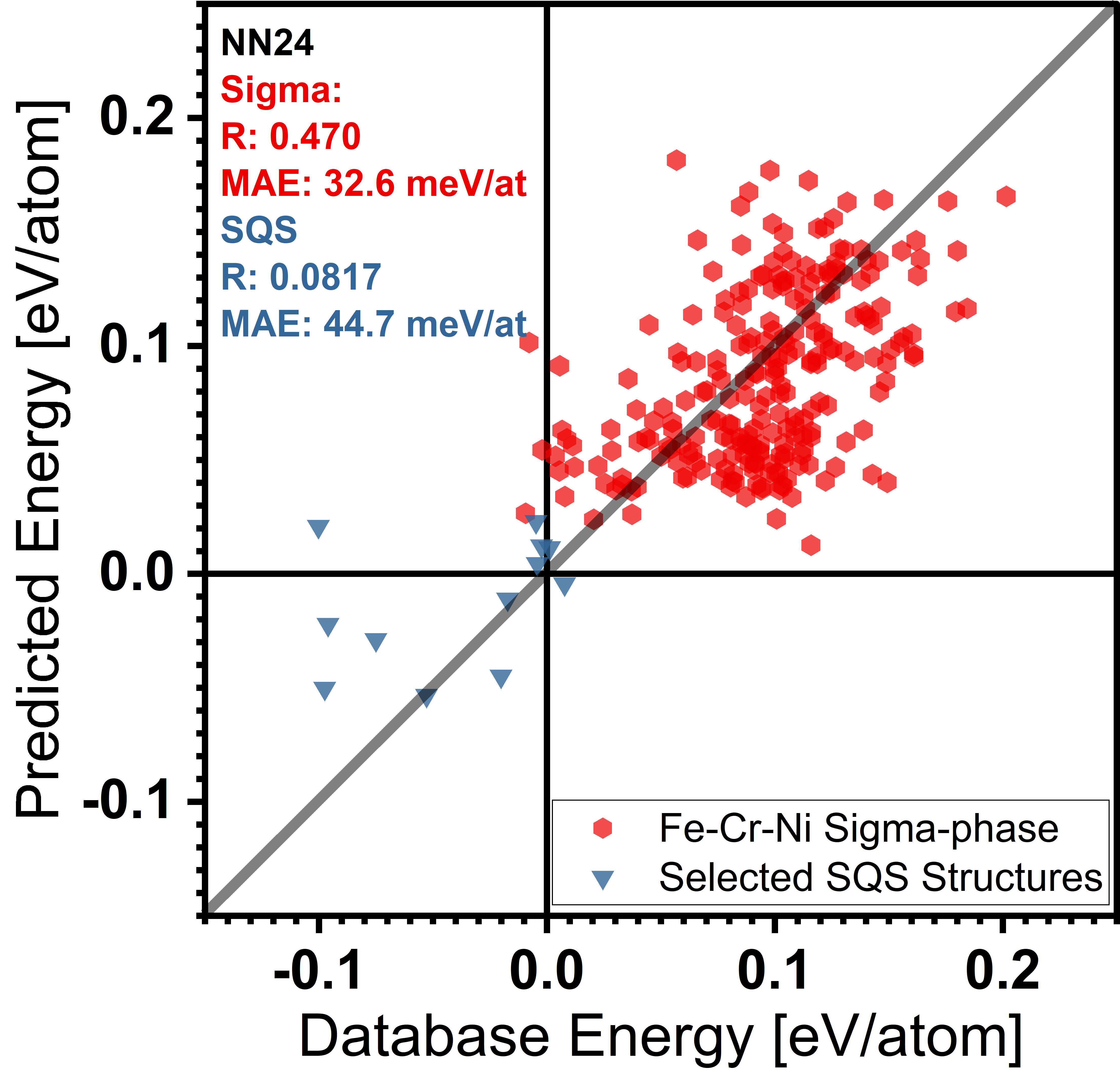}
    \hspace{0.03\textwidth}
    \includegraphics[width=0.3\textwidth]{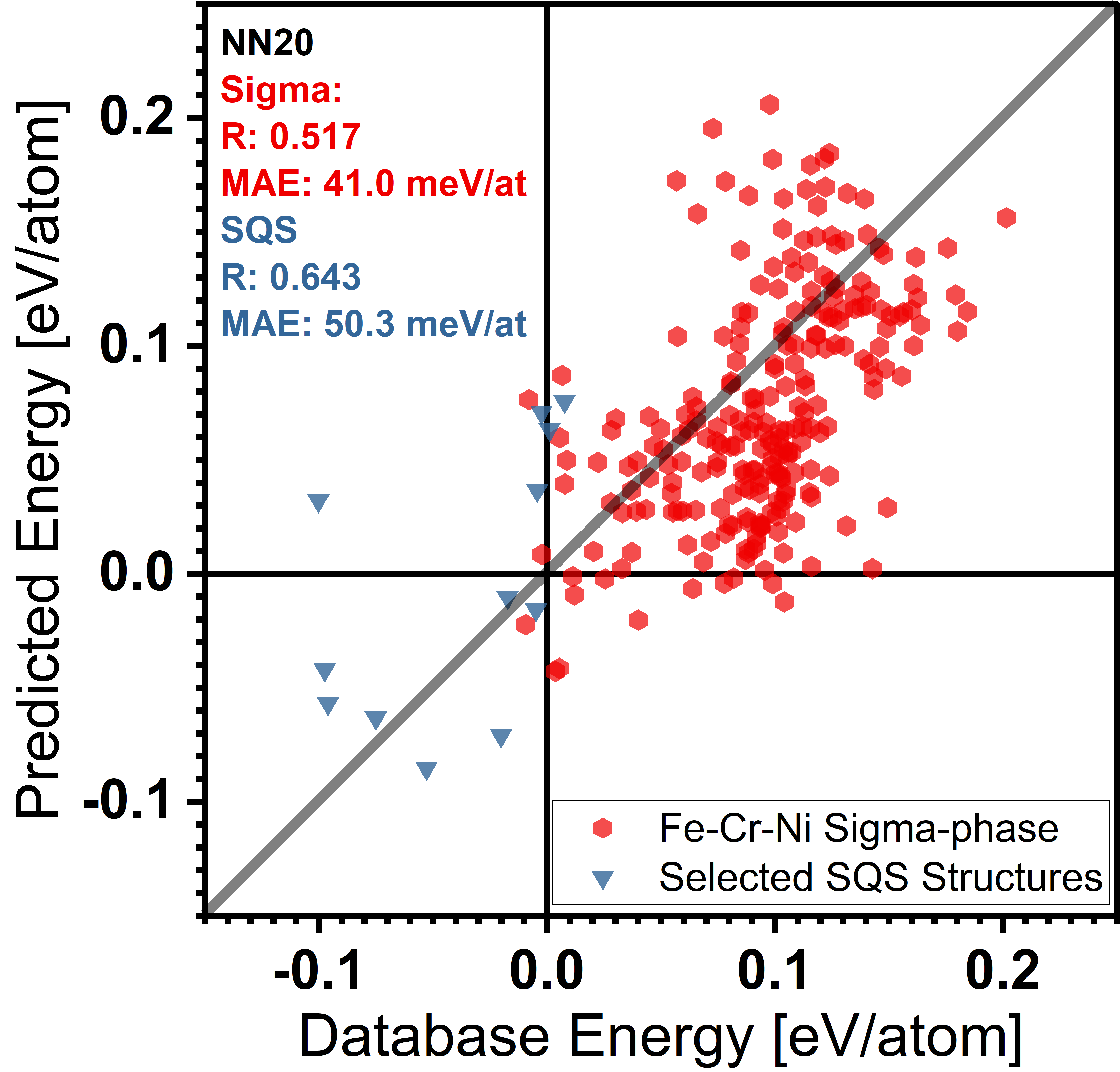}
    \caption{Performance of 3 selected neural networks on non-OQMD data described in \ref{sssec:Data}. Evaluated on (red) Fe-Cr-Ni $\sigma$-phase and (blue) SQS dataset. Networks organized by columns; optimized for (left) OQMD performance, (middle) predicting new materials, (right) size-constrained applications. Internally in the code, they are designated as NN9, NN20, and NN24 respectively.}
    \label{fig:sigmasqsperformance}
\end{figure}

In all cases, models created in the present paper were able to achieve approximately the same performance as on a random selection from the OQMD. To give a more in-depth analysis of the results, Figure \ref{fig:sigmasqsperformance} shows a magnified view of the predictions and basic statistics on the agreement between predictions and the database for the three models developed in the present work.

While all three models performed at around the same MAE level as for the OQMD, network optimized for new materials, the NN20 and NN24, performed better in the non-OQMD test cases of interest, providing major increases in correlations, significant for ranking of end-member configurations, except for 4 SQS configurations which were underestimated. The Pearson correlation slightly decreased in the first case and slightly increased in the second case. In both cases, the mean absolute error decreased by about 20\% compared to the OQMD-optimized model.

\subsection{Transfer Learning Capability} \label{ssec:transferlearningresults}
In this section, the technique of transfer learning is considered. It has been observed among deep learning models across a variety of domains \cite{tan2018survey,cirecsan2012transfer,chang2017unsupervised,george2018deep} and refers the to the ability of properly trained deep learning models to `transfer' their knowledge to related tasks. In the least complex approach, one does this by simply 'fine-tuning' the parameters of the model using new training data (from the new task). This methodology has shown in practice that deep neural networks are often able to transfer knowledge between different but related tasks. Such a problem is analogous to many others in materials science, where general knowledge is used to make meaningful statements without statistically significant patterns in locally available data. 

It is shown that a network trained on the OQMD database, which covers a broad yet limited spectrum of materials, can be quickly adjusted to materials outside of this spectrum with very little additional cost relative to the initial training. Specifically, the transfer learning capability of a network trained in this way on the set of all (243) Fe-Ni-Cr $\sigma$-phase 5-sublattice model endmembers, described in \ref{sssec:Data}, was tested. The ML model was first trained on a broad and general material dataset (OQMD) and then further trained (i.e., re-trained) for a given number of rounds on the new data (Fe-Ni-Cr $\sigma$-phase dataset) to adapt to the new system, while still conserving its broad knowledge, and can be thought of as fine-tuning a model to improve extrapolation outside of a prior knowledge space. 

\begin{figure}[H]
    \centering
    \includegraphics[width=0.6\textwidth]{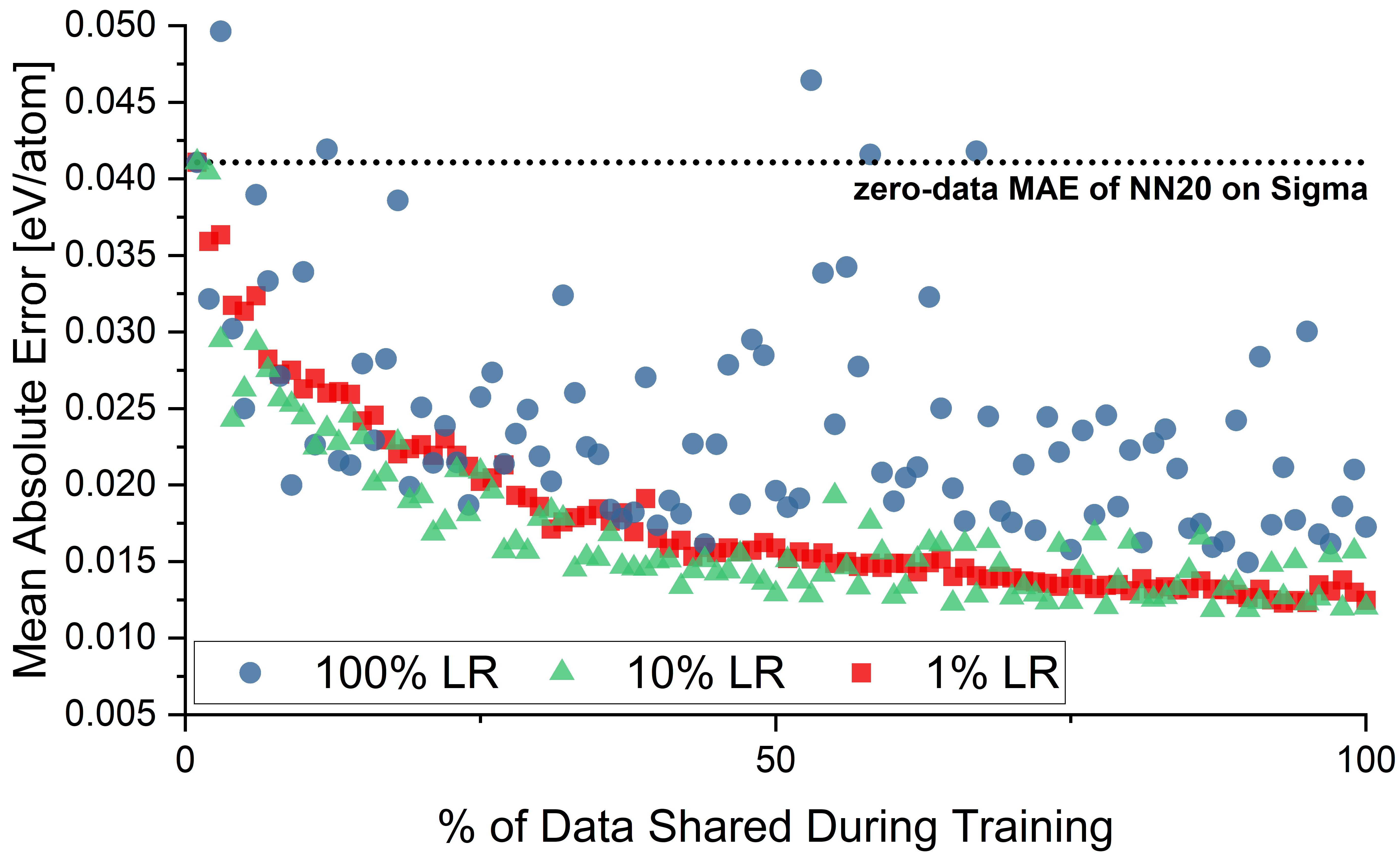}
    \caption{MAE evolution of NN20 model re-trained for 25 additional rounds on an increasing fraction of data from Fe-Cr-Ni $\sigma-$dataset. Presents the dependence of transfer learning from new data for different learning rates expressed as fractions of default ADAM learning rate (0.001).}
    \vspace{-12pt}
    \label{fig:transfersigmaLR}
\end{figure}

In order to achieve good performance, both the number of rounds and the learning rate have to be optimized. This can be accomplished by investigating the dependence of error on the fraction of available data while one of these parameters is fixed. Figure \ref{fig:transfersigmaLR} presents the dependence of transfer learning from new data for different learning rates expressed as fractions of default ADAM learning rate (0.001 shared across a vast majority of software). 

\begin{figure}[H]
    \centering
    \includegraphics[width=0.48\textwidth]{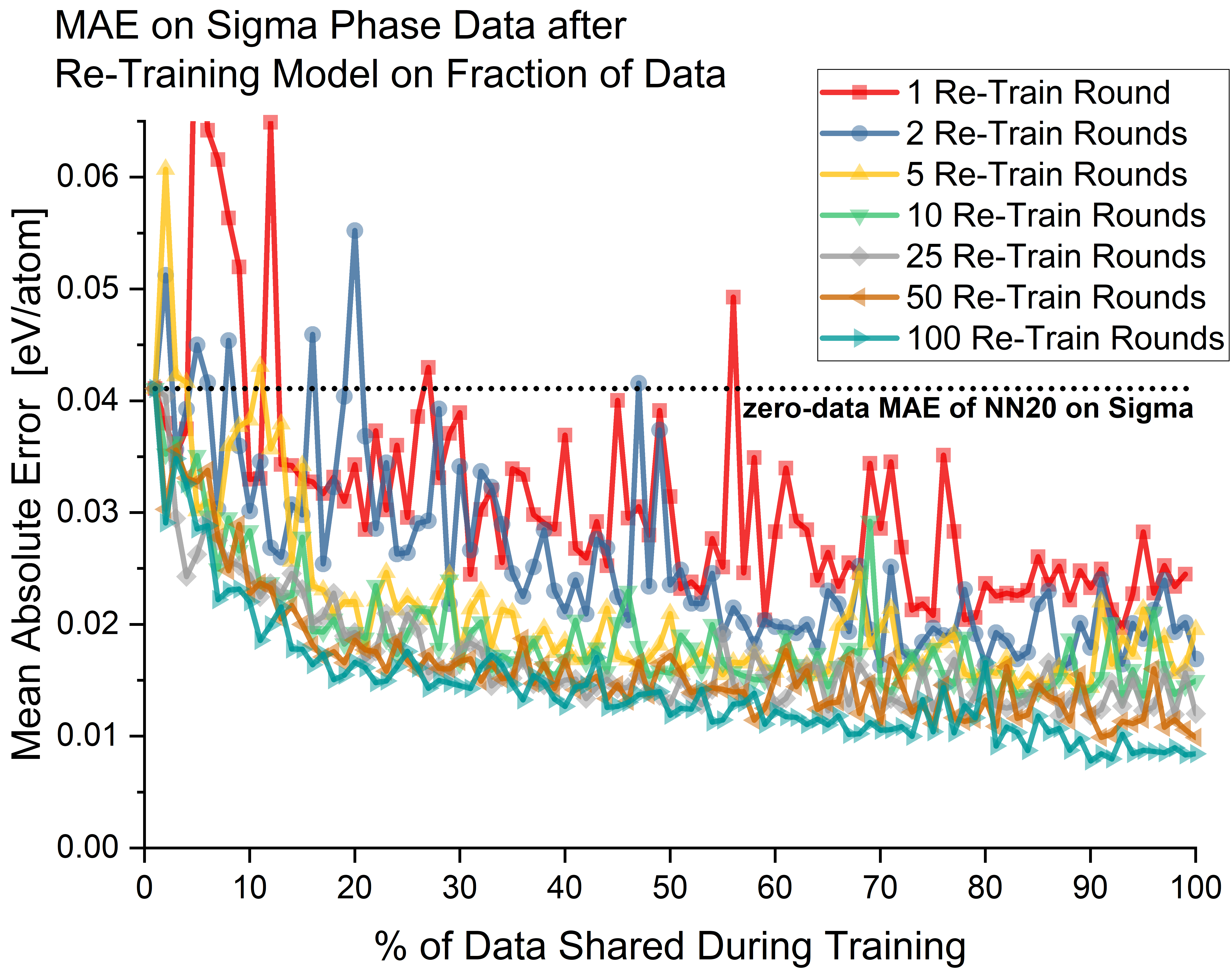}
    \includegraphics[width=0.48\textwidth]{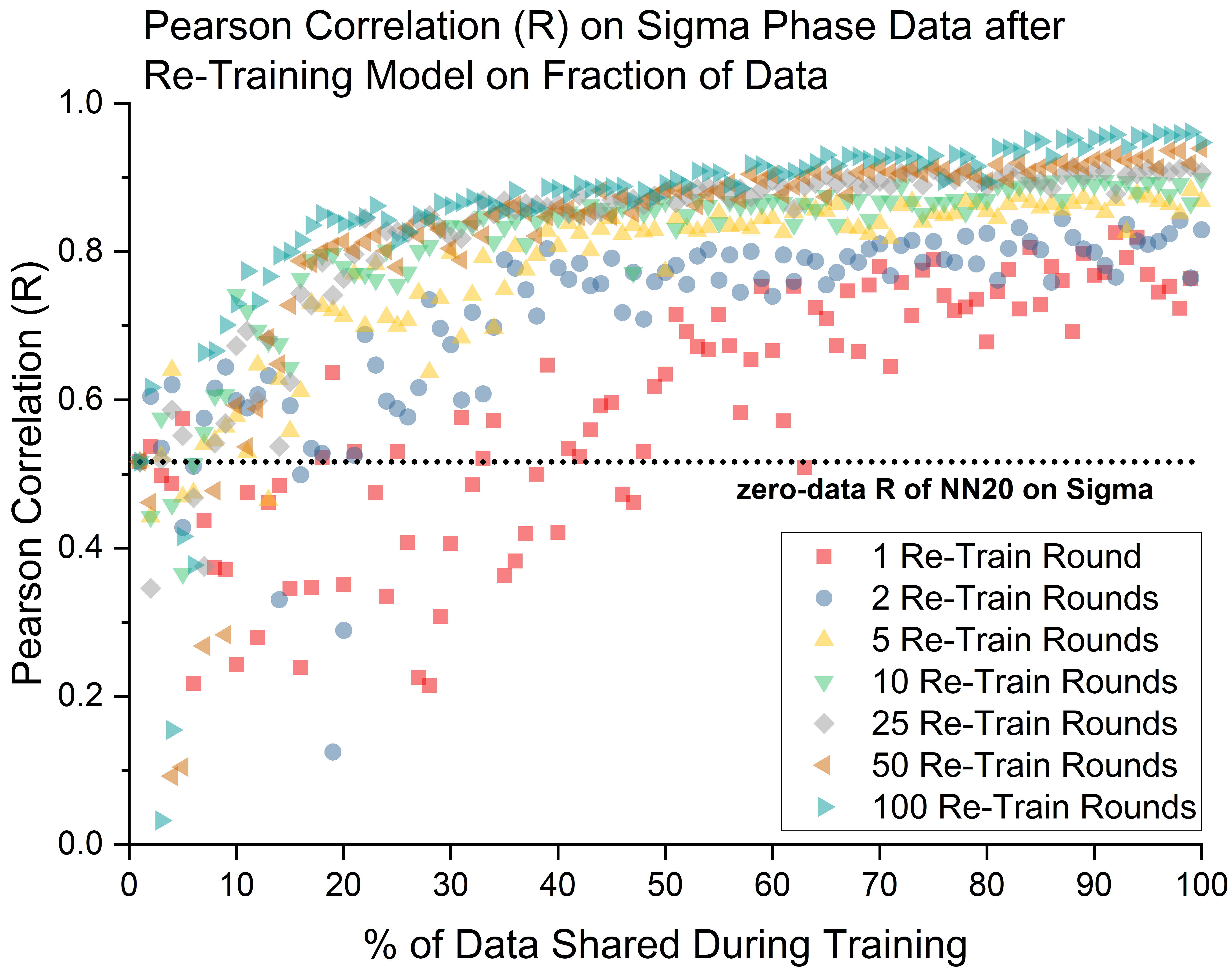}
    \caption{MAE and Person correlation (R) evolution of NN20 model re-trained at 10\% learning rate on an increasing fraction of data from Fe-Cr-Ni $\sigma-$dataset. Presents the dependence of transfer learning from new data for different re-training rounds numbers.}
    \vspace{-12pt}
    \label{fig:transfersigmaARR}
\end{figure}

As shown, in this case, the default learning rate (100\%)cannot be used for the transfer learning as it will adjust network parameters in both an unreliable and detrimental fashion, resulting in poor performance on the whole system of interest (both training and test sets) as shown in Figure \ref{fig:transfersigmaLR}. The same behavior would be observed if the process were conducted using an automated model design available in software such as MATLAB or Mathematica. The 10\% learning rate provided reliable enough outcomes and allowed a better performance improvement given little data, relative to using a 1\% learning rate (relative to the default). The second parameter to be optimized was the number of re-training rounds, as presented in Figure \ref{fig:transfersigmaARR}.

\begin{figure}[H]
    \centering
    \includegraphics[width=0.24\textwidth]{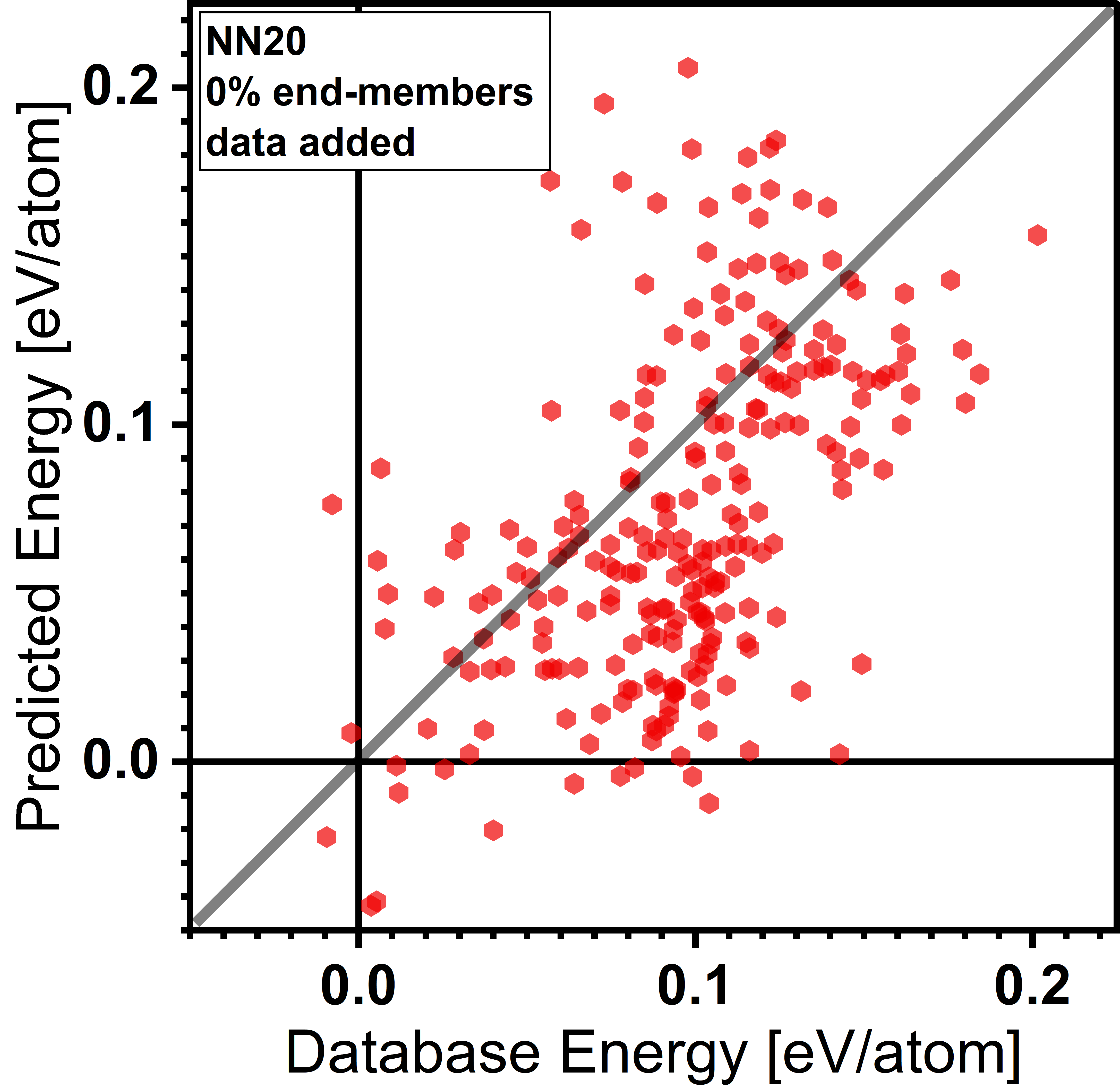}
    \includegraphics[width=0.24\textwidth]{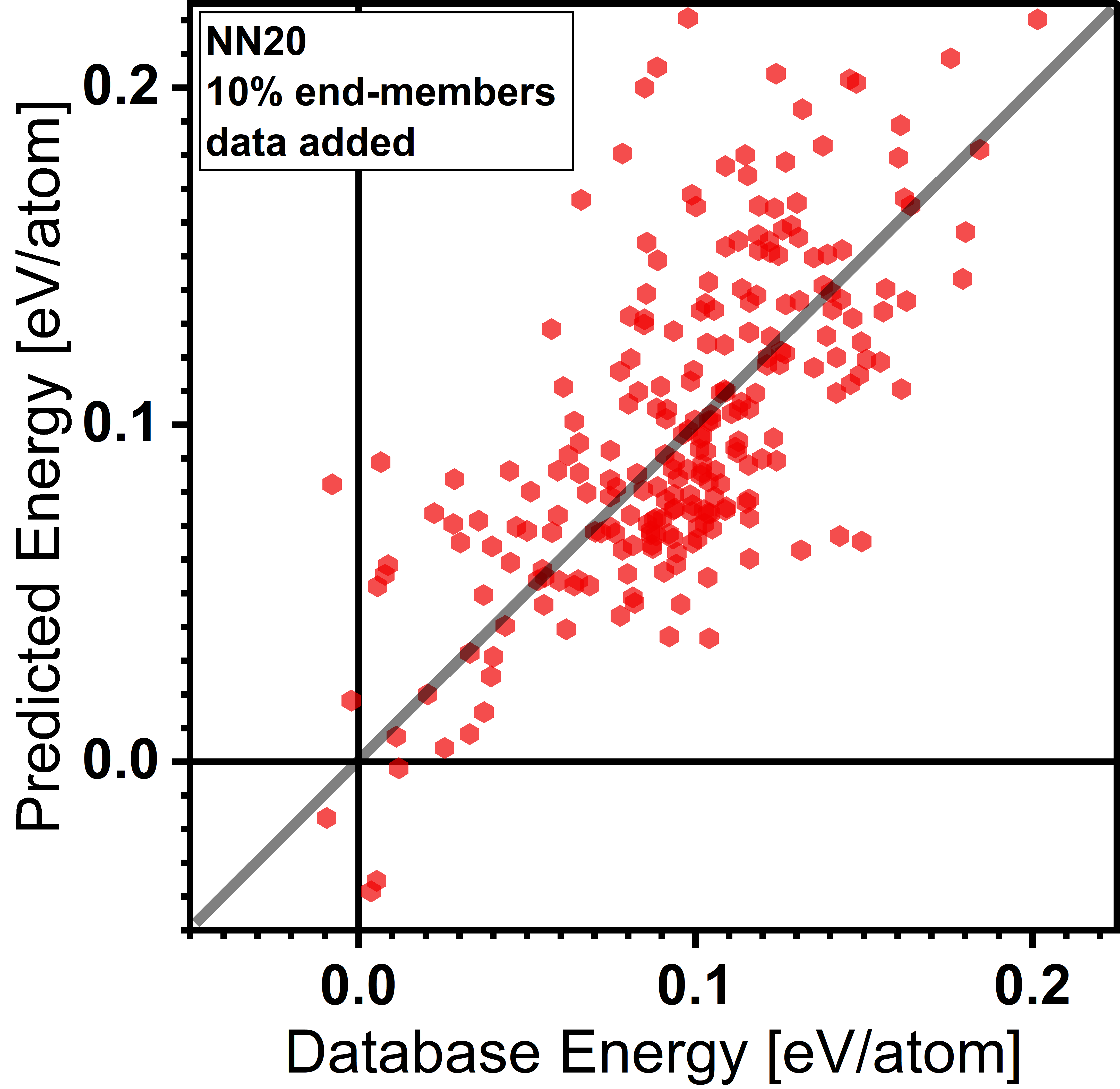}
    \includegraphics[width=0.24\textwidth]{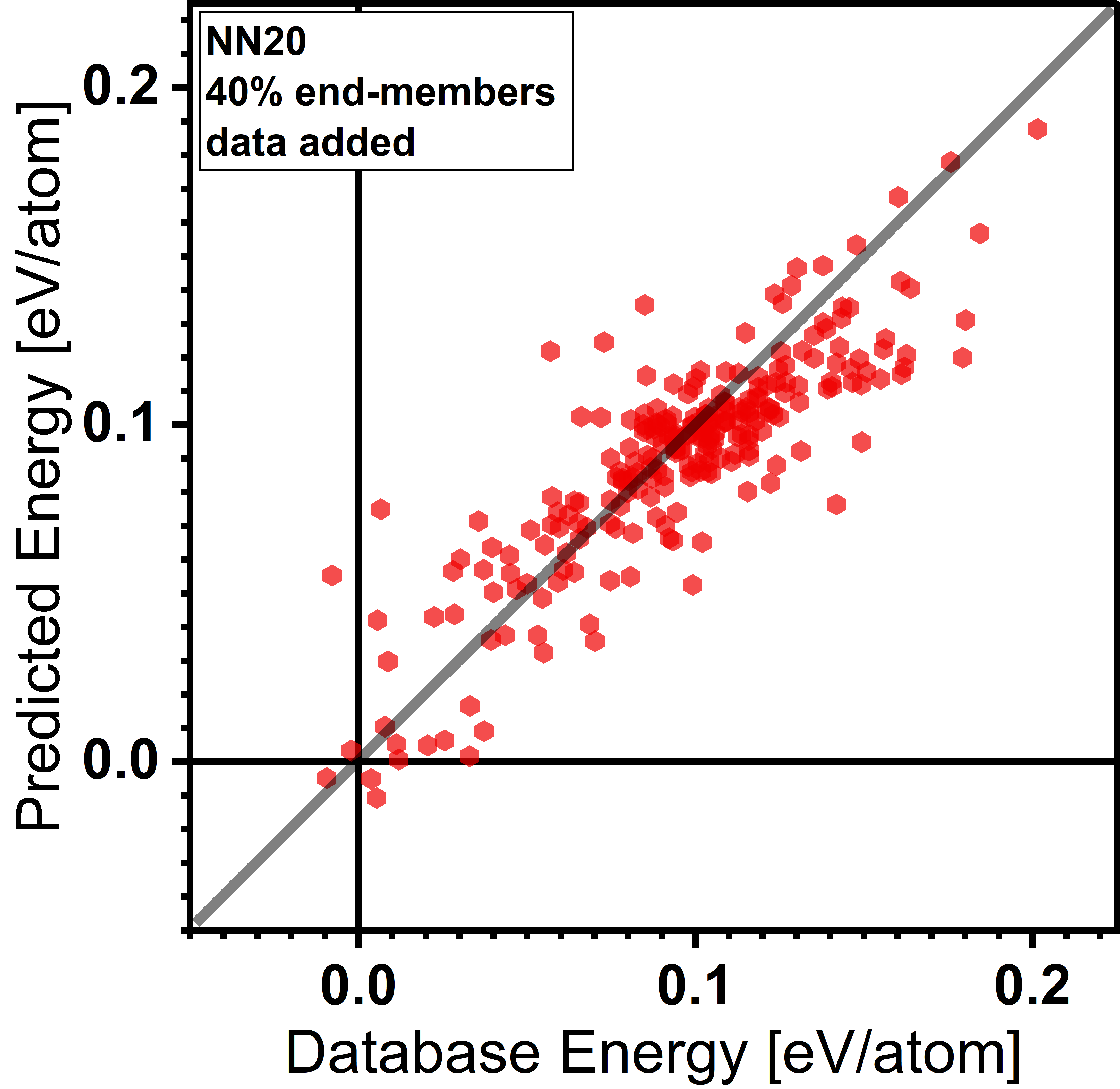}
    \includegraphics[width=0.24\textwidth]{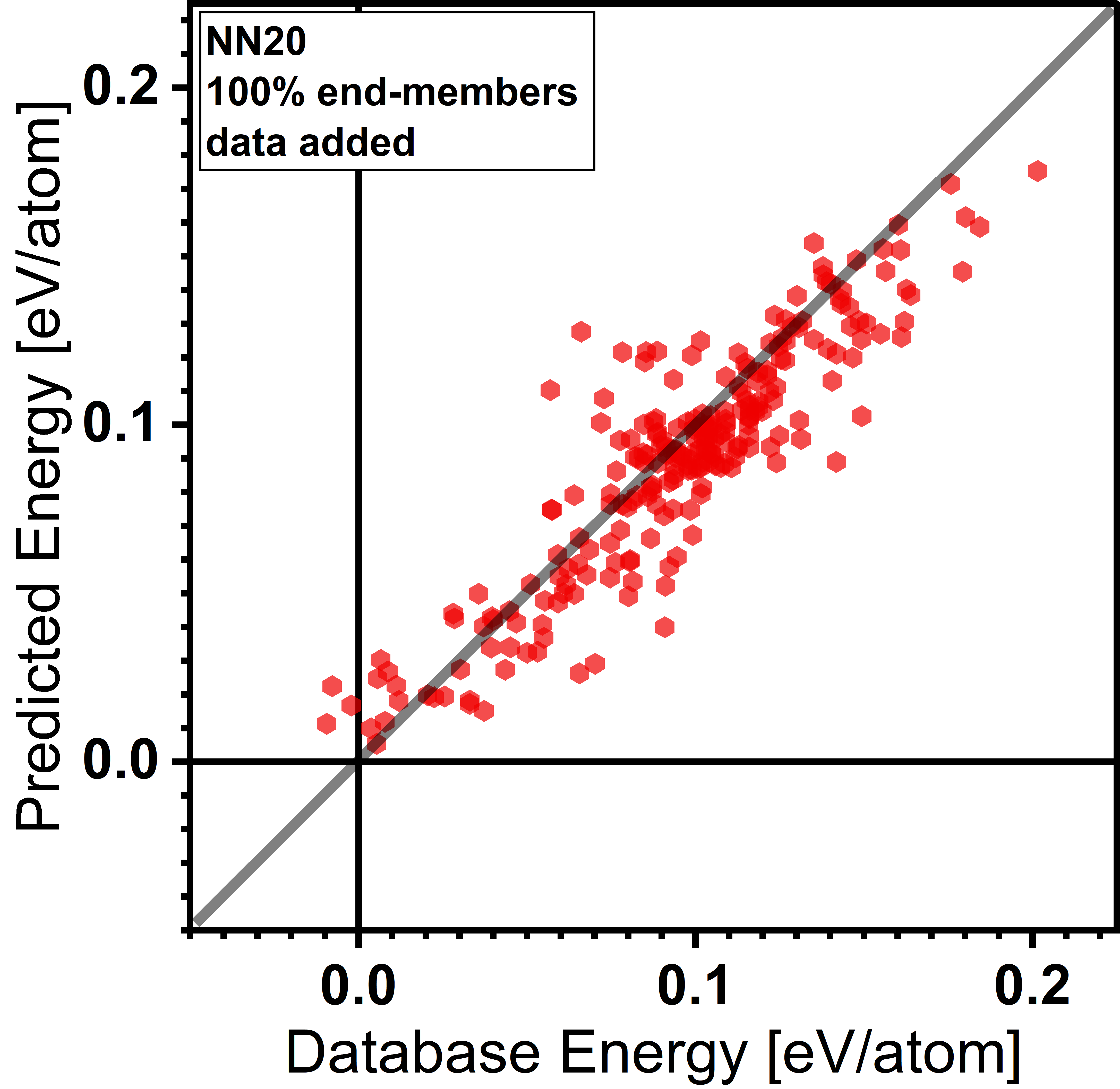}
    \caption{Performance of a new-materials-optimized network (NN20) on $\sigma$-phase data. Left-to-right: as trained on the OQMD, with additional training on 10\%, 40\%, and 100\% of the Fe-Cr-Ni $\sigma-$phase end-member data. The points on the figure correspond to all end-members (both training and testing data). Corresponding MAE and R are presented in Figure \ref{fig:transfersigmaARR} (gray rhombus points).}
    \vspace{-12pt}
    \label{fig:transfersigma}
\end{figure}

Figure \ref{fig:transfersigmaARR} shows that use of too few retraining rounds causes unreliable outcomes, while too many causes overfitting for low amounts of new data. In the case of Fe-Cr-Ni $\sigma-$dataset, retraining for 10 or 25 rounds provides balanced results across the whole dataset. With parameters for the process set to 10\% learning rate and 25 additional rounds, the performance can be evaluated graphically, as presented in Figure \ref{fig:transfersigma}.

As depicted, adding just 10\% of DFT-calculated data (24/243 endmembers) provided a significant improvement in the prediction quality over the system, including the other 90\%  was never shown to the model. This result indicates that the models in the present paper can be combined with partial data obtained through DFT calculations to create accurate predictive tools for a specific closed material system, such as sublattice endmembers, and potentially limit the number of calculations required within the study. This can then provide the ability to investigate broader material search spaces at a given computational cost.

\begin{figure}[H]
    \centering
    \includegraphics[width=0.48\textwidth]{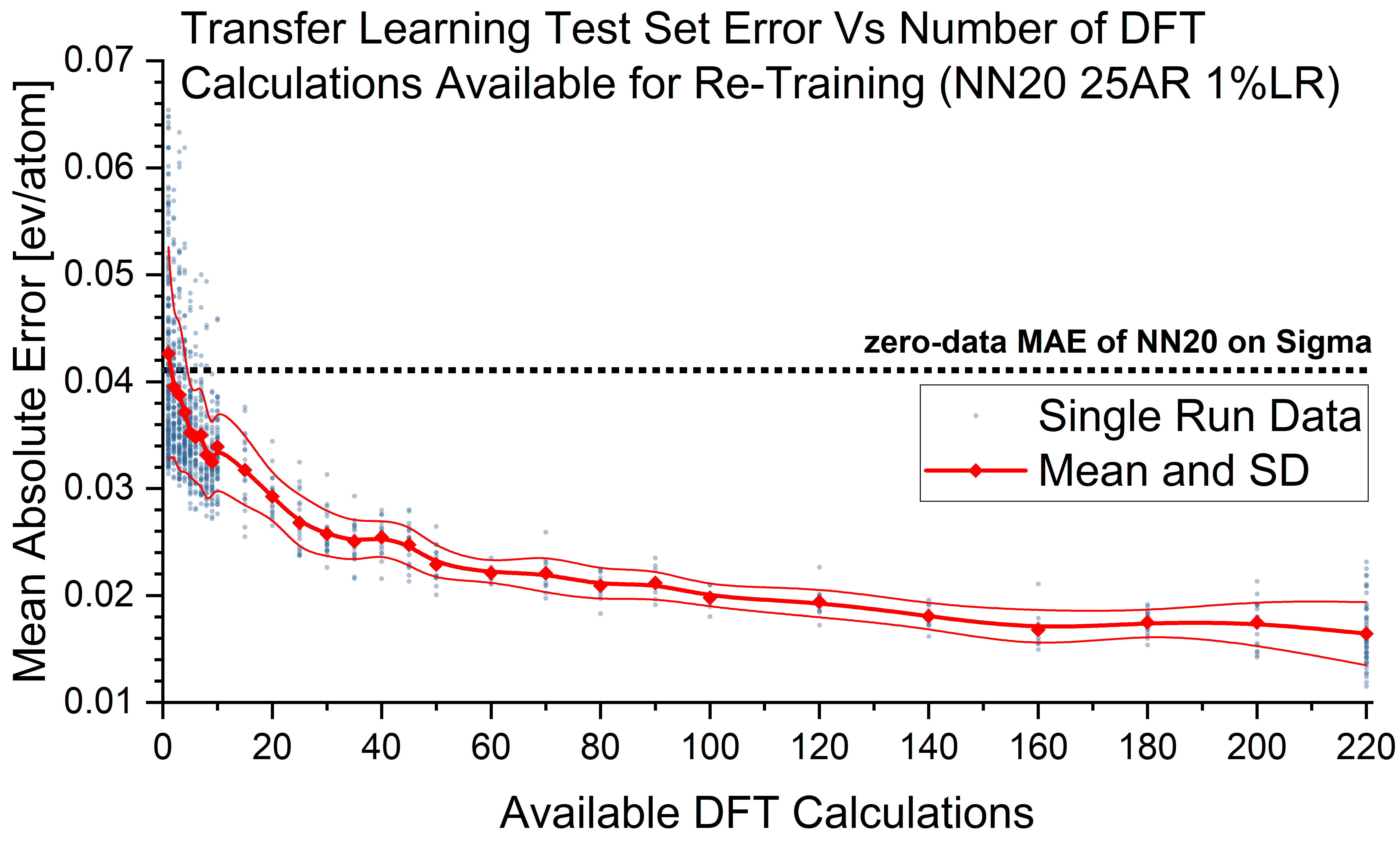}
    \includegraphics[width=0.48\textwidth]{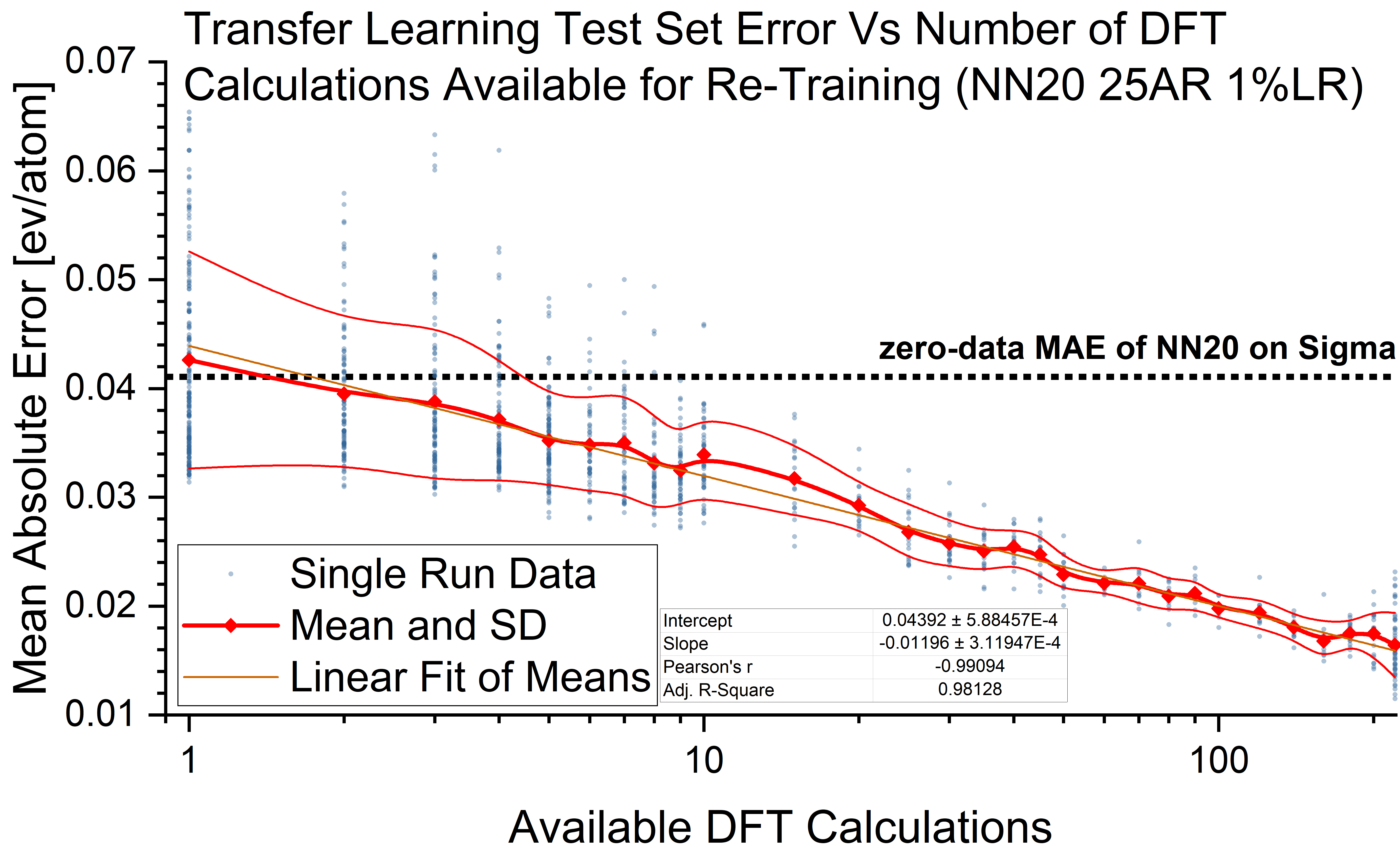}
    \caption{MAE of predictions evaluated on test set data vs number of newly available training datapoints. 1180 blue points correspond to single transfer learning processes. Red plot gives mean MAE and standard deviation. Both plots contain the same data.}
    \label{fig:transfersigmaVsDatapoints}
    \vspace{-12pt}
\end{figure}

Furthermore, the presented transfer learning capability could be used for a more broad materials exploration without a well-defined finite search space like the ternary Fe-Cr-Ni $\sigma-$phase. In such a case, it is better to evaluate and report the performance of the model on a test set that wasn't presented during the training and report, as a function of the number of added data points (new DFT calculations). With such a problem statement, the transfer learning process has been repeated 1180 for the statistical significance of the outcomes, which are presented in Figure \ref{fig:transfersigmaVsDatapoints}.

As presented in Figure \ref{fig:transfersigmaVsDatapoints}, adding just a small number of new data points allows to nearly half the MAE (around 20 datapoints). Furthermore, evident from the right plot, the mean performance increase is on average linear in log-lin scale and highly predictable ($R^2=0.98$).

\subsection{Model Limitations} \label{ssec:modellimitations}
As with any modeling tool, this modeling effort has some inherent limitations, coning from both data and methods used to create it. The most significant one comes from the type of data used for training of the model, where all data points correspond to DFT-relaxed structures, sitting in local minima in the configuration energy landscape. Thus, all energy predictions are given under an assumption that the input structure is fully relaxed with DFT settings inherited from the OQMD database \cite{Saal2013MaterialsOQMD}. At the same time, since the model was trained on many local energy minima configurations analyzed on the level of single-atom chemical environments, it should be able to approximate values for unrelaxed structures based on substitution from prototypes or similar compounds. Testing of this is performed by Ward 2017 \cite{Ward2017IncludingTessellations}, where it is shown that (a) in most of the test cases, the before-after relaxation energy difference is negligible in comparison to the DFT-ML difference for Ward 2017 model and usually much lower than the test MAE for models discussed in this work, and (b) in some particular cases (\ce{Li_6CaCeO_6}) can be very high.

When faced with a new configuration, the model can thus either be used to (1) give an accurate prediction if the configuration is already relaxed or (2) give an approximate result that needs to be validated with DFT if confidence in the result is needed. This is inherent to all structure-informed ML models. One possible solution to partially mitigate this limitation is to perform relaxation using the model, which should work reasonably well for most materials. 

\begin{figure}[H]
    \centering
    \includegraphics[width=0.65\textwidth]{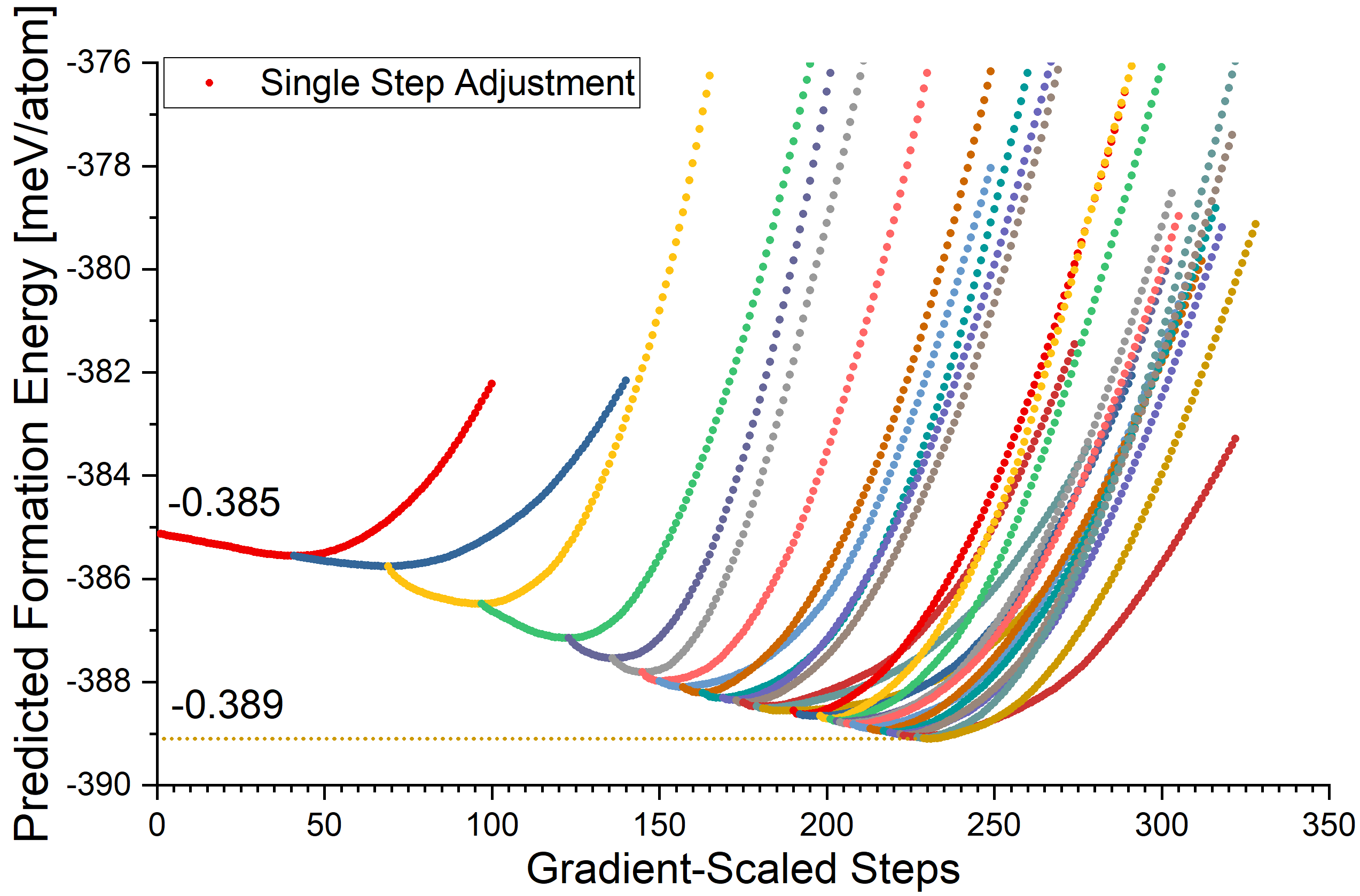}
    \caption{Local energy landscape relaxation of \ce{Pd_8Zn_5} $\gamma$-brass crystal structure guided by Novel Material Model (NN20).}
    \label{fig:localrelaxationpdzn}
    \vspace{-12pt}
\end{figure}

Discussion of such relaxation procedure in detail is extensive and beyond the scope of this work, yet a preliminary approach was constructed using the Novel Material Model (NN2) and deployed on all 16 end-members of Pd-Zn $\gamma$-brass crystal structure in an iterative fashion. At each iteration, first, the local energy gradient for each atom was calculated by comparing the starting configuration with perturbations in x, y, z directions. Then, all atoms were displaced proportionally to the gradient in 100 discrete steps, reaching some local minimum, which acted as a starting point for the next iteration. An example for \ce{Pd_8Zn_5} is presented in Figure \ref{fig:localrelaxationpdzn}.

As shown in \ref{fig:localrelaxationpdzn}, the resulting relaxation reduced predicted formation energy by 4 meV/atom for this particular end-member. In the other 15 cases, results were similar, ranging between near 0 and 15 meV/atom, converging into fine local minima, expected to correspond with true local relaxations; however, extensive research into the problem is needed before conclusions can be drawn.

\subsection{End-User Implementation - SIPFENN} \label{ssec:SIPFENN}

One of the main objectives of the present paper was to create a tool that is transparent, easy to use by the research community, and easily modifiable. This lead to the creation of SIPFENN (Structure-Informed Prediction of Formation Energy using Neural Networks) software. SIPFENN provides the user with near-instant access to the models presented in \ref{sssec:DesignedModels}. In the future, this selection will likely be further expanded. On the user side, the use of the software is as easy as selecting one of the models, specifying a folder containing structure information files like POSCARs \cite{POSCARFile} or CIFs \cite{Hall1991TheCrystallography}, running the predictions, and saving results.

\begin{figure}[H]
    \centering
    \includegraphics[width=0.67\textwidth]{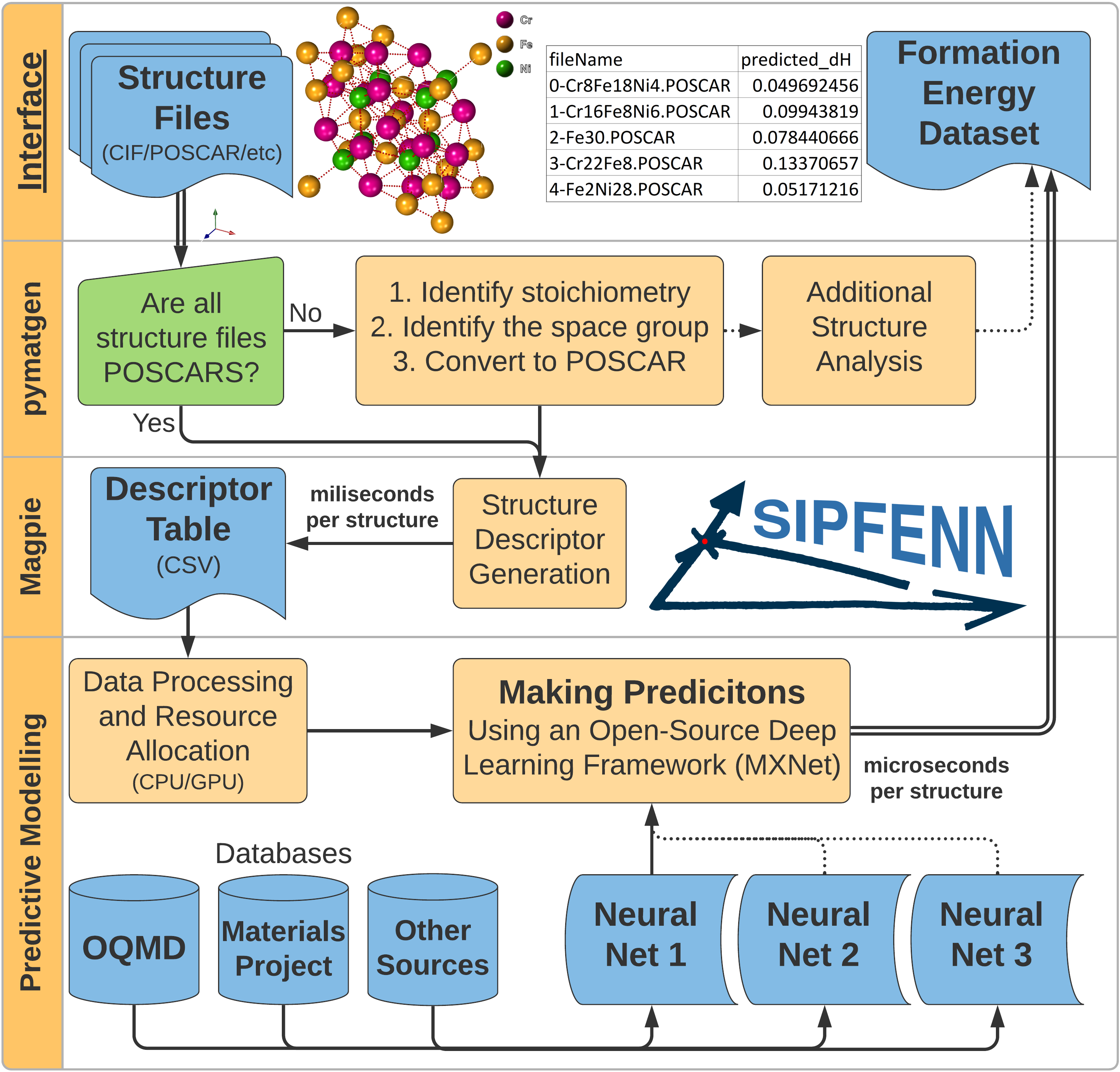}
    \caption{SIPFENN schematic description of operation.}
    \vspace{-12pt}
    \label{fig:sipfenn}
\end{figure}

\begin{wrapfigure}{r}{0.5\textwidth}
    \centering
    \vspace{-12pt}
    \frame{\includegraphics[width=0.49\textwidth]{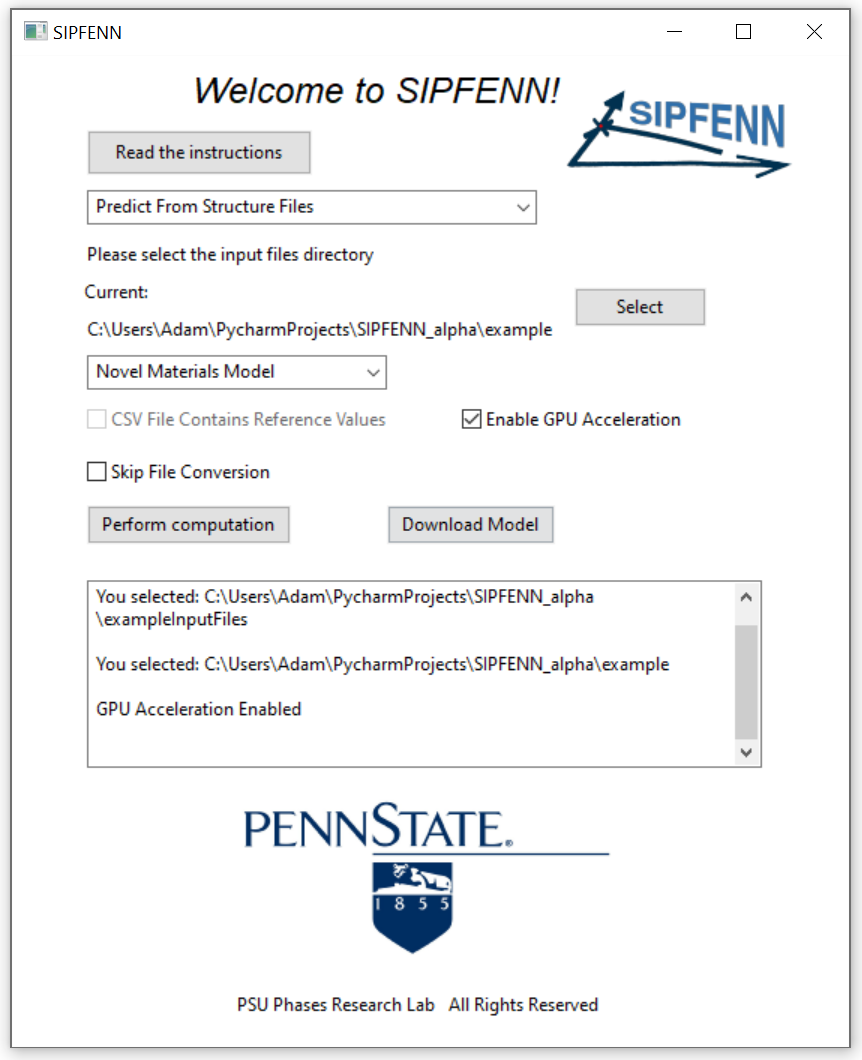}}
    \caption{A snapshot of the graphical user interface of SIPFENN.}
    \label{fig:sipfennGUI}
    \vspace{-24pt}
\end{wrapfigure}

SIPFENN was written entirely in Python to allow other researchers to easily modify it and adjust it to specific needs. Its schematic of operation is presented in Figure \ref{fig:sipfenn}. In broad scope, it first performs the structure analysis and modifications using the Python Materials Genomics library (pymatgen) \cite{Ong2013PythonAnalysis}. In the current implementation, it imports all structure files, analyzes the stoichiometry, creates unique names based on that, and exports them as POSCAR files. This is a rather simple task, however pymatgen is a powerful tool with a suit of more complex analytical tools that can be quickly implemented into SIPFENN by the user with even basic Python skills. Following the analysis, SIPFENN runs java-based Magpie \cite{Ward2016AMaterials} which calculates a descriptor for every imported structure and exports the result as a CSV file. This file is a descriptor table, where each row corresponds to a single material, and which can be stored and re-used later to run multiple predictive models at a fraction of the original computation time. It can also be used to create datasets for training procedures by replacing the last column with calculated or experimental values of formation energy.

Finally, the descriptor table is imported into the MXNet library framework, allocated into the CPU or GPU memory based on user selection, and evaluated using the selected predictive model. Once results are obtained, they are exported in CSV format and can be analyzed by any spreadsheet software such as Microsoft Excel.

SIPFENN was planned as a command-line tool, however, it was recognized that some users, especially those with little computational background, may find that difficult. Therefore, a simple graphical user interface (GUI) was created using wxPython library. It incorporates all the capabilities of the command line version. Furthermore, it lets the user download the predictive models from a repository in a single click. A sample snapshot of the GUI before performing calculations is presented in Figure \ref{fig:sipfennGUI}.

\section{Conclusions} \label{ssec:Conclusions}
In the present paper new machine learning models and a ready-to-use tool were created, based on the dataset and descriptor design by Ward et al. \cite{Ward2017IncludingTessellations}. Models reported in this work significantly improve upon existing methods, both in terms of performance and accessibility. For the most direct comparison, one of the designed models has been optimized for performing well on a random subset of the OQMD database and achieved an MAE of 28 meV/atom, compared to 80 meV/atom in the original  Ward et al. paper \cite{Ward2017IncludingTessellations}, and to 38 meV/atom in the most recent model called IRNet \cite{Jha2019IRNet}. Furthermore, it was shown that the error of the model is lowered when applied to the problem of finding the convex hull energy, achieving levels comparable with the current state-of-the-art approaches \cite{Jha2018ElemNet:Composition, Goodall2020PredictingStoichiometry}.

In addition, using appropriate overfitting mitigation efforts, such as Dropout and L2 regularization, models tuned for generalization to other types of materials datasets were developed. To test this, the models were evaluated on two datasets not contained within the OQMD, namely all end-members (243) of 5-sublattice topologically-close-packed Fe-Cr-Ni Sigma-phase \cite{Feurer2019Cr-Fe-NiCalculations, Hsieh2012OverviewSteels} and a few selected random-solution-approximating SQS \cite{Zunger1990SpecialStructures, Shin2006ThermodynamicStructures, Jiang2004First-principlesStructures}. The MAE values for these two test sets were found to be close to the values obtained on a test set from the OQMD. This exemplifies that the models are able to generalize to new datasets.

Furthermore, it was shown that models created within the present paper can be used for transfer learning, where vast knowledge of a broad spectrum of materials is combined with as little as a few DFT-datapoints from a specific materials system to provide excellent results within that specific system. Such at least partially process mitigates the issue of low data availability, present in numerous materials science problems, and consequently allows users to investigate a broader scope of materials at the same computational cost.

Finally, the three neural network models designed within the present paper were used, in conjunction with additional software, to create an end-user tool called SIPFENN. SIPFENN's capabilities extend far beyond allowing validation of the presented results. It is implemented to work without any intensive computations on the user side, using models accessible from a repository, requiring only a quick one-click model download  to run. It is very fast thanks to using one of the industry's leading ML frameworks capable of well-optimized computations on GPUs. Furthermore, it is an open-source tool written in Python, which can be easily modified to specific needs in a straightforward way without extensive changes in the code.

\section{Acknowledgements}
the present work was financially supported by the ICDS Seed Grant from the Pennsylvania State University, the Office of Naval Research (ONR) via Contract No. N00014-17-1-2567, the National Science Foundation (NSF) via Grant No. CMMI-1825538, and the Department of Energy (DOE) via Award Nos. DE-FE0031553 and DE-EE0008456. 

We would like to thank Zhengqi Liu for his help implementing the graphical user interface, Dr. ShunLi Shang for providing the Fe-Cr-Ni $\sigma$-phase dataset, and Brandon Bocklund for providing the SQS dataset.

\section{Software and Data Availability}
The most recent version of SIPFENN code is available through Penn State's Phases Research Lab website at 
\hyperlink{www.phaseslab.com/sipfenn}{www.phaseslab.com/sipfenn} in (1) a minimal version that can be run on pre-computed descriptors in CSV format as well as (2) ready-to-use version with pre-compiled Magpie \cite{Ward2016AMaterials}. SIPFENN contains hard-coded links to neural networks stored in the cloud that can be downloaded at a single-click (see Figure \ref{fig:sipfennGUI}). 

All neural networks are stored in both (1) open-source MXNet format maintained by Apache Foundation and used within SIPFENN, and in (2) closed-source WLNet format maintained by Wolfram Research and having the advantage of even easier deployment, as well as guaranteed forward compatibility with future versions of Wolfram Language.

For ensured longevity of results, SIPFENN neural networks are stored through the courtesy of Zenodo.org service under \hyperlink{doi:10.5281/zenodo.4006803}{doi:10.5281/zenodo.4006803} at the CERN’s Data Centre.

\pagebreak

\printbibliography

\pagebreak

\appendix

\section{Machine Learning Overview} \label{appedix1}
The class of deep learning methods has been remarkably successful in recent years in applications ranging from computer vision to natural language processing and simulations of quantum systems \cite{lecun2015deep,silver2016mastering,devlin2018bert,carleo2017solving}. Although deep neural networks have existed for a long time \cite{rosenblatt1958perceptron}, and had been successfully applied to computer vision tasks \cite{lecun1995comparison,lecun1990handwritten,lecun1998gradient}, a major breakthrough was the AlexNet network \cite{krizhevsky2012imagenet}, which dramatically improved the accuracy achievable on large-scale image classification. Following this success, deep neural networks have been very intensively studied and applied to a variety of problems \cite{lecun2015deep,silver2016mastering,devlin2018bert}. Deep neural networks are particularly effective when applied to regression problems, where one is learning a functional relationship between a feature and a prediction. For many problems, deep neural networks are able to achieve significantly better performance than competing machine learning methods, due to their ability to learn more complex relationships. With materials science being a field where many complex dependencies are intertwined, it is to be expected that this superior pattern recognition can carry over to the improvement in the prediction of material properties.

\subsection{Regression Problem Formulation and Artificial Neural Networks}
\label{ssec:regressionformulation}

The general formulation of a regression problem in statistical machine learning is to find a function $f:X\rightarrow Y$ which minimizes the risk \cite{vapnik1999overview}, also known as loss or expected error.
\begin{equation}\label{true_risk_app}
    R(f) = \mathbb{E}_{x,y\sim \mathcal{P}} l(y,f(x)).
\end{equation}
Here $X$ denotes a space of input features, $Y$ denotes an output space, the expectation above is taken over an unknown distribution $\mathcal{P}$ on $X\times Y$ (representing the true relationship between inputs and outputs), and $l$ is a given loss function. The goal is to find a function $f$ which accurately predicts the (potentially random) output $y$ given an input $x$.

In the present work, $x\in X$ represents the input features (descriptor) characteristic of the material, and $y\in Y$ represents the formation energy. The distribution $\mathcal{P}$ represents the true material-property relationship between given descriptor $x$ and corresponding formation energy. This relation may not be as simple as mapping a given structure to an energy since different DFT methodologies may give different results, based on many variables, such as employed functionals. \cite{CharlesW.BauschlicherJr.1995AFunctionals, Alturk2017ComparisonMaterial} Consequently it is useful to describe this relationship via a probability distribution. Furthermore, the loss function considered in the present paper is the commonly used $\ell^1$ or absolute error (AE) loss function $l(y_1,y_2) = |y_1-y_2|$. 

In practice, the distribution $\mathcal{P}$ is not known. Indeed it is this relationship that one is trying to learn in the first place. Instead, what is available is data $\{(y_i,x_i)\}_{i=1}^n$, which is sampled from $\mathcal{P}$. From this one forms the empirical risk \cite{hastie2009elements,vapnik2013nature}
\begin{equation}\label{empirical_risk_app}
    L(f) = \frac{1}{n}\displaystyle\sum_{i=1}^n l(y_i, f(x_i)),
\end{equation}
and seeks a function $f$ which minimizes the empirical risk, also known as the training error.

In addition, one must specify the type of relationship that is expected to be found between the inputs $x_i\in X$ and the predictions $y_i\in Y$. This is done by restricting the function $f$ to a specific class. For instance, by restricting $f$ to be linear, which corresponds to looking for a linear relationship between $x_i$ and $y_i$, one obtains a linear regression. On the other hand, choosing $\mathcal{F}$ to be a reproducing kernel Hilbert space of functions on $X$ with the same loss $l$ one obtains the kernel ridge regression method. Thus in order to fit the model, the training error is minimized over a specific class of function $\mathcal{F}$, i.e. one solves the optimization problem
\begin{equation}\label{empirical_risk_min_eq}
    f^* = \arg\min_{f\in \mathcal{F}} L(f) = \arg\min_{f\in \mathcal{F}} \frac{1}{n}\displaystyle\sum_{i=1}^n l(y_i, f(x_i)).
\end{equation}

\begin{figure}
    \centering
    \includegraphics[width=0.75\textwidth]{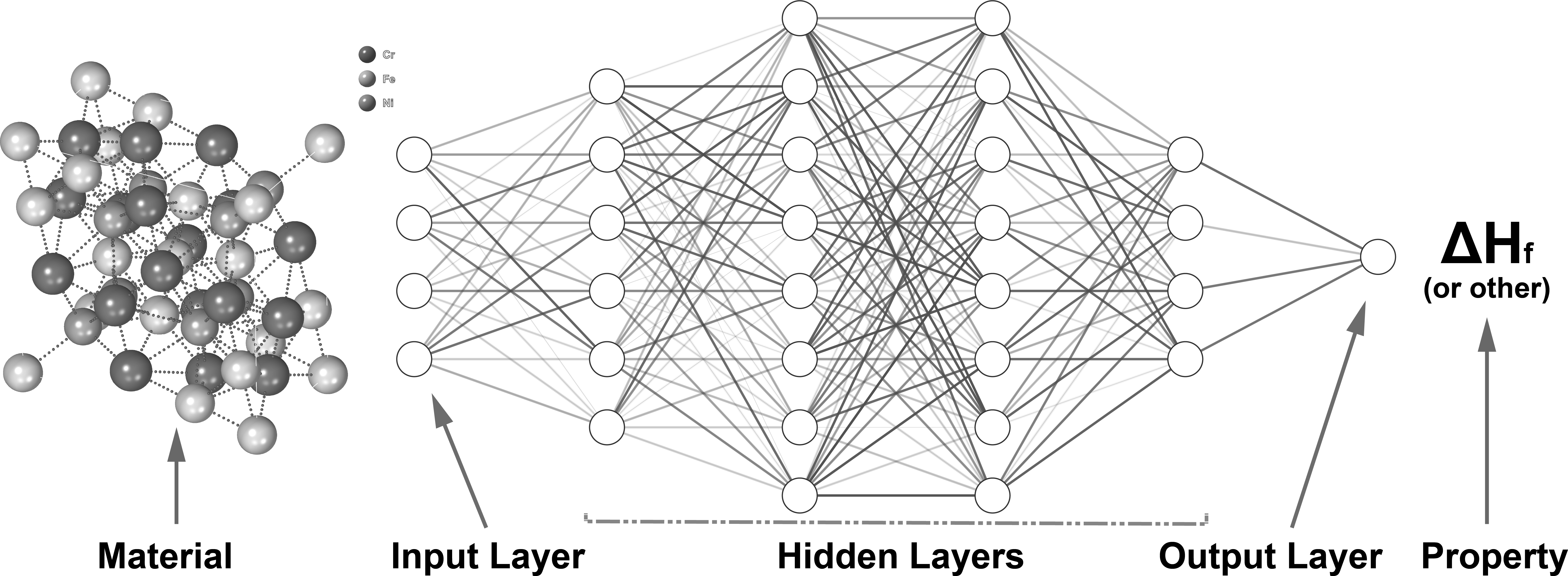}
    \caption{Simplified artificial neural network schematic}
    \label{fig:nnschematic}
\end{figure}

In this the class of functions $\mathcal{F}$ is chosen as the set of functions defined by a neural network architecture (schematic in Figure \ref{fig:nnschematic}), which leads to a deep learning method. A neural network architecture consists of a sequence of alternating linear functions and point-wise non-linear functions \cite{goodfellow2016deep}. In the figure \ref{fig:nnschematic} the nodes, or neurons, represent applications of a point-wise non-linear function, called an activation function, and the connections between nodes represent linear functions from the output of the nodes in one layer to the input of the next layer.

The class of functions represented by the neural network consists of the functions obtained by substituting different linear maps between each layer. Specifically, given weight matrices $W_1,...,W_n$ and biases $b_1,...,b_n$, which are parameters of the network, the corresponding neural network function is given by the composition

\begin{equation}
    f_{W_1,...,W_n,b_1,...,b_n}(x) = W_n\cdots\sigma(W_3\sigma(W_2\sigma(W_1x+b_1)+b_2)+b_3)\cdots + b_n
\end{equation}
where $\sigma$, called the activation function, is applied pointwise to each entry of the vector input (previous layer output). The neural network architecture is determined by the type, dimensionality, activation function $\sigma$, and arrangement of intermediate layers. This can potentially introduce some additional restrictions on the linear maps $W_i$, see for instance convolutional neural networks, where the linear maps $W_i$ are restricted to be convolutions with small kernels \cite{krizhevsky2012imagenet,lecun1995comparison,lecun1998gradient}.

Once the neural network architecture has been set, one must fit the values of the parameters $W_1,...,W_n$ and $b_1,...,b_n$ by optimizing the training loss $L$,

\begin{equation}
    \arg\min_{W_1,...,W_n,b_1,...,b_n} L(f_{W_1,...,W_n,b_1,...,b_n}).
\end{equation}

This optimization problem is typically solved using stochastic gradient descent \cite{lecun1998gradient}, or a more robust method such as ADAM \cite{kingma2014adam}, which was used in the present work. To solve the problem faster and to mitigate overfitting, which is discussed in the next sections, these methods form an estimate of the loss function gradient by considering a small subset of the data, called a batch. Each training step is done over all of the data in the batch, so parameters ($w$ and $b$) are updated based on many data points, rather than a single one. Most of the models created in the present work used a batch size of 2,048 data points.

This methodology has been successfully applied to a variety of practical machine learning problems \cite{krizhevsky2012imagenet,goodfellow2013multi,dahl2011context}. Specifically relevant to the present work, neural networks have been applied to problems in computational materials science \cite{Huang2019Machine-learningAlloys,Feng2019UsingDefects}. For example, in \cite{Huang2019Machine-learningAlloys} neural networks are used to classify the phases of high-entropy alloys. For this application, their neural network models compare favorably to other machine learning algorithms such as $k$-nearest neighbor (KNN) and support vector machines (SVM). Furthermore, in \cite{Feng2019UsingDefects} it is shown that even when training on small datasets which are typical of certain materials science problems, specifically in the prediction of solidification defects from optical microscopy data, deep neural networks can achieve better performance than other machine learning models. This is enabled by using a stacked auto-encoder (shallow neural network) to pre-train the deep neural network, whose weights are then fine-tuned on the small dataset. the present work complements these studies by applying deep neural networks to the prediction of thermodynamic quantities from atomic structure descriptors.

\subsection{Overfitting and its Mitigation}
\label{ssec:overfitting}

\begin{figure}[h]
    \centering
    \includegraphics[width=0.5\textwidth]{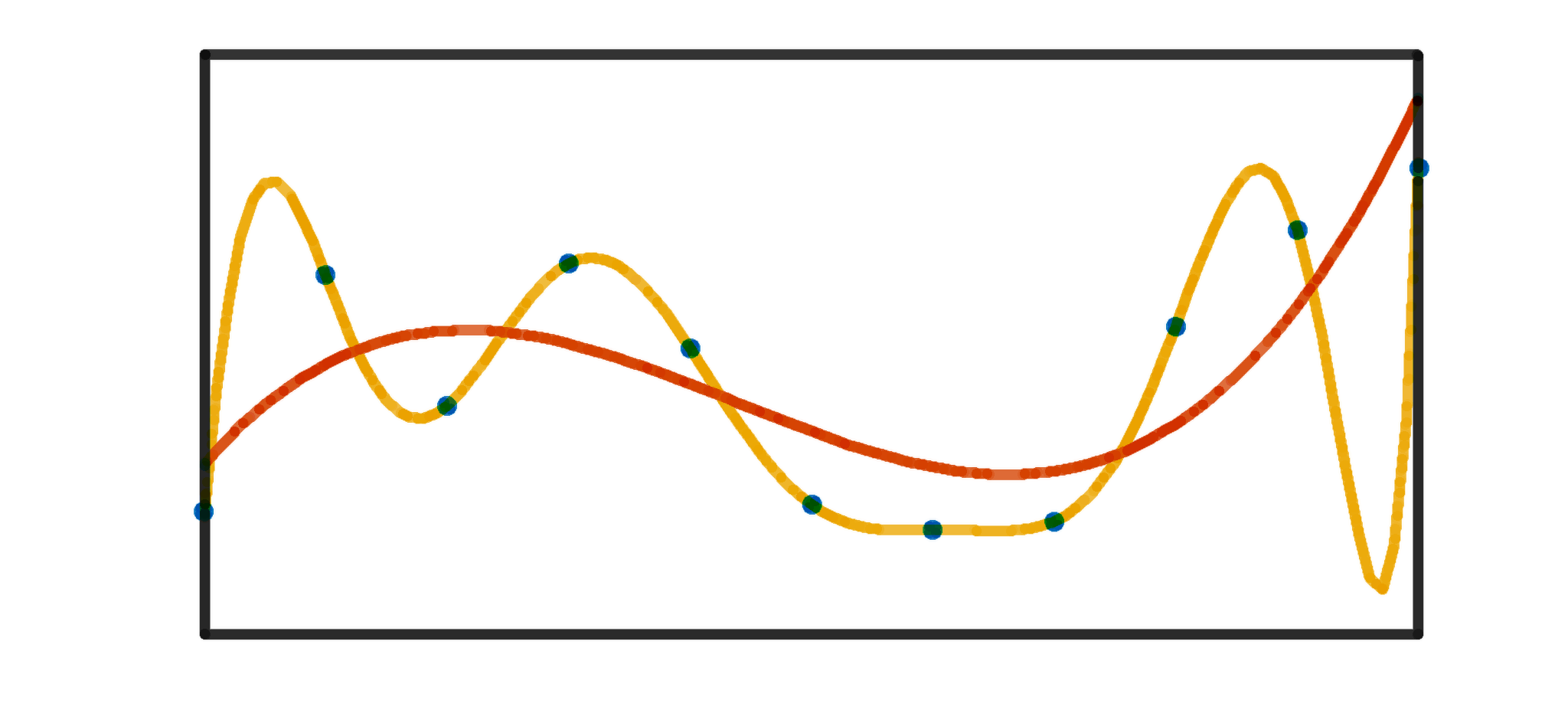}
    \caption{A schematic of overfitting. The overfit model (yellow) is too complex and memorizes the training data. This results in very low training error, but also very poor performance when predicting hidden data (test error) that follows the underlying phenomena (red).}
    \label{fig:overfitting}
\end{figure}

A major problem in statistical learning is avoiding overfitting \cite{hastie2009elements}, which, in simple terms, signifies that the model memorizes the training data instead of learning the true  relationship between descriptors $x$ and predictions $y$. This occurs when the class of functions $\mathcal{F}$ is too large, and at the optimal function $f^*$ in \eqref{empirical_risk_min_eq} the empirical \eqref{empirical_risk_app} and true risk \eqref{true_risk} diverge sharply. This results in very low training error, but poor performance on data that was not presented to the network.

Overfitting is typically detected by separating the training data into two sets, the data used in \eqref{empirical_risk_min_eq} to learn the function $f^*$, called the training data, and a separate set of data used to evaluate the performance of $f^*$, called the validation set. Consequently, in addition to the training loss in \eqref{empirical_risk_min_eq}, the validation error

\begin{equation}\label{validation_loss}
    L_{val} = \frac{1}{m}\displaystyle\sum_{i=1}^m l(\tilde{y}_i, f(\tilde{x}_i)),
\end{equation}
where $(\tilde{y}_i,\tilde{x}_i)$ for $i=1,...,m$ is the validation set, which was not presented to the network when adjusting its parameters, is used to detect overfitting. 
The fraction of the data set aside for validation set should be large enough to be representative of the whole dataset to provide statistically significant conclusions, yet small enough so that knowledge loss in the process is minimized. In the present work, a randomly selected 15\% of every dataset has been used as validation sets for all training. This corresponded to 65,300 data points in the case of the OQDM dataset described in \ref{sssec:Data}.

\begin{figure}{H}
    \centering
    \includegraphics[width=0.65\textwidth]{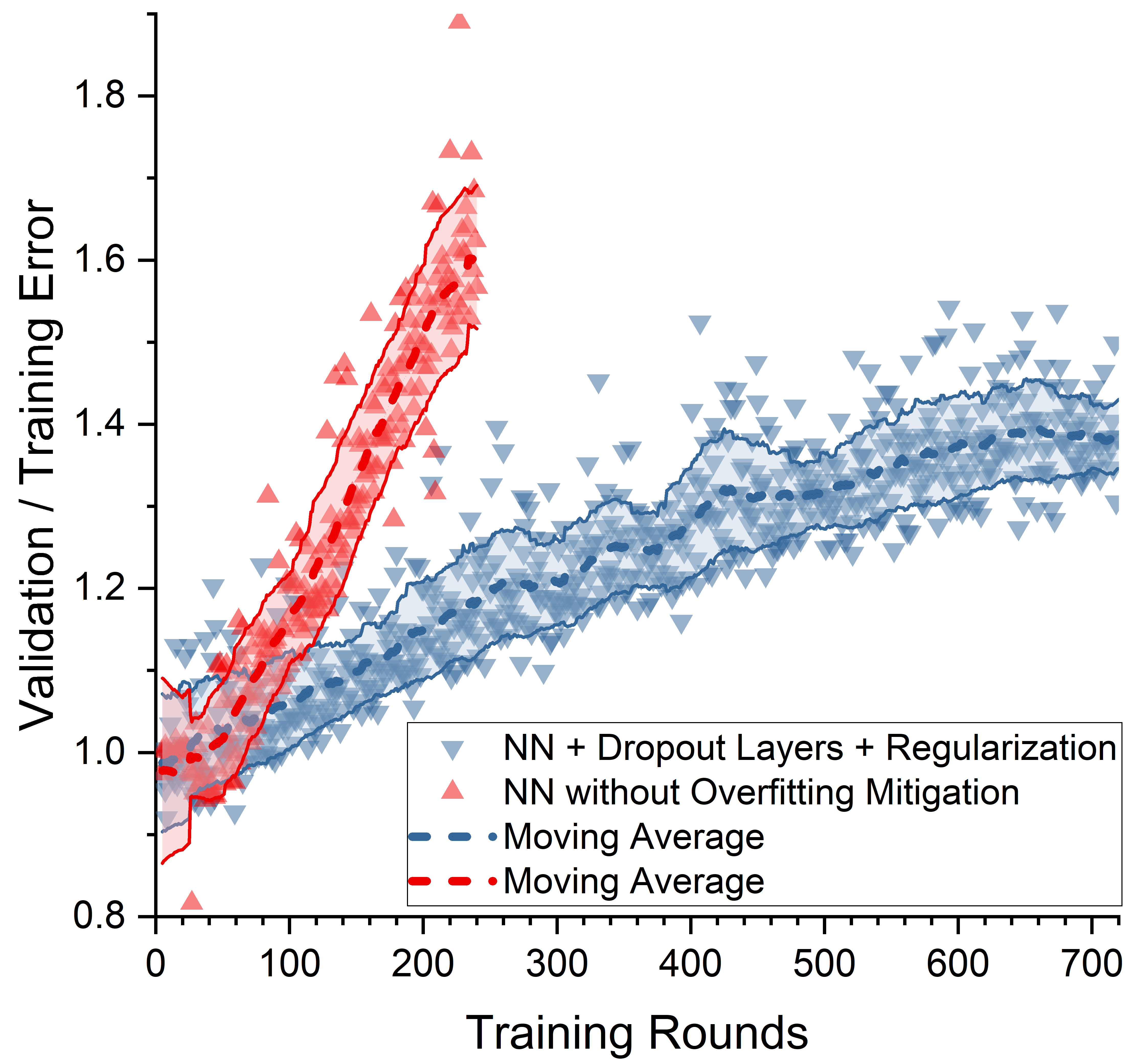}
    \caption{Training Loss to Validation Loss in a model that does without (NN9) and with overfitting mitigation (NN20), plotted versus training progress.}
    \label{fig:trainingvalidation}
\end{figure}

Typically, the validation loss will be greater than the training loss, as the validation set is not available for training. This is illustrated in Figure \ref{fig:trainingvalidation}, where the ratio between the validation loss \eqref{validation_loss} and test loss \eqref{empirical_risk_min_eq} during the course of two trainings of similar NN architectures on the same data with the same learning rate schedule has been plotted. This figure indicates that as the training proceeds, the gap between the training and validation errors widens and then increases. The size of this gap is an estimated measure of how much the model has overfitted to the data. In one of the models in this figure, extensive techniques to mitigate overfitting have been used, and for this model, the figure shows that the rate at which the model overfits to the data is much lower. At the same time both models exhibit similar performance on the test set.

There are numerous techniques used to prevent the issue of overfitting \cite{hastie2009elements,everitt2002cambridge}. These include utilization of a regularization term $\lambda R(\theta)$ added to the training error \eqref{empirical_risk_min_eq} to give the regularized empirical loss function

\begin{equation}
    f^* = \arg\min_{f\in \mathcal{F}} R_{emp}(f) + \lambda R(\theta).
\end{equation}

A standard regularizer typically added to the linear regression is the $\ell^2$-norm $R(\theta) = \|\theta\|_2^2$, which is often called Tikhonov regularization \cite{tikhonov1963solution} or ridge regression \cite{hoerl1970ridge}. The $\ell^2$-norm is also a popular regularizer in deep learning problems, where it is referred to as weight decay \cite{goodfellow2016deep}. In the context of the present work, it is implemented as a part of the training process, rather than network architecture, and causes rejection of some features in the descriptor that are not contributing to pattern recognition. Results of its implementation are shown throughout Section \ref{sssec:NetDesign}.

Another important method used to prevent overfitting in machine learning is the Dropout technique \cite{srivastava2014dropout}. The concept behind Dropout is to prevent neurons in the network from becoming overly dependent on the output from a specific neuron in the previous layer, often referred to as hard-wiring neuron paths. A Dropout layer, placed within a neural network, is implemented as a function operating during the training process and randomly discarding a specified fraction $p$ of previous layer outputs and multiplying the remaining values by $1/(1-p)$. This forces the pattern recognition ability to be dispersed across the network, as during evaluation of every training step, a random part of the network is acting as if it was not gone. Once the training is completed, all Dropout layers are deactivated and simply pass all information forward, so that the model returns to its deterministic character.

In the experiments performed in the present work, as later discussed in \ref{sssec:NetDesign}, both Dropout and weight decay were used to mitigate overfitting, with good effects shown in particular in Figure \ref{fig:trainingvalidation}.

Methods for avoiding overfitting typically come with one or more "hyperparameters" (i.e. parameters which control the training process) that can represent how much confidence is given to the training data versus prior knowledge. For instance, if a regularizer is used, the strength of the regularizer, $\lambda$, would be a hyperparameter. In the terms of the present work, it generally corresponds to how many features in the material descriptor can be considered non-essential to making predictions and therefore discarded systematically throughout the training. Furthermore, when using Dropout, the probability $p$ is also a hyperparameter. 

One typically trains the model on the training dataset using a number of different hyperparameters and then subsequently chooses the best set of them using the validation error. This allows the determination of hyperparameter values that are appropriate to the problem at hand. However, in order to ensure that the determined hyperparameter values are not overly specific to the validation set, the final accuracy of the model is evaluated on a test set that was not used at all during training \cite{hastie2009elements}.

\begin{figure}{h}
    \centering
    \includegraphics[width=0.65\textwidth]{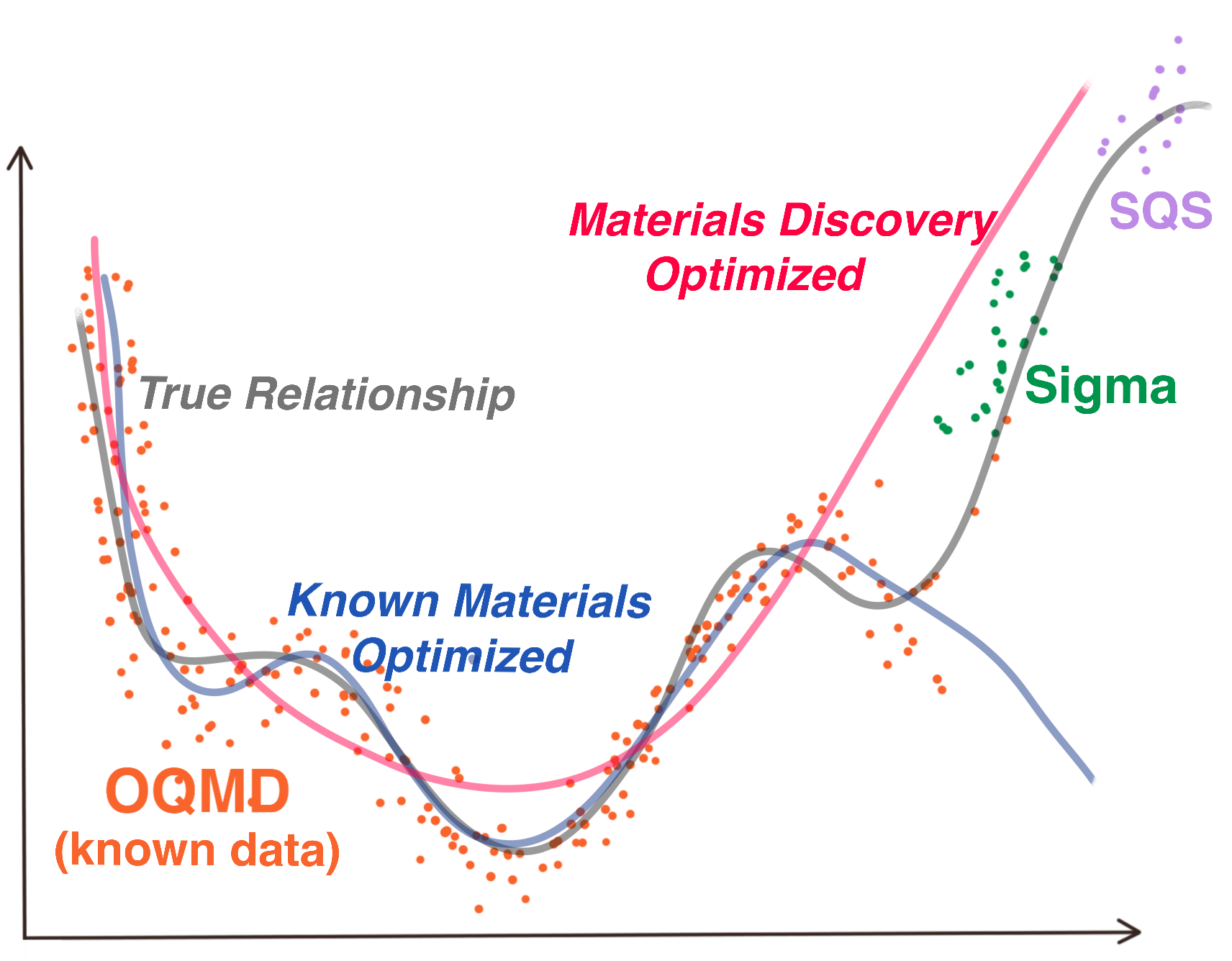}
    \caption{A conceptual drawing depicting how overfitting mitigation effort can improve performance beyond regions with high known data density.}
    \label{fig:overfitting_newregions}
\end{figure}

An additional advantage of mitigating overfitting to known data can be increased performance during extrapolation, as depicted conceptually in Figure \ref{fig:overfitting_newregions}. This is thanks to reduced model complexity, which forces recognition of stronger and more broadly exhibited patterns rather than small deviations present in the training data, whether real or due to noise, that can significantly degrade the extrapolation capability of the ML model. It is important to recognize that cost of such model simplification is often reduced performance on previously unseen data that lays within the known region.

\subsection{Transfer Learning} \label{ssec:transferlearning}
Finally, one should consider the technique of transfer learning, which has been observed among deep learning models across a variety of domains \cite{tan2018survey,cirecsan2012transfer,chang2017unsupervised,george2018deep}. Transfer learning refers the to the ability of properly trained deep learning models to `transfer' their knowledge to related tasks. In the least complex approach, one does this by simply `fine-tuning' the parameters of the model using new training data (from the new task). This has to be done using a small learning rate and a small number of iterations on a loss function defined by the new training data. It has been observed that this often produces accurate results on the new task for a relatively small amount of additional data. 

As an illustrative example, in \cite{cirecsan2012transfer}, a network is first trained to recognize lower case handwritten characters. It is then shown that with minimal `fine-tuning,' such a network can be made to accurately recognize upper case characters. The same phenomenon was also observed with a network that was first trained to recognize Chinese characters. Considering that this behavior has been widely observed \cite{tan2018survey,chang2017unsupervised,george2018deep}, this shows that deep neural networks are often able to transfer knowledge between different but related tasks. 

the present work adds to this evidence by showing that a network trained on the knowledge from the OQMD database covering a broad yet limited spectrum of material, can be easily adjusted to materials outside this spectrum with very little cost relative to the initial training. Specifically, the set of all (243) Fe-Ni-Cr $\sigma$-phase endmembers, described in \ref{sssec:Data}, is shown in \ref{ssec:transferlearningresults} to require transfer of only a few examples from that set to dramatically improve model performance on the rest.

\pagebreak

\section{Intermediate Neural Network Models} \label{appendix2}

\begin{wrapfigure}{R}{0.4\textwidth}
    \centering
    \includegraphics[width=0.3\textwidth]{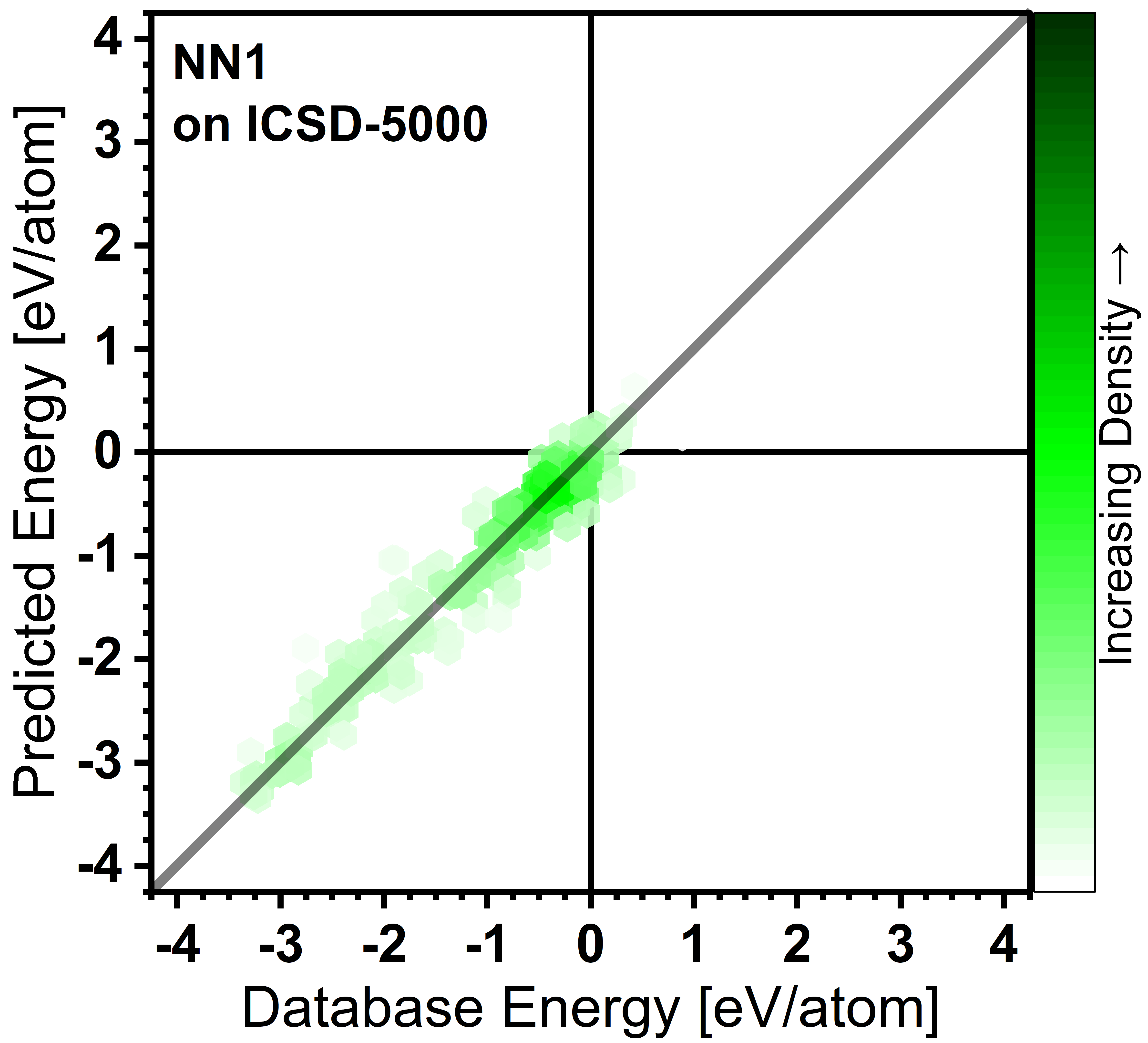}
    \caption{Test of perceptron trained on the data from the first 5000 entries in the ICSD dataset and evaluated on the test set of 230 randomly selected entries ($\approx5\%$)}
    \vspace{-12pt}
    \label{fig:nn1performance}
\end{wrapfigure}

The neural network design process was conducted in incremental fashion, starting from a perceptron, which is the simplest type of neural network proposed by Frank Rosenblatt in 1957 \cite{Rosenblatt1957TheAutomaton}. It effectively operates as a linear function $f(\vec{d}) = A(w_1 d_1 + w_2 d_2 + ... + w_n d_n)$ where $d_i$ is i-th element of the descriptor $\vec{d}$, $w_i$ is the weight associated with it, and $A$ is an activation function that can introduce non-linearity or turn it into a classifier. Here, the popular Sigmoid activation function was used. 

\begin{wrapfigure}{R}{0.55\textwidth}
    \centering
    \includegraphics[width=0.53\textwidth]{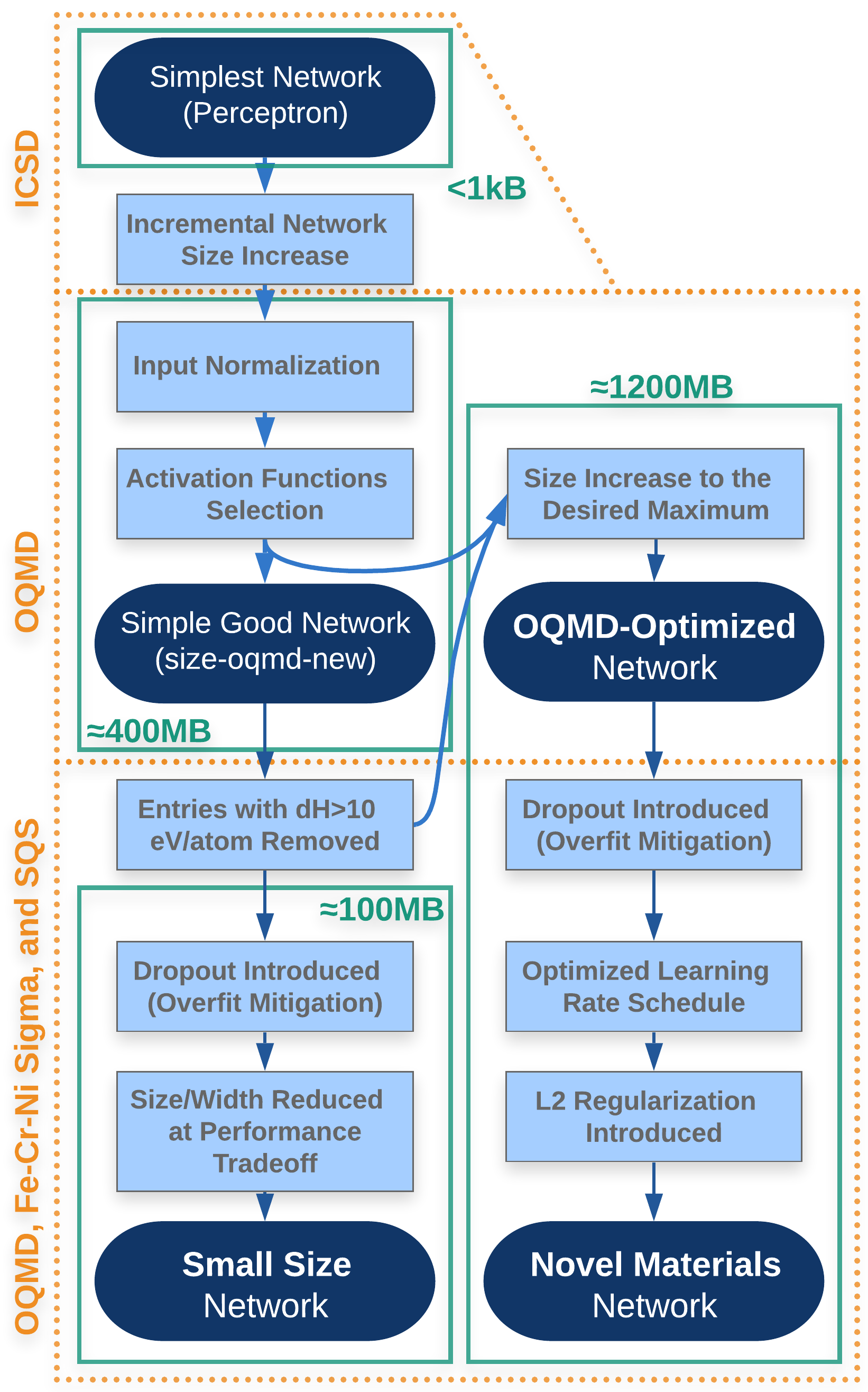}
    \caption{The network design process schematic leading to the three final models.}
\end{wrapfigure}

The perceptron was first trained on the data from the first 5000 entries in the ICSD, to check whether the training was set up correctly. It achieved a MAE of 195 meV/atom on the test set of 230 randomly selected entries ($\approx 5\% \text{ from } 5000$). Results are shown in Figure \ref{fig:nn1performance}. When trained on the data from all entries in the ICSD, it achieved an MAE of 364 meV/atom on the test set ($\approx5\% \text{ from } 32116$). This error is comparable to the performance of a random forest model based on PRDF (370 meV/atom), is slightly worse than a CM (250 meV/atom), and is significantly worse than a random-forest model trained on the same descriptor (90 meV/atom), as reported by Ward et al. \cite{Ward2017IncludingTessellations}. Part of the significance of these results is the evident quality of the descriptor, as the model achieved performance that would be considered excellent just a few years prior to the present work while being much less complex and computationally costly. Furthermore, it is important to note the time- and space-complexity of the perceptron model. Training the final network took less than 8 seconds compared to around 10,000 seconds reported for the aforementioned random-forest methods, and the resulting model occupied less than 1kb of memory. Following the testing of a perceptron, which allowed rough estimation of the a good size of the network (i.e. number of weights), the design of the actual architecture began. All of these steps are schematically depicted in Figure \ref{fig:designprocess}.

Next, in a few steps, the size of the network was incrementally increased. First, a layer of 1000 neurons was introduced. This reduced the performance on the first 5000 entries in the ICSD, likely due to overfitting issues, as the data was very limited. Performance on the ICSD was improved, reducing the test MAE to 305 meV/atom on the test set, however. The introduction of the next two 1000-width layers further reduced the MAE to 215 meV/atom. Based on these results, it was estimated that introducing 4 hidden layers with Sigmoid activation function and widths of 10000, 10000, 1000, and 100 would provide good results when trained on the much larger OQMD.

After switching to OQMD, the network exhibited issues with convergence, often predicting a single value for all of the entries. To mitigate this, the descriptor (i.e. network input) was normalized by dividing every element by its maximum value across the whole dataset. This solved the issue. Next, to improve the training behavior, the activation functions were changed from only the Sigmoid function to a mix of Soft Sign, Exponential Linear Unit, and Sigmoid, which was found to work well. These steps improved both the predictive performance and reduced the time required to converge. The network architecture resulting from these steps (internally designated NN8 / Simple Good Network in Figure \ref{fig:designprocess}) was the first to improve performance compared to the Ward et. al approach \cite{Ward2017IncludingTessellations}, achieving an MAE of 42 meV/atom on the test set of random subset 5\% of the OQMD dataset. When testing this network, a small fraction of around 0.03\% of likely incorrect entries in the OQMD was found, as described in \ref{sssec:Data}, and was removed from the dataset used later in the design process.

Once a network with desired performance was obtained, the network size was increased until it either exceeded 1GB or showed signs of extensive overfitting. At the first step of this process, two layers of width 10,000 were added, resulting in a network size of 1.2GB and reduced overfitting, as indicated by the ratio of validation-to-training error lowered from 2.2 to 1.6, relative to NN8. The resulting network (internally designated NN9 / OQMD-Optimized Network in Figure \ref{fig:designprocess}), achieved an MAE of 28 meV/atom on the test set of random subset 5\% of OQMD, which was the best performance on OQMD out of all the networks created in this project. If the 0.03\% of abnormal data wasn't removed as described in \ref{sssec:Data}, it would correspond to, on average, 6 data points which in one tested instance increased the MAE to 35 meV/atom. Important to point out, the training of this network was prone to staying in local minima at the beginning. The reported instance of the trained network exhibited no training progress between around rounds 5 and 25, after which it's performance quickly increased.  Detailed analysis of the performance is given in \ref{ssec:oqmdperformance}.

Once the main objective of the design process was obtained, i.e. the performance on the OQMD has improved appreciably beyond existing methods, the design process was focused on creating a tool for modeling materials that were not reported in the OQMD. Therefore, the objective changed from achieving the lowest MAE on a random subset 5\% of OQMD to (1) reducing the mismatch between training and validation sets errors (i.e. difference between training accuracy and validation accuracy) during the training process, (2) keeping the test MAE on the OQMD below 50 meV/atom, and (3) improving performance on two material groups significantly different from the OQMD data, namely Special Quasirandom Structures (SQS) and Fe-Cr-Ni $\sigma$-phase (see \ref{sssec:Data}).

\begin{wrapfigure}{R}{0.37\textwidth}
\centering
    \includegraphics[width=0.35\textwidth]{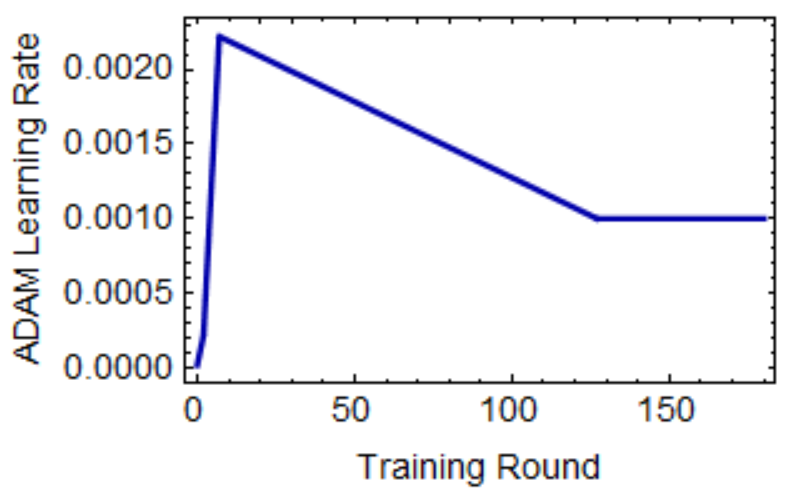}
    \caption{The learning rate schedule used for training of more complex networks in the later stage of the design process (e.g., NN18).}
    \label{fig:learningrate}
\end{wrapfigure}

With these new objectives, two Dropout layers in the middle part of the network were introduced to promote the distribution of pattern recognition abilities across the network. \cite{Srivastava2014Dropout:Overfitting} This introduced a problem with convergence as the network became more likely to fall into local minima at the initial stages of the training, which was solved by introducing custom learning rate schedules. Specifically, the learning rate was initially set to a value orders of magnitude lower than during the default initial training and then ramped up to the previous (ADAM default setting in the majority of frameworks) learning rate of 0.001 (or above) after around 2 rounds of training. This type of learning rate schedule is known as a warm-up in the deep learning literature \cite{gotmare2018closer}. The schedule found to perform the best is presented in Figure \ref{fig:learningrate}.

The next step was the introduction of $\ell^2$ regularization, which is a technique that favors simplification of the descriptor and effectively rejects features of the descriptor that do not contribute to prediction performance \cite{L2Regularization}. An overview on it is given in Section \ref{ref:machinelearningoverview}. In the models reported in the present work an $\ell^2$ value of $10^{-6}$ was used. Higher values were found to stop the training at early stages, impairing the pattern recognition, or in extreme cases (above $10^{-3}$) force the network to discard the input completely, resulting in constant or near-constant output (i.e. mean value from the training dataset predicted for any structure).

The final step was small curation of the training data based on the OQMD-reported structure stability, i.e. the energy difference between the formation energy and the energy convex hull. The motivation for that was the notion that DFT results are inherently less accurate for unstable phases. In this step, all entries with energies of more than 2000 meV/atom above the convex hull were removed from the training set. Importantly, the validation and testing sets were not modified for consistent performance reporting.

All of these changes resulted in a neural network that has been optimized for predicting new materials. In the code and Supplementary materials, it is designated as NN20 (Novel Materials Network in Figure \ref{fig:designprocess}). Compared to the OQMD-optimized network it was derived from, the test MAE on the OQMD increased from 28 to 49 meV/atom. However, at the same time, the mismatch between the training and validation set was reduced from 1.57 to 1.38. Or, as presented earlier in Figure \ref{fig:trainingvalidation}, reduced to about 1.15 for the same training duration. Furthermore, a relatively large portion of this error can be attributed to some unstable structures that were removed from the training set, but not from the test set. Once entries with formation energies of more than 1000 meV/atom above the convex hull were removed, the test MAE decreased to only 38 meV/atom. Restricting the test set further to only somewhat stable structures (stability below 250 meV/atom) resulted in an MAE of 30 meV/atom.

While the new-material-optimized network presented an increased MAE across a random subset of the OQMD, performance has significantly improved on the Fe-Cr-Ni $\sigma-$phase described in \ref{sssec:Data}. The MAE has decreased from 55 to 41 meV/atom, indicating that the model based on this neural network is more capable of making predictions for new materials.

Once two performance-oriented models were developed, increasing the performance-to-cost ratio has been explored with the motivation that some studies would benefit from many times higher throughput at minor accuracy decrease. Architecture design started from the selection of a network with a balanced size-to-performance ratio (NN8) and the introduction of an overfitting mitigation technique (Dropout \cite{srivastava2014dropout}) used for the network optimized for new materials, as depicted in Figure \ref{sssec:NetDesign}. Next, the network was gradually narrowed (fewer neurons in layers) until the performance started to noticeably deteriorate (41.9 meV/atom for 5000- and 4000-width vs 42.1 for 3000-width). This approach allowed a significant reduction of the network size (and the computational intensity to run it) from around 1,200MB of the two other models to around 145MB. If an application demands even more of a reduction in model size and computational cost, the same procedure could be continued until some minimum required performance is retained. 

\section{Feature Ranking } \label{appendix3}

\begin{longtable}{|l|l|}
\hline
\multicolumn{1}{|c|}{\textbf{Descriptor Feature}} & \multicolumn{1}{c|}{\textbf{Normalized Squared Weights Sum}} \\ \hline
\endfirsthead
\endhead
mean\_NeighDiff\_shell1\_MeltingT & 1 \\ \hline
mean\_MeltingT & 0.97502 \\ \hline
max\_MeltingT & 0.73512 \\ \hline
mean\_NeighDiff\_shell1\_NdUnfilled & 0.69157 \\ \hline
MaxPackingEfficiency & 0.68889 \\ \hline
most\_MeltingT & 0.67373 \\ \hline
dev\_GSvolume\_pa & 0.61042 \\ \hline
var\_NeighDiff\_shell1\_Column & 0.58782 \\ \hline
var\_NeighDiff\_shell1\_CovalentRadius & 0.57826 \\ \hline
var\_NeighDiff\_shell1\_MeltingT & 0.57259 \\ \hline
maxdiff\_GSvolume\_pa & 0.55156 \\ \hline
dev\_MeltingT & 0.5286 \\ \hline
mean\_SpaceGroupNumber & 0.51761 \\ \hline
min\_MeltingT & 0.50437 \\ \hline
var\_CellVolume & 0.49467 \\ \hline
var\_NeighDiff\_shell1\_MendeleevNumber & 0.492 \\ \hline
min\_NeighDiff\_shell1\_MeltingT & 0.47853 \\ \hline
mean\_NeighDiff\_shell1\_Column & 0.45566 \\ \hline
maxdiff\_CovalentRadius & 0.42998 \\ \hline
var\_NeighDiff\_shell1\_Electronegativity & 0.42642 \\ \hline
var\_EffectiveCoordination & 0.40506 \\ \hline
min\_NeighDiff\_shell1\_Column & 0.39822 \\ \hline
dev\_NdUnfilled & 0.39739 \\ \hline
dev\_CovalentRadius & 0.36935 \\ \hline
range\_NeighDiff\_shell1\_Column & 0.35956 \\ \hline
range\_NeighDiff\_shell1\_CovalentRadius & 0.34585 \\ \hline
mean\_WCMagnitude\_Shell1 & 0.34275 \\ \hline
mean\_NeighDiff\_shell1\_MendeleevNumber & 0.33911 \\ \hline
mean\_EffectiveCoordination & 0.33899 \\ \hline
mean\_Number & 0.33769 \\ \hline
mean\_NdUnfilled & 0.33408 \\ \hline
maxdiff\_MeltingT & 0.33348 \\ \hline
mean\_AtomicWeight & 0.33149 \\ \hline
mean\_NeighDiff\_shell1\_NdValence & 0.33142 \\ \hline
range\_NeighDiff\_shell1\_MeltingT & 0.33107 \\ \hline
max\_NfUnfilled & 0.33041 \\ \hline
dev\_Electronegativity & 0.33001 \\ \hline
mean\_NeighDiff\_shell1\_CovalentRadius & 0.32999 \\ \hline
var\_NeighDiff\_shell1\_NdUnfilled & 0.31973 \\ \hline
dev\_Column & 0.31662 \\ \hline
var\_NeighDiff\_shell1\_NdValence & 0.31481 \\ \hline
mean\_WCMagnitude\_Shell2 & 0.31359 \\ \hline
most\_NfUnfilled & 0.30916 \\ \hline
MeanIonicChar & 0.30732 \\ \hline
mean\_NeighDiff\_shell1\_Electronegativity & 0.30277 \\ \hline
min\_EffectiveCoordination & 0.29705 \\ \hline
min\_NeighDiff\_shell1\_CovalentRadius & 0.29392 \\ \hline
max\_NeighDiff\_shell1\_GSvolume\_pa & 0.2875 \\ \hline
most\_SpaceGroupNumber & 0.28472 \\ \hline
max\_NdUnfilled & 0.28424 \\ \hline
maxdiff\_NdUnfilled & 0.28405 \\ \hline
var\_NeighDiff\_shell1\_GSvolume\_pa & 0.28008 \\ \hline
min\_BondLengthVariation & 0.27922 \\ \hline
var\_MeanBondLength & 0.2768 \\ \hline
dev\_NdValence & 0.27566 \\ \hline
max\_NeighDiff\_shell1\_MeltingT & 0.27097 \\ \hline
max\_BondLengthVariation & 0.26565 \\ \hline
mean\_NfValence & 0.26558 \\ \hline
mean\_NsUnfilled & 0.2612 \\ \hline
max\_NeighDiff\_shell1\_CovalentRadius & 0.26026 \\ \hline
max\_GSvolume\_pa & 0.25985 \\ \hline
min\_GSvolume\_pa & 0.25895 \\ \hline
mean\_NdValence & 0.25573 \\ \hline
mean\_NeighDiff\_shell1\_GSvolume\_pa & 0.25299 \\ \hline
max\_NValance & 0.24749 \\ \hline
range\_NeighDiff\_shell1\_NdUnfilled & 0.24643 \\ \hline
max\_CovalentRadius & 0.23136 \\ \hline
CanFormIonic & 0.23135 \\ \hline
min\_NeighDiff\_shell1\_Electronegativity & 0.22873 \\ \hline
min\_SpaceGroupNumber & 0.22766 \\ \hline
max\_Electronegativity & 0.22609 \\ \hline
max\_NdValence & 0.22576 \\ \hline
most\_NdUnfilled & 0.22198 \\ \hline
min\_NeighDiff\_shell1\_MendeleevNumber & 0.21991 \\ \hline
var\_NeighDiff\_shell1\_NpValence & 0.21609 \\ \hline
min\_NeighDiff\_shell1\_NdUnfilled & 0.2114 \\ \hline
dev\_SpaceGroupNumber & 0.2099 \\ \hline
most\_NfValence & 0.20888 \\ \hline
min\_MeanBondLength & 0.2086 \\ \hline
mean\_BondLengthVariation & 0.20507 \\ \hline
var\_NeighDiff\_shell1\_Row & 0.20454 \\ \hline
max\_NeighDiff\_shell1\_NdUnfilled & 0.20318 \\ \hline
min\_NeighDiff\_shell1\_NdValence & 0.20123 \\ \hline
min\_CovalentRadius & 0.19974 \\ \hline
range\_NeighDiff\_shell1\_MendeleevNumber & 0.19591 \\ \hline
min\_NeighDiff\_shell1\_GSvolume\_pa & 0.19565 \\ \hline
most\_NpUnfilled & 0.19457 \\ \hline
maxdiff\_NUnfilled & 0.19316 \\ \hline
max\_NeighDiff\_shell1\_NdValence & 0.19307 \\ \hline
max\_NpValence & 0.1929 \\ \hline
range\_NeighDiff\_shell1\_GSvolume\_pa & 0.19166 \\ \hline
most\_NdValence & 0.1904 \\ \hline
max\_MeanBondLength & 0.19021 \\ \hline
maxdiff\_NfUnfilled & 0.18897 \\ \hline
max\_NeighDiff\_shell1\_Column & 0.18518 \\ \hline
range\_NeighDiff\_shell1\_Electronegativity & 0.18322 \\ \hline
var\_NeighDiff\_shell1\_SpaceGroupNumber & 0.18313 \\ \hline
dev\_NpValence & 0.18099 \\ \hline
mean\_NpUnfilled & 0.18091 \\ \hline
range\_NeighDiff\_shell1\_SpaceGroupNumber & 0.17858 \\ \hline
dev\_MendeleevNumber & 0.17753 \\ \hline
MaxIonicChar & 0.176 \\ \hline
mean\_Column & 0.17206 \\ \hline
min\_Electronegativity & 0.17164 \\ \hline
mean\_WCMagnitude\_Shell3 & 0.17077 \\ \hline
mean\_Row & 0.17035 \\ \hline
min\_NeighDiff\_shell1\_SpaceGroupNumber & 0.17031 \\ \hline
most\_NsUnfilled & 0.16714 \\ \hline
var\_BondLengthVariation & 0.16653 \\ \hline
var\_NeighDiff\_shell1\_NfUnfilled & 0.16223 \\ \hline
range\_NeighDiff\_shell1\_NdValence & 0.16094 \\ \hline
frac\_fValence & 0.1609 \\ \hline
maxdiff\_Column & 0.16083 \\ \hline
max\_NUnfilled & 0.15916 \\ \hline
mean\_NpValence & 0.15639 \\ \hline
maxdiff\_NpValence & 0.15637 \\ \hline
mean\_MendeleevNumber & 0.15491 \\ \hline
most\_Electronegativity & 0.15469 \\ \hline
mean\_Electronegativity & 0.15458 \\ \hline
max\_SpaceGroupNumber & 0.15429 \\ \hline
dev\_Row & 0.15382 \\ \hline
maxdiff\_MendeleevNumber & 0.15373 \\ \hline
var\_NeighDiff\_shell1\_NpUnfilled & 0.15135 \\ \hline
max\_NeighDiff\_shell1\_Electronegativity & 0.15115 \\ \hline
most\_NUnfilled & 0.14955 \\ \hline
max\_GSbandgap & 0.14945 \\ \hline
mean\_NeighDiff\_shell1\_NUnfilled & 0.14891 \\ \hline
maxdiff\_NValance & 0.14819 \\ \hline
mean\_NeighDiff\_shell1\_NpValence & 0.14768 \\ \hline
maxdiff\_NdValence & 0.14735 \\ \hline
max\_NpUnfilled & 0.14647 \\ \hline
maxdiff\_Electronegativity & 0.14523 \\ \hline
min\_MendeleevNumber & 0.14119 \\ \hline
mean\_CovalentRadius & 0.14049 \\ \hline
mean\_NeighDiff\_shell1\_Row & 0.13945 \\ \hline
maxdiff\_GSbandgap & 0.13891 \\ \hline
max\_NeighDiff\_shell1\_MendeleevNumber & 0.13858 \\ \hline
most\_Number & 0.13823 \\ \hline
most\_AtomicWeight & 0.13798 \\ \hline
max\_NeighDiff\_shell1\_NpValence & 0.13757 \\ \hline
Comp\_L10Norm & 0.13598 \\ \hline
min\_Row & 0.13596 \\ \hline
range\_NeighDiff\_shell1\_NpValence & 0.13524 \\ \hline
mean\_GSvolume\_pa & 0.1331 \\ \hline
max\_NeighDiff\_shell1\_NUnfilled & 0.13205 \\ \hline
mean\_NeighDiff\_shell1\_NfValence & 0.12888 \\ \hline
min\_NeighDiff\_shell1\_NpUnfilled & 0.12778 \\ \hline
mean\_NeighDiff\_shell1\_SpaceGroupNumber & 0.12722 \\ \hline
mean\_NsValence & 0.12642 \\ \hline
most\_CovalentRadius & 0.12616 \\ \hline
var\_NeighDiff\_shell1\_NUnfilled & 0.12525 \\ \hline
mean\_NeighDiff\_shell1\_Number & 0.12466 \\ \hline
Comp\_L7Norm & 0.12293 \\ \hline
mean\_NeighDiff\_shell1\_AtomicWeight & 0.12229 \\ \hline
min\_NeighDiff\_shell1\_NpValence & 0.12026 \\ \hline
max\_EffectiveCoordination & 0.11995 \\ \hline
min\_NdValence & 0.11984 \\ \hline
maxdiff\_NpUnfilled & 0.11976 \\ \hline
mean\_NeighDiff\_shell1\_NsUnfilled & 0.11836 \\ \hline
max\_NeighDiff\_shell1\_GSbandgap & 0.11657 \\ \hline
min\_NUnfilled & 0.11648 \\ \hline
most\_Column & 0.1164 \\ \hline
var\_NeighDiff\_shell1\_Number & 0.11483 \\ \hline
most\_MendeleevNumber & 0.11312 \\ \hline
max\_NeighDiff\_shell1\_SpaceGroupNumber & 0.11292 \\ \hline
var\_NeighDiff\_shell1\_AtomicWeight & 0.11234 \\ \hline
most\_NpValence & 0.11231 \\ \hline
frac\_dValence & 0.11126 \\ \hline
NComp & 0.11097 \\ \hline
min\_Number & 0.11062 \\ \hline
range\_NeighDiff\_shell1\_NpUnfilled & 0.11002 \\ \hline
dev\_NValance & 0.10868 \\ \hline
min\_Column & 0.10846 \\ \hline
max\_NeighDiff\_shell1\_NpUnfilled & 0.10837 \\ \hline
maxdiff\_Row & 0.10735 \\ \hline
Comp\_L5Norm & 0.10726 \\ \hline
mean\_NeighDiff\_shell1\_NpUnfilled & 0.10682 \\ \hline
maxdiff\_SpaceGroupNumber & 0.10604 \\ \hline
dev\_GSbandgap & 0.10604 \\ \hline
max\_AtomicWeight & 0.10495 \\ \hline
max\_GSmagmom & 0.10416 \\ \hline
maxdiff\_GSmagmom & 0.1039 \\ \hline
dev\_NUnfilled & 0.10336 \\ \hline
var\_NeighDiff\_shell1\_NfValence & 0.10059 \\ \hline
dev\_GSmagmom & 0.10046 \\ \hline
most\_GSbandgap & 0.09997 \\ \hline
var\_NeighDiff\_shell1\_NValance & 0.09842 \\ \hline
min\_NeighDiff\_shell1\_Row & 0.09798 \\ \hline
min\_NeighDiff\_shell1\_NUnfilled & 0.09563 \\ \hline
most\_Row & 0.09538 \\ \hline
max\_Number & 0.0925 \\ \hline
most\_GSvolume\_pa & 0.09166 \\ \hline
mean\_GSbandgap & 0.09097 \\ \hline
range\_NeighDiff\_shell1\_Row & 0.09081 \\ \hline
mean\_NValance & 0.0889 \\ \hline
mean\_NeighDiff\_shell1\_NsValence & 0.08449 \\ \hline
min\_NsValence & 0.08408 \\ \hline
frac\_pValence & 0.08403 \\ \hline
mean\_NUnfilled & 0.08244 \\ \hline
mean\_NfUnfilled & 0.08194 \\ \hline
dev\_NpUnfilled & 0.0818 \\ \hline
dev\_Number & 0.08065 \\ \hline
max\_NeighDiff\_shell1\_GSmagmom & 0.08049 \\ \hline
max\_Column & 0.07989 \\ \hline
min\_AtomicWeight & 0.07959 \\ \hline
Comp\_L3Norm & 0.07913 \\ \hline
max\_NeighDiff\_shell1\_Row & 0.0776 \\ \hline
mean\_NeighDiff\_shell1\_NValance & 0.07619 \\ \hline
mean\_NeighDiff\_shell1\_NfUnfilled & 0.07413 \\ \hline
range\_NeighDiff\_shell1\_NfUnfilled & 0.07381 \\ \hline
min\_NValance & 0.07297 \\ \hline
max\_NeighDiff\_shell1\_NValance & 0.0726 \\ \hline
range\_NeighDiff\_shell1\_NfValence & 0.07163 \\ \hline
min\_NdUnfilled & 0.07145 \\ \hline
most\_NsValence & 0.07114 \\ \hline
mean\_NeighDiff\_shell1\_GSbandgap & 0.06709 \\ \hline
max\_NfValence & 0.06661 \\ \hline
dev\_AtomicWeight & 0.06581 \\ \hline
maxdiff\_Number & 0.06576 \\ \hline
max\_NeighDiff\_shell1\_NfUnfilled & 0.06523 \\ \hline
dev\_NfUnfilled & 0.06477 \\ \hline
dev\_NfValence & 0.06373 \\ \hline
range\_NeighDiff\_shell1\_GSmagmom & 0.06305 \\ \hline
var\_NeighDiff\_shell1\_NsUnfilled & 0.06288 \\ \hline
min\_NeighDiff\_shell1\_Number & 0.0623 \\ \hline
frac\_sValence & 0.06099 \\ \hline
min\_NeighDiff\_shell1\_NfValence & 0.06033 \\ \hline
max\_Row & 0.05998 \\ \hline
min\_NeighDiff\_shell1\_NValance & 0.05844 \\ \hline
range\_NeighDiff\_shell1\_NUnfilled & 0.05819 \\ \hline
var\_NeighDiff\_shell1\_GSbandgap & 0.05683 \\ \hline
range\_NeighDiff\_shell1\_AtomicWeight & 0.0568 \\ \hline
Comp\_L2Norm & 0.05638 \\ \hline
min\_NeighDiff\_shell1\_NsUnfilled & 0.05541 \\ \hline
most\_NValance & 0.0553 \\ \hline
maxdiff\_NsValence & 0.05459 \\ \hline
range\_NeighDiff\_shell1\_NValance & 0.0537 \\ \hline
min\_NeighDiff\_shell1\_AtomicWeight & 0.05369 \\ \hline
max\_NsValence & 0.05329 \\ \hline
range\_NeighDiff\_shell1\_GSbandgap & 0.05299 \\ \hline
min\_NeighDiff\_shell1\_NfUnfilled & 0.05266 \\ \hline
maxdiff\_NfValence & 0.05147 \\ \hline
dev\_NsUnfilled & 0.04884 \\ \hline
max\_MendeleevNumber & 0.04844 \\ \hline
maxdiff\_AtomicWeight & 0.04814 \\ \hline
max\_NeighDiff\_shell1\_NsUnfilled & 0.04675 \\ \hline
max\_NeighDiff\_shell1\_NsValence & 0.04663 \\ \hline
var\_NeighDiff\_shell1\_GSmagmom & 0.04635 \\ \hline
range\_NeighDiff\_shell1\_Number & 0.04416 \\ \hline
max\_NeighDiff\_shell1\_NfValence & 0.04376 \\ \hline
mean\_NeighDiff\_shell1\_GSmagmom & 0.0433 \\ \hline
most\_GSmagmom & 0.04239 \\ \hline
range\_NeighDiff\_shell1\_NsUnfilled & 0.03954 \\ \hline
min\_NeighDiff\_shell1\_NsValence & 0.03932 \\ \hline
max\_NeighDiff\_shell1\_AtomicWeight & 0.03905 \\ \hline
max\_NeighDiff\_shell1\_Number & 0.03815 \\ \hline
min\_NfValence & 0.03794 \\ \hline
dev\_NsValence & 0.0373 \\ \hline
maxdiff\_NsUnfilled & 0.03558 \\ \hline
min\_NfUnfilled & 0.03537 \\ \hline
min\_NeighDiff\_shell1\_GSmagmom & 0.03353 \\ \hline
var\_NeighDiff\_shell1\_NsValence & 0.02948 \\ \hline
min\_NpValence & 0.02946 \\ \hline
max\_NsUnfilled & 0.02933 \\ \hline
min\_NeighDiff\_shell1\_GSbandgap & 0.02735 \\ \hline
mean\_GSmagmom & 0.02402 \\ \hline
min\_NpUnfilled & 0.02233 \\ \hline
range\_NeighDiff\_shell1\_NsValence & 0.02171 \\ \hline
min\_NsUnfilled & 0.02051 \\ \hline
min\_GSbandgap & 0.01299 \\ \hline
min\_GSmagmom & 0.00132 \\ \hline
\end{longtable}

\end{document}